%% file: main-manuscript.tex
\documentclass[format=acmsmall, review=false, screen=true]{acmart}
\pdfoutput=1
\AtBeginDocument{%
  \providecommand\BibTeX{{%
    \normalfont B\kern-0.5em{\scshape i\kern-0.25em b}\kern-0.8em\TeX}}}
\setcopyright{acmcopyright}
\acmJournal{TOSEM}
\acmDOI{10.1145/3488245}

\usepackage{color}
\usepackage{colortbl}
\usepackage{amsmath}
 
\usepackage{amssymb,amsfonts}
\usepackage{graphicx}
\usepackage{textcomp}
\usepackage{xcolor}
\usepackage{multirow}
\usepackage{subfigure}
\usepackage{makecell}
\usepackage{booktabs}  
\usepackage{array} 
\usepackage{balance}
\begin{document}

\title{A Survey of Large Language Models for Code: Evolution, Benchmarking, and Future Trends}
\fancyfoot{} 
        \renewcommand{\headrulewidth}{0pt} 
        \fancypagestyle{titlepage}{
        \fancyhf{} 
        \renewcommand{\headrulewidth}{0pt} 
        }

\author{Zibin Zheng}
\affiliation{%
  \institution{Sun Yat-sen University}
  \city{Zhuhai}
  \country{China}}
\email{zhzibin@mail.sysu.edu.cn}

\author{Kaiwen Ning}
\affiliation{%
  \institution{Sun Yat-sen University}
  \city{Zhuhai}
  \country{China}}
\email{ningkw@mail2.sysu.edu.cn}

\author{Yanlin Wang}
\authornote{Yanlin Wang is the corresponding author (wangylin36@mail.sysu.edu.cn).}
\affiliation{%
  \institution{Sun Yat-sen University}
  \city{Zhuhai}
  \country{China}}
\email{wangylin36@mail.sysu.edu.cn}

\author{Jingwen Zhang}
\affiliation{%
  \institution{Sun Yat-sen University}
  \city{Zhuhai}
  \country{China}}
\email{zhangjw273@mail2.sysu.edu.cn}

\author{Dewu Zheng}
\affiliation{%
  \institution{Sun Yat-sen University}
  \city{Zhuhai}
  \country{China}}
\email{zhengdw5@mail2.sysu.edu.cn}

\author{Mingxi Ye}
\affiliation{%
  \institution{Sun Yat-sen University}
  \city{Zhuhai}
  \country{China}}
\email{yemx6@mail2.sysu.edu.cn}

\author{Jiachi Chen}
\affiliation{%
  \institution{Sun Yat-sen University}
  \city{Zhuhai}
  \country{China}}
\email{chenjch86@mail.sysu.edu.cn}

\begin{abstract}
  General large language models (LLMs), represented by ChatGPT, have demonstrated significant potential in software engineering tasks such as code generation. This led to the development of specialized LLMs for software engineering, called Code LLMs. Further, Code LLMs are often derived from general LLMs through fine-tuning and their performance can be influenced by the base LLMs. However, there is a lack of systematic investigation into Code LLMs. In this study, we conduct a comprehensive survey of Code LLMs to address three questions: (1) What LLMs are specifically designed for software engineering tasks, and their relationship? (2) Do Code LLMs outperform general LLMs in software engineering tasks? (3) Which LLMs are more proficient in different software engineering tasks? To answer these questions, we first collect relevant literature and work from four databases and categorize Code LLMs based on their publishers. Next, we investigate the performance differences between general LLMs and Code LLMs in software engineering tasks to demonstrate future trends. Finally, we comprehensively maintained the performance of LLMs to identify the best-performing LLMs for each software engineering task. Our research helps researchers understand the evolution and performance of Code LLMs and provides insights for practitioners to improve Code LLMs.
\end{abstract}





\maketitle

\input{Introduction}
\input{Methodology}

\input{RQ1}

\input{RQ2}

\input{RQ3}

\input{RelatedWork}

\input{Conclusion}

\balance
    \bibliographystyle{ACM-Reference-Format}
    \bibliography{text}

\end{document}

%% file: Introduction.tex
\section{Introduction}

Large Language Models (LLMs) are blurring the boundaries between human languages and machine languages with their powerful text understanding and generation capabilities~\cite{N138}. LLMs have also had a significant impact on various fields, including software engineering~\cite{N136}. Currently, a considerable amount of work focuses on evaluating and optimizing the performance of LLMs on software engineering tasks such as code generation~\cite{N139}, code summarization~\cite{N140}, vulnerability mining~\cite{N141}, and more~\cite{N142, N143}. However, many studies adopt a negative stance on the current performance of LLMs in the field of software engineering~\cite{N144, N145, N146}. This is largely due to the fact that LLMs may not effectively comprehend the structure and semantics of code, lacking domain-specific knowledge in software engineering~\cite{N147, N148}. Fortunately, we can enhance the performance and adaptability of LLMs in specific tasks within the software engineering domain through fine-tuning and additional training~\cite{N149}, thereby designing LLMs specifically tailored for software engineering tasks. We refer to these LLMs designed for software engineering tasks as Code LLMs~\cite{N16}.

Code LLMs can not only be generated through fine-tuning but also through traditional pre-training methods~\cite{N38}. As language models specifically developed for the field of software engineering, ideally, LLMs can have their capabilities further enhanced and possess stronger abilities in code generation, vulnerability mining, and other tasks~\cite{N54}. Therefore, there is currently a considerable amount of work focusing on this area, resulting in the development of many milestone Code LLMs such as Codex~\cite{N5}, Code Llama~\cite{N52}.

However, due to the uncertainties brought by fine-tuning, training data, and other factors~\cite{N12,N124}, as well as the remarkable performance of state-of-the-art general LLMs like GPT-4 in various software engineering domains~\cite{N36}, it is challenging to determine whether the current state-of-the-art Code LLMs outperform the current state-of-the-art general LLMs in software engineering tasks. Moreover, there is a wide range of Code LLMs available, many of which are derived through fine-tuning on general LLMs or other Code LLMs~\cite{N131, N117,N114,N113}. They may possess different structures and fine-tuning techniques~\cite{N7, N18}, and it is likely that they excel in different software engineering tasks. However, there is currently a lack of systematic investigation into Code LLMs and their performance, particularly in comparison to general LLMs and among different Code LLMs.

In this study, our goal is to provide a comprehensive review of Code LLMs and their performance. To achieve this objective, we initially collected relevant literature and works from four major databases known for publishing new models: GitHub\footnote{https://github.com/}, dblp\footnote{https://dblp.uni-trier.de/}, Google Scholar\footnote{https://scholar.google.com}, and arXiv\footnote{https://arxiv.org/}. We employed a card sorting method~\cite{N137} to remove duplicate and irrelevant papers and further expanded our list of research targets through a snowballing approach. After the screening process, we obtained 149 relevant and valid papers. Through this research, we aim to address the following three questions:

\noindent \textbf{RQ1: What LLMs are specifically designed for software engineering tasks, and what is the relationship between them?}

We classified Code LLMs based on the types of institutions to which their main developers belong, such as companies, universities, etc. We not only conducted a chronological review of Code LLMs but also provided a comprehensive summary of their development relationships. These development relationships include but are not limited to, iterations, fine-tuning, and improvements.

\noindent \textbf{RQ2: Do Code LLMs really outperform general LLMs in software engineering tasks?}

To address this question, we selected relevant content from the aforementioned literature, specifically focusing on experimental sections or evaluation work that compared general LLMs and Code LLMs. We conducted a detailed statistical analysis and presentation of the experimental findings or evaluation results. Our findings indicate that, for the same model, the newly fine-tuned models specifically tailored for software engineering tasks outperform their base models. Furthermore, when the number of parameters is comparable, Code LLMs tend to outperform general LLMs. Overall, the current state-of-the-art Code LLMs, such as CodeFuse-CodeLlama-34B, show better performance than the current state-of-the-art general LLMs, such as GPT-4, in code generation tasks. However, GPT-4 still demonstrates strong competitiveness in other tasks.

\noindent \textbf{RQ3: Which LLMs are more proficient in different software engineering tasks? }

We have comprehensively summarized the experimental part of the aforementioned work, manually annotating the performance of 130 Code LLMs on major benchmarks such as HumanEval~\cite{N150}. Additionally, a considerable amount of research has utilized alternative evaluation methods or proposed new benchmarks and evaluation metrics. We have also organized and presented the experimental results from these studies. We found that different Code LLMs do exhibit certain performance differences across various software engineering tasks. Some of these findings can assist developers of Code LLMs in making better choices regarding base models and fine-tuning approaches to develop more advanced LLMs tailored to different software engineering tasks.

The main contributions of this paper are:
\begin{itemize}
   \item  To the best of our knowledge, this is the first systematic review of Code LLMs. We manually selected 149 relevant works from a large number of articles in four major open-source communities or databases\footnote{https://github.com/KevinHeiwa/LLM\_SE\_Papers\_List};
   \item We classified Code LLMs based on the types of institutions to which their main developers belong and organized them chronologically and in terms of their development relationships. This can help practitioners better understand and utilize Code LLMs;
   \item We conducted a comprehensive analysis and compilation of the performance of general LLMs and Code LLMs in software engineering tasks. We provided statistical results and analyzed interesting phenomena. This can assist developers in understanding key improvement directions for Code LLMs;
   \item We maintained the scores of 126 Code LLMs on major benchmarks and conducted a detailed analysis of their performance across different software engineering tasks. This can aid Code LLM developers in making informed decisions regarding base models and fine-tuning approaches.
\end{itemize}

The organization of this paper is as follows. In  Section~\ref{sec:methodogy}, we describe the methods of literature collection and the criteria for screening; In Section~\ref{sec:RQ1}, we summarize and categorize the Code LLMs and propose an answer to research question one (RQ1); In Section~\ref{sec:RQ2}, we address research question two (RQ2); In Section~\ref{sec:RQ3}, we compiled the performance of different LLMs across various benchmarks and literature to address research question there (RQ3); In Section~\ref{sec:RW}, we elucidate related work; Finally, in Section~\ref{sec:conclusion1}, we summarize the entire paper.

%% file: Methodology.tex
\section{Methodology}
\label{sec:methodogy}
\begin{figure}
\centering
\includegraphics[width=0.75\textwidth]{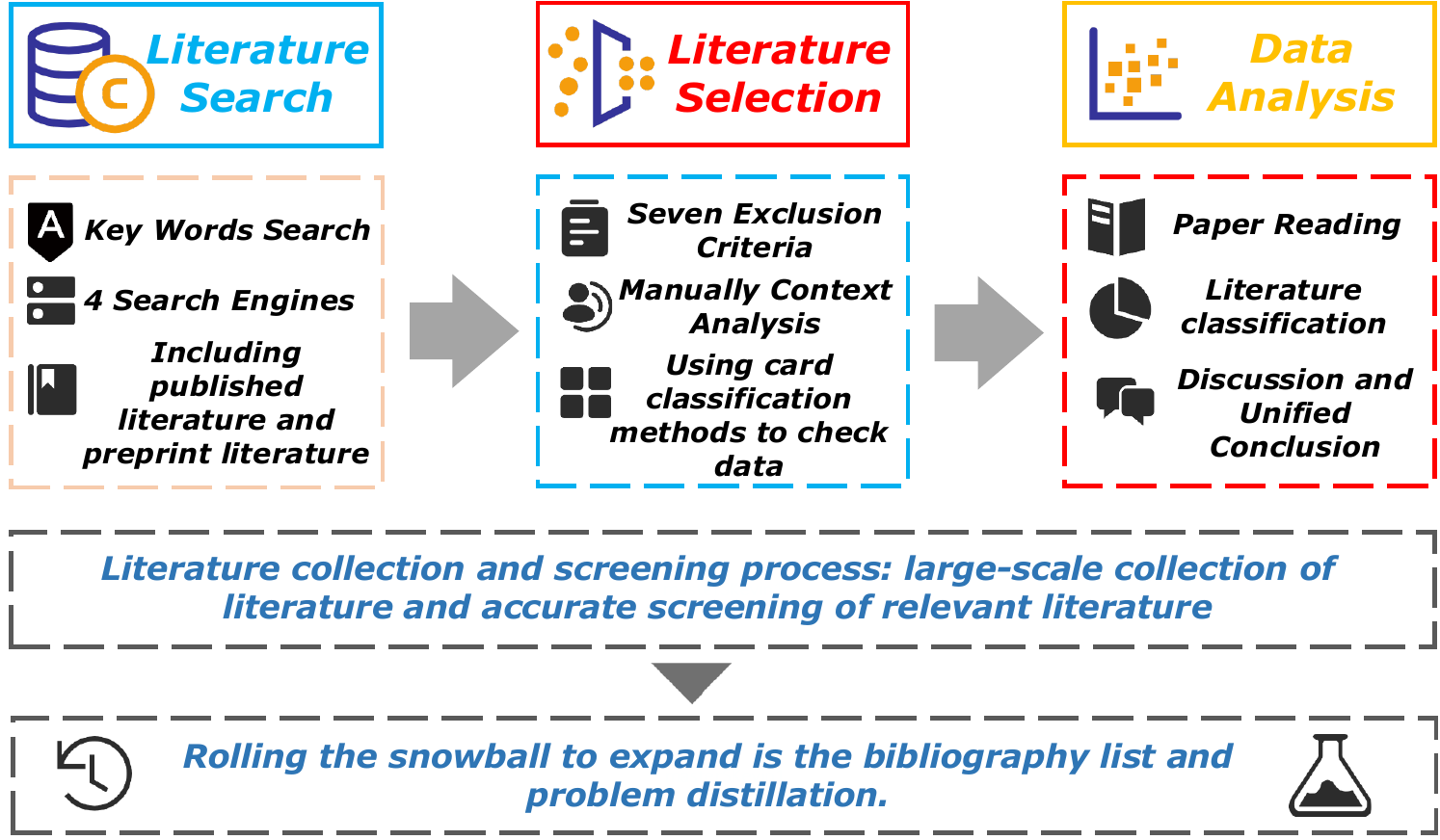}
\caption{Overview of methodology design.}
\label{fig:dataprocess}
\end{figure}

In this section, we introduce the detailed steps of conducting a literature review. We follow the method provided by~\cite{N135} for the literature review with consists of three steps: literature search, literature screening, and data analysis. As shown in Fig.~\ref{fig:dataprocess}.

\subsection{Literature Search}
Based on the previous literature review~\cite{N135}, we selected four databases and communities: dblp, Google Scholar, GitHub and arXiv. From these search engines, we can not only find publications in journals, conferences, workshops, and numerous preprint papers but also access the latest advancements in the industry and open-source projects.

We conducted searches using the following six keywords: ``SE LLM", ``Software Engineering Large Language Model", ``Software Engineering LLM", ``SE Large Language Model", ``Code LLM", and ``Code Large Language Model" on the aforementioned four databases. The obtained results are presented in Table~\ref{tab:paperdata}. It is worth noting that there might be a significant number of duplicate papers and irrelevant articles resulting from different keyword searches within the same engine or the same keyword across different engines. Therefore, we need to manually screen and select these papers, which is known as literature screening or literature selection. 

\begin{table}[t]
\centering
\caption{Number of keyword searches returned by each search engine.}
\setlength{\tabcolsep}{1.5mm}
\renewcommand\arraystretch{1}
\label{tab:paperdata}
\begin{tabular}{p{35pt} p{35pt} p{50pt} p{50pt}p{50pt}p{50pt}p{50pt}}
\toprule
 & \textbf{SE LLM} & \textbf{Software Engineering Large Language Model} & \textbf{Software Engineering LLM} & \textbf{SE Large Language Model} & \textbf{Code LLM} & \textbf{Code Large Language Model} \\ \midrule

GitHub & 399      & 272                                       & 4                        & 85                      & 711       & 328                                            
\\ 
dblp   & 22      & 92                                         & 1                        & 139                     & 15        & 62                                             
\\ 
Google Scholar              & 128000  & 4480000                                    & 24100                    & 532000                   & 75000    & 5280000                                          
\\ 
arXiv                       & 7      & 364                                       & 26                       & 17                      & 705      & 2484                                           \\ \bottomrule
\end{tabular}
\end{table}

\subsection{Literature Selection}
During this stage, we not only need to remove duplicate papers but also filter out irrelevant ones. For example, there are papers that primarily focus on LLMs or the field of software engineering but do not include LLM/Large Language Model keywords in their abstracts. Additionally, Google Scholar returns a large number of unrelated results. According to the industry's definition of LLMs~\cite{N136}, we excluded works published before 2021. As for GitHub, simple keyword matching may result in numerous irrelevant issues and pull requests. Therefore, we only focused on the work done in repositories. In summary, we applied the following eight exclusion criteria to filter the literature:

\textbf{Exclusion Criteria}
\begin{itemize}
    \item Studies are not written in English.
    \item Master or Ph.D. theses.
    \item Keynote papers.
    \item Studies not related to LLMs.
    \item Studies not related to software engineering.
    \item Duplicate papers.
    \item Studies up to 2021 (not including 2021).
    \item Results in GitHub except repositories.
\end{itemize}

In this study, we only focus on the performance of Code LLMs and General LLMs in software engineering tasks. Therefore, many works that do not involve the demonstration and evaluation of LLMs' performance in software engineering tasks are beyond the scope of our research. We specifically focus on the following topics:

\begin{itemize}
    \item Code LLMs.
    \item Application of Different LLMs in Software Engineering.
    \item Evaluation of LLMs on Software Engineering Tasks
\end{itemize}

To improve the accuracy of the literature screening process, we will use the card sorting method to evaluate the collected data. Card sorting is a common technique used to assess data and derive categories from it. There are three types of card sorting methods: open card sorting, closed card sorting, and hybrid card sorting. Among these three types, closed card sorting involves predefined categories~\cite{N137}. In our case, we applied closed card sorting to select relevant papers since we have only two categories: relevant and irrelevant. Each card will have a title (paper title) and a description (paper abstract). By using this method, we can systematically evaluate and categorise the papers based on their relevance to our research.

Three experienced researchers, including one non-coauthors, independently conducted a thorough review of the search engine results from the four databases. After individually organizing the papers, they engaged in a collaborative discussion to align their findings. Through this rigorous process, we ultimately identified 121 relevant works that met the criteria for inclusion in our study.

Furthermore, we expanded the paper list using a snowballing approach~\cite{N137}. Specifically, we manually checked the references of the identified 121 papers and found an additional 13 papers that met our selection criteria. Therefore, we ultimately selected 134 papers for analysis. The list of papers can be found at https://github.com/.

\subsection{Data Analysis}
We used an open card sorting approach to help find the answers to these three research questions. We carefully read the articles and actively searched for answers related to the two questions shown in Table~\ref{tab:RQdata}. If we couldn't find any answers in a particular paper, it was removed from our list.

\begin{figure*}[]
\centering
\includegraphics[width=1\textwidth]{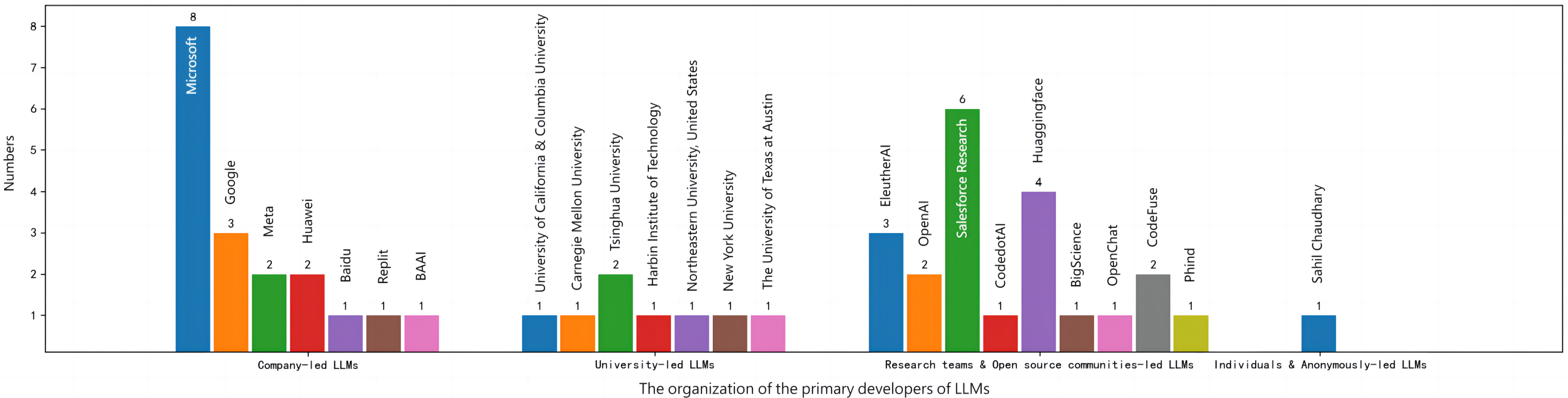}
\caption{Number of Code LLMs issued by organizations.}
\label{fig:paperdata}
\end{figure*}

For the answers to (1), we primarily examined whether and observed whether the work released brand new Code LLMs or mentioned other unknown Code LLMs.We organized this information and categorized the work according to the type of organization (e.g., company, university) in which the main developers are located, as shown in Fig.~\ref{fig:paperdata}. We can see that there are 18 corporate-led Code LLMs, with Microsoft publishing the most, accounting for 8. The report publishes a total of 18 Code LLMs, while research teams and communities publish the most Code LLMs in comparison, with 21. The specifics of these tasks are presented in \ref{sec:RQ1}.
For the answers to (2), We mainly read whether the articles include in the experimental section a comparison of the performance of Code LLMs with general LLMs on software engineering tasks. Especially those example studies, and research articles.

For the answers to (3), we organized the performance of LLMs according to different software engineering tasks, including code generation, code summarization, code translation, and vulnerability detection. We primarily looked for experimental sections in various papers that involved the comparison of different LLMs. Additionally, we compiled the performance of different LLMs on benchmarks commonly used to assess software engineering tasks. In this section, we focused on evaluating articles and papers introducing new Code LLMs.

\begin{table}[t]
\centering
\caption{Data Collection for Each RQ.}
\setlength{\tabcolsep}{1.8mm}
\renewcommand\arraystretch{1}
\label{tab:RQdata}
\begin{tabular}{p{25pt} p{350pt}}
\toprule
\textbf{RQs} & \textbf{Type of Data We Collected}                                                                                                                                 \\ \midrule
RQ1 & What LLMs are specifically designed for software engineering tasks, and what is the relationship between them? For example, which LLMs are fine-tuned based on other LLMs? \\  \midrule
RQ2 & Do Code LLMs really outperform general LLMs in software engineering tasks? For instance, do Code LLMs exhibit better performance than general LLMs in code generation tasks?                   \\ 
\midrule
RQ3 & Which LLMs are more proficient in different software engineering tasks? Different LLMs may excel in different software engineering tasks.                  \\ \bottomrule
\end{tabular}
\end{table}

%% file: RQ1.tex
\section{LLMs for software engineering}
\label{sec:RQ1}
In this section, we answer and summarize mainly for \textbf{RQ1}. We organize our collection of LLMs designed specifically for software engineering tasks and show the relationships between them.

\begin{figure*}
\centering
\includegraphics[width=0.95\textwidth]{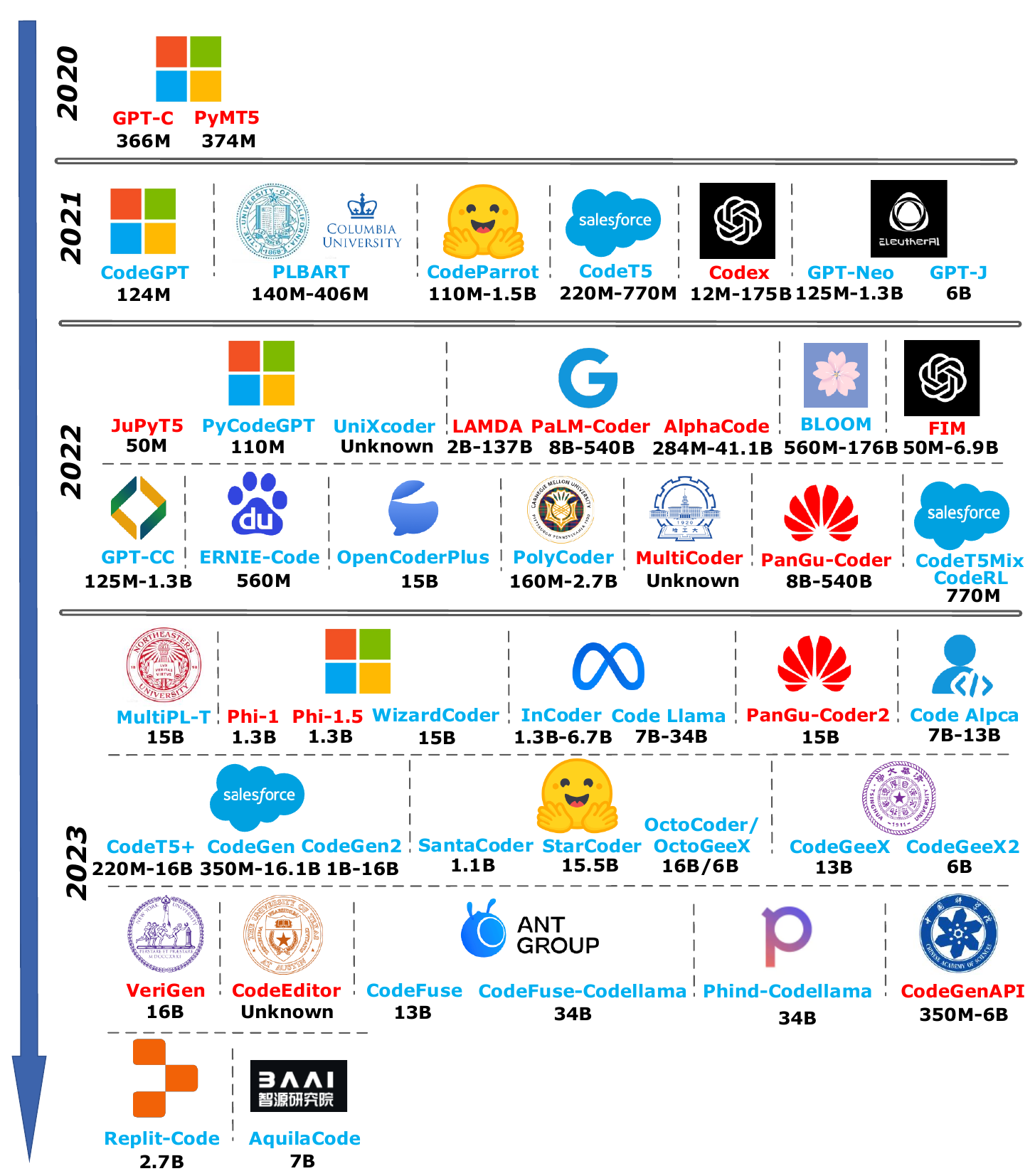}
\caption{Evolution of Code LLMs}
\label{fig:LLMInSE}
\end{figure*}

According to the different release times, we can obtain Fig.~\ref{fig:LLMInSE}. We can see that with the increase in years, more and more companies, organizations, and research teams have been involved in the development of Code LLMs~\cite{N120,N121,N61,N55}. This fully demonstrates the high level of attention from both the industry and academia to the performance of LLMs in software engineering tasks. In order to better illustrate the development process of different LLMs, we will elaborate on the types of affiliations of the main developers of Code LLMs in the following text. These include companies, organizations, research teams \& open-source communities, and individuals \& anonymous contributors.

\subsection{Company-led LLMs}

\textbf{Microsoft: }Microsoft has been at the forefront of exploring numerous LLMs in the field of software engineering and has produced several LLMs specifically designed for software engineering tasks. In 2020, Microsoft introduced the first LLM for programming, GPT-C~\cite{N57}, followed by the release of PyMT5~\cite{N63} and CodeGPT~\cite{N58}. GPT-C is a variant of GPT-2~\cite{N68}, belonging to the class of multilayer generative Transformer models. GPT-C underwent retraining on a large-scale unsupervised multilingual source code dataset and achieved an average edit similarity of 86.7\% on code completion tasks in the Python programming language. On the other hand, PyMT5 is a code-based LLM tailored for Python and natural language. Its architecture is built on the encoder-decoder transformer framework. PyMT5 was trained on a massive parallel corpus containing 26 million Python methods and 7.7 million method-docstring (documentation strings) pairs. Its primary focus is on translation tasks between Python methods and method-docstrings, such as predicting Python methods based on docstrings or generating summaries for Python code. The article also showcases the performance of PyMT5 on the CodeSearchNet~\cite{N64}. CodeGPT is another LLM  developed by Microsoft for software engineering. It shares the same model architecture and training objectives as GPT-2. CodeGPT is pretrained on the CodeSearchNet dataset and primarily used for code completion and code generation tasks. There are two variants of CodeGPT. The first variant, CodeGPT, involves retraining the model with randomly initialized parameters. The second variant, CodeGPT-adapted, involves training directly on top of the GPT-2 model. The article evaluates CodeGPT's performance in code completion tasks on two datasets: PY150~\cite{N69} and Github Java Corpus~\cite{N70}. The results indicate that both CodeGPT and CodeGPT-adapted outperform methods such as LSTM~\cite{N66} and Transformer~\cite{N67}, with CodeGPT-adapted achieving the highest scores.

Phi-1~\cite{N98} is another code-based LLM proposed by Microsoft Research. Phi-1 utilizes a decoder-only transformer architecture and has a size of 1.3B. It was trained on a high-quality dataset of 7B samples. Despite its smaller model size and dataset, Phi-1 outperforms some models trained with larger parameters and larger-scale datasets on multiple benchmarks. This demonstrates that a high-quality dataset can significantly enhance model performance.

Given the strong potential demonstrated by phi-1, Microsoft continues to explore the power of small-scale Transformer-based language models with the subsequent version, phi-1.5~\cite{N99}. Phi-1.5 is a 1.3 billion parameter LLM trained primarily on a specially curated ``textbook-quality" synthetic dataset. The paper indicates that phi-1.5 achieves performance on par with models five times its size in natural language tasks and surpasses most non-cutting edge LLMs in more complex inference tasks like elementary math and basic coding. Regarding the impressive performance of phi-1.5, the article emphasizes the importance of data quality. While the model still lags behind the capabilities of the largest LLMs, it showcases features previously only seen in larger models and demonstrates the feasibility of achieving high-level functionality in smaller LLMs.

Microsoft also released a code-based LLM called JuPyT5~\cite{N73}, which was trained on all publicly available Jupyter Notebook GitHub repositories. Chandel et al.\cite{N73} aimed to explore a Data Science assistant based on the Seq2Seq transformer architecture and introduced a new metric called Data Science Problems (DSP) in this paper. This metric evaluates the model's proficiency in using Python for mathematics and data science by utilizing a curated set of 1,119 questions from educational notebooks. The results of the study show that JuPyT5 was able to solve 77.5\% of the DSP problems in 100 sampling tests.

PyCodeGPT~\cite{N62} is a library-oriented code generation model developed by Microsoft using unlabeled code corpora for training. However, the main focus of the article is on introducing CERT, a novel two-stage approach for library-oriented code generation. On the HumanEval benchmark, PyCodeGPT (110M) achieves a competitive pass@1 rate of 8.33\%.

WizardCoder~\cite{N97} is an open-source large model co-developed by Microsoft and Hong Kong Baptist University. It applies the Evol-Instruct method mentioned in WizardLM~\cite{N96} to the field of code and utilizes the Code Alpaca's 20K instruction-following code dataset to obtain 78K fine-tuning data containing code instructions of varying complexity. WizardCoder uses StarCoder~\cite{N38} as the base model and fine-tunes it on the aforementioned 78K instruction-based fine-tuning dataset. The article also tests WizardCoder on four widely recognized code generation benchmark metrics.

\textbf{Google: }Google has been exploring ways to improve the safety and quality of LLMs' output. They propose addressing this challenge by fine-tuning models using annotated data and enabling models to query external sources of knowledge. Based on this approach, Google introduced LaMDA~\cite{N80}, a series of decoder-only Transformer-based LLMs ranging from 2B to 137B parameters. LaMDA undergoes pretraining on 1.56 trillion words from public dialogue data and web text. It is then fine-tuned using annotated data and equipped with the ability to invoke external retrieval tools' APIs.

Google also explores the impact of model scale on few-shot learning~\cite{N81}. They introduce a new machine learning system called Pathways, which allows efficient training of large models across multiple TPU Pods. Utilizing this system, they trained a dense activation Transformer language model with 540B parameters called Pathways Language Model (PaLM). PaLM uses a standard decoder-only Transformer model architecture with some modifications, such as using the SwiGLU activation function in MLPs, employing parallel forms in each Transformer block, and utilizing multi-query attention.

For code-related tasks, researchers collected a Python code dataset called ExtraPythonData. They fine-tuned PaLM on this dataset for code tasks, resulting in the model PaLM-Coder. PaLM-Coder achieves excellent performance on multiple code generation benchmarks and the DeepFix code repair test~\cite{N81}.

However, current LLMs do not perform well when it comes to evaluating more complex and unknown problems that require surpassing the mere translation of instructions into code~\cite{N83}. For instance, understanding algorithmic and complex natural language competition-style programming problems. To address this, Google DeepMind proposed AlphaCode~\cite{N83}, which aims to tackle these programming problems that require deep reasoning to find novel solutions.

AlphaCode utilizes an asymmetric Encoder-Decoder Transformer architecture, consisting of a shallow encoder and a deep decoder. This modification significantly improves training efficiency without compromising inference speed. When evaluated on Codeforces, AlphaCode's performance is considered average, roughly on par with the median competitor. Interestingly, the authors note that scaling upsampling and filtering samples to cluster them into a small set, as well as introducing new efficient transformers that support large-scale sampling, can effectively enhance the performance of LLMs.

\textbf{Meta: }Facebook AI Research has introduced InCoder~\cite{N84}, a unified generative model that can perform program synthesis through left-to-right generation and code editing through masking and infilling. InCoder is the first large-scale generative code model capable of filling arbitrary code regions. It is trained on a large corpus of code files with permissive licenses, where random code regions are masked and moved to the end of each file, allowing bidirectional context for code completion.

InCoder adopts the MoE (Mixture of Experts) model architecture~\cite{N85}. The paper highlights that training code generation models with causal masking objectives can achieve strong zero-shot performance on challenging and practically meaningful code completion and editing tasks. It also lays the groundwork for future work on supervised filling and editing via model fine-tuning and iterative decoding.

Meta has also introduced Code Llama~\cite{N52}, a large-scale code language model series based on Llama 2. Code Llama offers multiple versions, including a base model (Code Llama), a Python specialization model (Code Llama - Python), and an instruction-following model (Code Llama - Instruct), each available with 7B, 13B, and 34B parameters. All models are trained on 16k token sequences and demonstrate improvements on inputs with up to 100k tokens. The paper provides extensive experimental evidence to demonstrate the superiority of Code Llama. Meta has also made Code Llama available for research and commercial use under licensing terms.

\textbf{Huawei: }PanGu-Coder~\cite{N86} is a pre-trained decoder language model proposed by Huawei Noah's Ark Lab. It adopts the PANGU-$\alpha$ architecture~\cite{N87} and is specifically designed for text-to-code generation. Pangu-Coder utilizes a two-stage training strategy: in the first stage, it undergoes causal language modelling (CLM) pre-training on raw programming language data, while in the second stage, a combination of CLM and masked language modelling (MLM) training objectives is used, focusing on downstream tasks of text-to-code generation. It is trained on loosely filtered natural language program definitions and code-function pairs, using a dataset of 147GB of code data collected from GitHub.

The results of the study show that the two-stage training approach helps achieve comparable or even better performance than models of similar scale, with smaller context windows and fewer training data requirements. Additionally, the experiments with PANGU-CODER-FT demonstrate that by carefully selecting and fine-tuning with data closely related to the target task, the base model's scores can be improved more quickly, as the model is sensitive to data distribution mismatch and shifts during fine-tuning. Building upon PanGu-Coder, Huawei Cloud has launched the Huawei Cloud Intelligent Programming Assistant CodeArts Snap to enhance developers' programming efficiency.

PanGu-Coder2~\cite{N23} is an iterative upgrade of PanGu-Coder. PanGu-Coder2 achieved a pass@1 rate of 62.20\% in the OpenAI HumanEval benchmark test. Furthermore, through extensive evaluations on CoderEval and LeetCode benchmarks, the paper demonstrates that PanGu-Coder2 consistently outperforms all previous Code LLMs. Interestingly, PanGu-Coder2 is developed based on a novel framework called RRTF (Rank Responses to align Test \& Teacher Feedback). The paper claims that this framework can efficiently enhance large-scale pre-trained language models for code generation.

\textbf{Baidu: }Currently, LLMs in the field of software engineering tend to be English-centric, which leads to limited comprehension of other natural languages. Therefore, Baidu has proposed a large code model called ERNIE-Code N90 that supports multiple natural languages and programming languages. ERNIE-Code is based on the T5 architecture and utilizes two approaches for universal cross-lingual pre-training. Firstly, it undergoes unsupervised learning using a monolingual programming language (PL) and natural language (NL) data. Then, it undergoes supervised learning on cross-lingual NL-PL or NL-NL pairs to enable the model to learn cross-lingual/modal alignment and zero-shot capabilities. The paper acknowledges that while ERNIE-Code demonstrates strong performance on various tasks involving computer programs and natural language, the lack of corresponding benchmarks prevents the authors from evaluating ERNIE-Code's performance systematically across a wide range of multi-language NLs.

\textbf{Replit: }Replit-Code~\cite{N116} is an open-source code model developed by Replit, primarily targeting code completion tasks. Replit-Code has a scale of 2.7B and is trained on a subset of the Stack Dedup v1.2 dataset~\cite{N100}. It offers code completion capabilities for 20 programming languages. However, the current lack of technical and testing reports for Replit-Code makes it difficult for us to know the specific performance of Replit-Code on software engineering tasks.

\textbf{BAAI: }AquilaCode-multi~\cite{N124} is a bilingual code model developed by BAAI (Beijing Academy of Artificial Intelligence). It is trained from scratch on a high-quality corpus of both Chinese and English languages, with approximately 40\% of the data consisting of Chinese text. This approach ensures that the model accumulates native Chinese knowledge during the pre-training phase, rather than relying on translation. The training data for code generation is derived from high-quality filtered code samples that adhere to open-source licenses. Currently, there is also a lack of comprehensive analytical data on the capabilities of AquilaCode-multi. Further exploration is needed to uncover the specific performance of AquilaCode-multi.

\subsection{University-led LLMs}
PLBART~\cite{N65} N65 is a code-based large language model jointly released by the University of California and Columbia University. It is built upon the BART architecture and is pre-trained using denoising autoencoding on a large corpus of Java and Python functions along with their associated natural language texts. Experimental results demonstrate that PLBART outperforms methods such as Seq2Seq~\cite{N105}, Transformer~\cite{N67}, RoBERTa~\cite{N104}, and CodeBERT~\cite{N103} on tasks such as code summarization, code generation, and code translation across seven programming languages.

Carnegie Mellon University conducted a systematic evaluation of code-based large models such as Codex, GPT-J, GPT-Neo, GPT-NeoX-20B, and CodeParrot~\cite{N82}. The article also highlights the lack of a large-scale open-source model exclusively trained on a multilingual code corpus among these models. As a result, Carnegie Mellon University introduced PolyCoder, an LLM trained on a single machine using 249GB of code from 12 programming languages. PolyCoder is available in three versions: 160M, 400M, and 2.7B. The experimental results in the article demonstrate that PolyCoder outperforms all the models at that time, including Codex, in terms of performance on the C programming language. Moreover, it is noteworthy that PolyCoder is open-source, which adds to its value.

CodeGeeX~\cite{N51-1} is an open-source multilingual LLM led by Tsinghua University. CodeGeeX adopts a decoder-only autoregressive (programming) language modelling approach, with a core architecture consisting of a 39-layer Transformer decoder. The training corpus includes open-source code datasets such as the Pile, and CodeParrot, as well as publicly available code scraped directly from GitHub repositories, totaling 850 billion tokens from 23 different programming languages. The experimental results in the article demonstrate that CodeGeeX consistently outperforms other open-source LLMs of similar scale in code generation and translation tasks. Furthermore, the extension capabilities built by CodeGeeX bring significant benefits in improving coding efficiency.

Tsinghua University has released the second generation of CodeGeeX, called CodeGeeX2~\cite{N115}. Unlike the first generation, CodeGeeX2 is built on the ChatGLM2 architecture and incorporates code data for pretraining. Leveraging the improved performance of ChatGLM2, CodeGeeX2 demonstrates performance improvements across multiple metrics. With just 6 billion parameters, CodeGeeX2 achieves nearly a 10\% improvement over StarCoder-15B, which has over 15 billion parameters. Due to inheriting the characteristics of the ChatGLM2-6B model, CodeGeeX2-6B provides better support for both Chinese and English inputs. It also supports a maximum sequence length of 8192 and significantly improves inference speed compared to the first-generation CodeGeeX-13B model. After quantization, CodeGeeX2-6B only requires 6GB of GPU memory to run and supports lightweight local deployment.

MultiCoder~\cite{N9} is a collaborative development by Harbin Institute of Technology and Huawei Noah's Ark Lab. It addresses the challenge of code completion on low-resource programming languages (PL) that are widely used by developers but present difficulties for data-driven approaches. To enhance code completion capabilities for low-resource PLs, Gong et al.~\cite{N9} proposes MultiCoder, which leverages MultiPL pre-training and a MultiPL Mixture-of-Experts (MoE) layer. The article also introduces a novel PL-level MoE routing strategy (PL-MoE) to improve code completion across all PLs. Performance analysis of MultiCoder is conducted on CodeXGLUE and MultiCC datasets. The results indicate that MultiCoder significantly outperforms the MonoPL baseline in low-resource programming languages. Importantly, the PL-MoE module further enhances performance across six programming languages.

Cassano et al.~\cite{N16} also focus on low-resource languages and propose an effective method called MultiPL-T to improve the performance of low-resource language Code LLM using semi-synthetic data. By utilizing the data generated by MultiPL-T, the article presents a fine-tuned version of StarCoderBase and achieves better performance on benchmark problems for Racket, OCaml, and Lua languages.

Thakur et al.~\cite{N21} explored the ability of LLMs to generate high-quality Verilog code and fine-tuned a new LLM, called VeriGen, based on the CodeGen-16B architecture using Verilog datasets collected from GitHub and Verilog textbooks. The experimental results of the paper demonstrate that VeriGen outperforms the state-of-the-art commercial GPT-3.5-turbo model in terms of generating Verilog code, showing an overall improvement of 1.1\%. Although the improvement is not significant, it showcases the potential of small-scale LLMs in hardware design automation.

Due to the frequent usage of private library APIs by programmers when writing proprietary code, LLMs lack the capability in this aspect. This is because, during pre-training, LLMs have limited exposure to these private libraries. To address this issue, Zan et al.~\cite{N22} propose a novel framework to simulate the process of programmers writing private code. The framework consists of two modules: APIFinder, which retrieves potentially useful APIs from API documentation, and APICoder, which utilizes these retrieved APIs to generate private code. Furthermore, Zan et al.~\cite{N22} introduce an enhanced version of the APICoder called CodeGenAPI, developed based on GodeGen, an LLM. The paper also creates four benchmark metrics for private libraries, including TorchDataEval, TorchDataComplexEval, MonkeyEval, and BeatNumEval. Test cases are written for each benchmark test to support the comprehensive evaluation of CodeGenAPI.

Zhang et al.~\cite{N24} focuses on the task of code translation, which involves converting code changes from one programming language to another. A language model called Codeeditor is designed and implemented for this purpose. Codeeditor explicitly models code changes as an editing sequence and learns cross-language associations for the changes. To evaluate Codeeditor, the paper collects 6,613 consistent code changes from 8 pairs of open-source software projects that have similar functionality in two programming languages: Java and C\#. The results of the study show that Codeeditor outperforms both Codex and CodeT5 by a significant margin across various commonly used metrics. The performance of Codeeditor on a single software engineering task provides new insights for the development of Code LLMs and complements the capabilities of existing Code LLMs. The combination of these models can ensure higher performance overall.

\subsection{Research teams \& Open source communities-led LLMs}

\textbf{EleutherAI: }EleutherAI has conducted equivalent reproductions of the non-open-source GPT-3 N71 model and released two open-source equivalents: GPT-Neo (125M~2.7B) N59 and GPT-J (6B) N60, corresponding to GPT-3 Ada (2.7B) and GPT-3 Babbage (6.7B), respectively. To reproduce GPT-3, EleutherAI first proposed a 825GB English text corpus called ``the Pile" N72, which includes 95GB of code data sourced from GitHub. GPT-Neo is a GPT2-like causal language model trained on the Pile dataset and utilizes the mesh-tensorflow library for parallelism. GPT-J is an autoregressive text generation model with 6 billion parameters trained on the Pile using Mesh Transformer JAX.

Although GPT-Neo and GPT-J outperform GPT-3 models with equivalent parameters on the tested NLP reasoning benchmarks, they still have a significant gap when compared to the largest GPT-3 model, GPT-3 Davinci (175B). Additionally, due to the inclusion of code data in the training set, both GPT-Neo and GPT-J show some performance in code-related tasks. GPT-NeoX~\cite{N132}, an open-source LLM by EleutherAI, is a 20B model and currently the largest publicly available model in terms of weight. It is trained on the Pile dataset with minor modifications to the GPT-J architecture. However, the paper does not provide experimental evaluations of the programming capabilities of GPT-NeoX. Instead, it primarily showcases the impressive performance of GPT-NeoX on mathematical tasks.

\textbf{OpenAI: }In 2021 and 2022, OpenAI introduced two LLMs, Codex~\cite{N61} and FIM~\cite{N74}, designed to assist with programming tasks. Codex~\cite{N61} is a GPT-3 model fine-tuned using publicly available code from GitHub, with a maximum parameter count of 12B. The research paper also introduces a new evaluation benchmark called HumanEval, which involves synthesizing programs from docstrings. HumanEval consists of 164 handwritten programming questions and has become an important tool for evaluating the performance of large models on software engineering tasks. Experimental results show that in a single sample from HumanEval, Codex solved 28.8\% of the problems, while GPT-3 solved 0\% and GPT-J solved 11.4\%.

Building upon Codex, GitHub collaborated with OpenAI to develop a code generation tool called GitHub Copilot. GitHub Copilot leverages context and prompts to automatically generate code snippets, comments, and more. It provides programming suggestions and code completion functionality to assist developers in their coding tasks.

Indeed, all model classes, including Codex and GPT-3, have limitations when it comes to infilling, which involves generating text at a specific location within a prompt while conditioning on both a prefix and a suffix~\cite{N61}. Left-to-right models can only condition on the prefix and cannot handle this type of task effectively. To address this limitation, OpenAI proposed the addition of fill-in-the-middle (FIM) capability within the paradigm of LLMs~\cite{N74}.

The FIM framework allows for simple modifications to the training data without altering the model structure, enabling the models to handle fill-in-the-middle tasks more effectively. The research paper showcases eight causal transformer decoder models trained on the FIM framework, which have architectures similar to Codex and GPT-3. To prevent contamination from the HumanEval dataset, which was used for evaluation, FIM utilized the same 159GB code dataset as Codex for pre-training. The models were initialized with random parameters similar to GPT-3. Due to the efficiency of the FIM training framework, subsequent models such as Incoder and SantaCoder also adopted this approach.

\textbf{Salesforce Research: }Salesforce Research has released three LLM models in the field of software engineering: CodeT5~\cite{N75}, CodeRL~\cite{N77}, and CodeGen~\cite{N76}.

CodeT5~\cite{N75} is a pre-trained encoder-decoder model based on the T5 architecture. The developers of CodeT5 recognized that previous code models treated source code as token sequences similar to natural language, overlooking the rich structural information present in code. To address this, the authors proposed a novel identifier-aware mechanism that enables the model to distinguish which tokens are identifiers and recover them when they are masked. CodeT5 was trained on the CodeSearchNet dataset and further fine-tuned on the CodeXGLUE~\cite{N58} benchmark. Experimental results demonstrate that CodeT5 outperforms previous works on most tasks within the CodeXGLUE benchmark, showcasing its improved performance in the software engineering domain.

CodeRL~\cite{N77} introduces a training framework that leverages reinforcement learning to improve the performance of pre-trained language models in program synthesis tasks. Based on this framework, the authors used CodeT5~\cite{N75} as the base model and trained the CodeRL model on the GCPY dataset, which focuses on Python code. During training, the code generation language model is treated as an actor-network, and a critic network is introduced to predict the functional correctness of the generated programs and provide dense feedback signals to the actor. Experimental results presented in the paper demonstrate that CodeRL surpasses CodeT5 not only on the APPS benchmark test and MBPP benchmark test but also exhibits strong zero-shot transfer learning capabilities. This indicates that CodeRL can effectively generalize its learning to new tasks and datasets without explicit training on them.

By increasing the model size and data scale, the language modelling capability can be improved, enabling LLMs to understand long contexts and generate coherent responses. Therefore, utilizing this capability can lead to a better understanding of user intent and achieve improved program synthesis results~\cite{N76}. To validate this approach's effectiveness, the authors propose a multi-step program synthesis method and introduce a new LLM called CodeGen~\cite{N76} for verification. CodeGen is also used for program synthesis tasks, adopting the form of an autoregressive transformer. CodeGen is trained on three datasets: the Pile, BigQuery, and BigPython. 

The authors initially trained CodeGen on the Pile dataset. The Pile dataset is a large-scale English text corpus open-sourced by EleutherAI, which includes a substantial amount of GitHub code data. This training results in the first version of CodeGen-NL. Subsequently, developers utilize the parameters of CodeGen-NL as initial parameters for training on BigQuery. BigQuery is a publicly available multi-programming language code dataset provided by Google, and six programming languages are selected for this purpose, leading to the second version, CodeGen-MULTI. Finally, developers employ the parameters of CodeGen-MULTI as initial parameters for training on BigPython. BigPython is a single-language code dataset focused on Python. Upon completion of training, CodeGen-MONO is obtained. 

The experiments in the article demonstrate that multi-step program synthesis can enhance the model's program synthesis capability. Furthermore, the multi-step program synthesis capability exhibits a linear relationship with the model size and data scale.

Salesforce Research also explored improving the efficiency of LLMs for program synthesis training by unifying four key components: model architecture, learning methods, infill sampling, and data distribution. In this exploration, researchers trained four models with parameter sizes of 1B, 3.7B, 7B, and 16B, referred to as CodeGen2~\cite{N133}. Regarding model architecture, CodeGen2 attempted to unify the encoder and decoder models into a single prefix-based language modelling (Prefix-LM)~\cite{N91}. This approach aims to unify bidirectional encoder representations and unidirectional decoder representations.

In terms of learning methods, CodeGen2 employed an algorithm that combines a causal language modelling objective with span corruption. And it also explored the ``free lunch" hypothesis. In the aspect of data distribution, CodeGen2 investigated the impact of mixing programming language and natural language distributions on model performance.

The experimental results showed the following findings:
\begin{itemize}
    \item The Prefix-LM architecture did not demonstrate measurable improvements in this task.
    \item Training a model with infill sampling did not provide a free lunch.
    \item A simple mixture of causal language modeling and span corruption, limited to within-file spans, was sufficient.
    \item A mixture distribution of programming and natural languages showed promising results.
\end{itemize}

Salesforce Research also introduced an upgraded version of CodeT5 called CodeT5+~\cite{N36}. CodeT5+ allows flexible combinations of its component modules to adapt to various downstream code tasks. The authors of the article proposed the use of mixed pretraining objectives to alleviate the discrepancy between pretraining and fine-tuning stages. They covered multiple pretraining tasks, including span denoising, contrastive learning, text-code matching, and causal language modeling, across both single-modal and multimodal code corpora. Furthermore, the authors suggested initializing CodeT5+ with frozen pre trained LLMs, without the need for training from scratch. This approach effectively scales up these models and explores instruction conditioning to align with natural language instructions.

CodeT5Mix~\cite{N89} is based on CodeT5. Building upon CodeT5, CodeT5Mix introduces an encoder component, RoBERTa, and two decoder components, GPT-2 and CodeGPT. By combining these components (or using them individually), CodeT5Mix is able to handle various code-related tasks. The training incorporates diverse pre-training tasks and employs an efficient weight-sharing strategy. CodeT5Mix comes in two variants: base (220M) and large (770M). The results of the study demonstrate that it can fully support a semi-parametric retrieval-augmented code generation approach, as the model effectively performs in both code retrieval and generation tasks.

\textbf{CodedotAI: }CodedotAI attempted to replicate GitHub Copilot and proposed the GPT-CC (GPT-Code-Clippy)~\cite{N78} model. GPT-CC utilized the GPT-Neo model as its base model and was trained on Code Clippy Data, with fine-tuning performed on APPS and CodeSearchNet Challenge Data. Code Clippy Data is a 132GB dataset created by developers by deduplicating and merging code data collected from GitHub with code data from the Pile. However, GPT-CC did not provide a comparison with other LLMs in terms of performance on software engineering tasks, nor did it further analyze the effectiveness and superiority of Code Clippy Data.

\textbf{Huaggingface: }Huaggingface has made significant contributions to the open-source landscape of LLMs in the field of software engineering. They have provided a training tutorial similar to Codex, allowing users to build a large model called CodeParrot)~\cite{N79} from scratch. CodeParrot is based on the GPT-2 architecture, and users can train both 110M and 1.5B versions of CodeParrot using just 150 lines of code. Huaggingface has also introduced the BigCode project, which is a scientific collaborative effort dedicated to developing large code models. Within the BigCode project, they proposed SantaCoder~\cite{N7}, an LLM based on the FIM (Filling in the Missing) model. SantaCoder's Tokenizer draws inspiration from InCoder and uses the Hugging Face Tokenizer~\cite{N109}. It employs the Byte-Pair Encoding (BPE) algorithm to train the original bytes. The Stack v1.1 is a dataset released by Huaggingface in the BigCode community, containing 6.4TB of source code data from 384 programming languages. This dataset has been filtered to remove sensitive information such as usernames while generating it.

In addition, the BigCode community has also released an open-sourced StarCoder~\cite{N38}, claiming it to be one of the most advanced LLMs for assisted programming currently available. StarCoder is divided into two versions: StarCoder and StarCoderBase. The architecture of StarCoderBase is similar to SantaCoder, with a parameter size of 15.5B and the utilization of techniques such as FIM (Filling in the Missing) and MQA (Multi-Question Answering). StarCoderBase is trained on 1 trillion tokens from the Stack dataset and then fine-tuned on 35B Python code data to obtain StarCoder. StarCoderBase outperforms other existing open-source code LLMs in popular programming benchmarks. Additionally, it has a context length of over 8,000 tokens, allowing StarCoder to handle larger inputs than any other open-source LLM.

In Muennighoff et al.~\cite{N18}, the BigCode community introduced COMMITPACK, a 4TB Git commit dataset covering 350 programming languages. This dataset includes paired information between code changes and human instructions, which aids in better instruction-based fine-tuning of LLMs. The article also introduces HUMANEVALPACK, which extends the HumanEval benchmark to three coding tasks (code fixing, code interpretation, and code synthesis) across six languages (Python, JavaScript, Java, Go, C++, Rust). Additionally, the article presents two Code LLM models, OctoCoder and OctoGeeX. Experimental results demonstrate that these two LLMs achieve optimal performance on HUMANEVALPACK and outperform GPT-4 in terms of performance.

\textbf{BigScience Workshop: }The BigScience Workshop introduced an open-source 176B language model called BLOOM~\cite{N88}. It is a decoder-only Transformer language model trained on the ROOTS corpus. ROOTS is a composite collection consisting of 498 Hugging Face datasets~\cite{N111} with a total text size of 1.61TB, covering 46 natural languages and 13 programming languages. BLOOM achieves competitive performance in various benchmark metrics and shows stronger results after multi-task prompt fine-tuning. Unlike considering a mixture of expert models (MoE) or the state-space model~\cite{N112}, BLOOM employs causal decoder-only models. However, the article does not provide a comprehensive evaluation of BLOOM on software engineering tasks. Only a few test results indicate that BLOOM performs weaker than Codex in code generation. This suggests that multitasking fine-tuned BLOOMZ models do not significantly improve over BLOOM models.

\textbf{OpenChat: }OpenCoderPlus~\cite{N101}is a code-oriented large language model in the OpenChat series, fine-tuned on a dataset of code. The base model for OpenCoderPlus is StarCoderPlus. According to the data provided by the authors, OpenCoderPlus achieves 102.5\% of the ChatGPT score on the Vicuna GPT-4 evaluation and has a win rate of 78.7\% on AlpacaEval.

\textbf{CodeFuse: }CodeFuse-MFTCoder~\cite{N113}is an open-source project by CodeFuse that focuses on multi-task code large language models (Code-LLMs). It includes models, datasets, training code repositories, and inference guidelines. MFTCoder supports most mainstream open-source large models, particularly those related to Code-LLMs, such as Code-LaMA and Starcoder. Currently, CodeFuse has released two high-quality code-related instruction fine-tuning datasets: Evol-instruction-66k and CodeExercise-Python-27k. They have also introduced two models: CodeFuse-13B and CodeFuse-CodeLlama-34B. According to the results presented in MFTCoder~\cite{N113}, CodeFuse-CodeLlama-34B outperforms GPT-4 in terms of performance on HumanEval.

\textbf{Phind: }Indeed, there are several Code LLMs that have been fine-tuned based on Code Llama, including Phind-CodeLlama~\cite{N114}. Phind-CodeLlama offers three versions that are fine-tuned from Code Llama: Phind-CodeLlama-34B-v1, Phind-CodeLlama-34B-v2, and CodeLlama-34B-Python. Among them, CodeLlama-34B and CodeLlama-34B-Python achieved 67.6\% and 69.5\% pass@1 on the HumanEval benchmark using an internal Phind dataset. Additionally, Phind-CodeLlama-34B-v2 achieved an even higher score of 73.8\% pass@1 on the HumanEval benchmark. These scores surpass the performance of GPT-4, which achieved 67\% pass@1 on the same benchmark.

\subsection{Individuals \& Anonymously-led LLMs}
CodeAlpaca~\cite{N92} is an instruction-guided LLaMA~\cite{N94} model trained based on code generation instructions. LLaMA is a set of efficient open-source large language models trained on publicly available datasets using the standard transformer architecture. Alpaca
on the other hand, combines the Self-Instruct~\cite{N95} method with LLaMA to create a dataset for instruction-based fine-tuning. LLaMA is then fine-tuned on this instruction-based dataset to create an LLM. Furthermore, CodeAlpaca focuses on code generation, editing, optimization tasks, and similar code-related tasks by placing emphasis on instruction-based fine-tuning using the Alpaca model. The LLaMA model is fine-tuned on instruction-based data in these code domains, resulting in a large model capable of following code instructions.

\begin{figure*}
\centering
\includegraphics[width=1\textwidth]{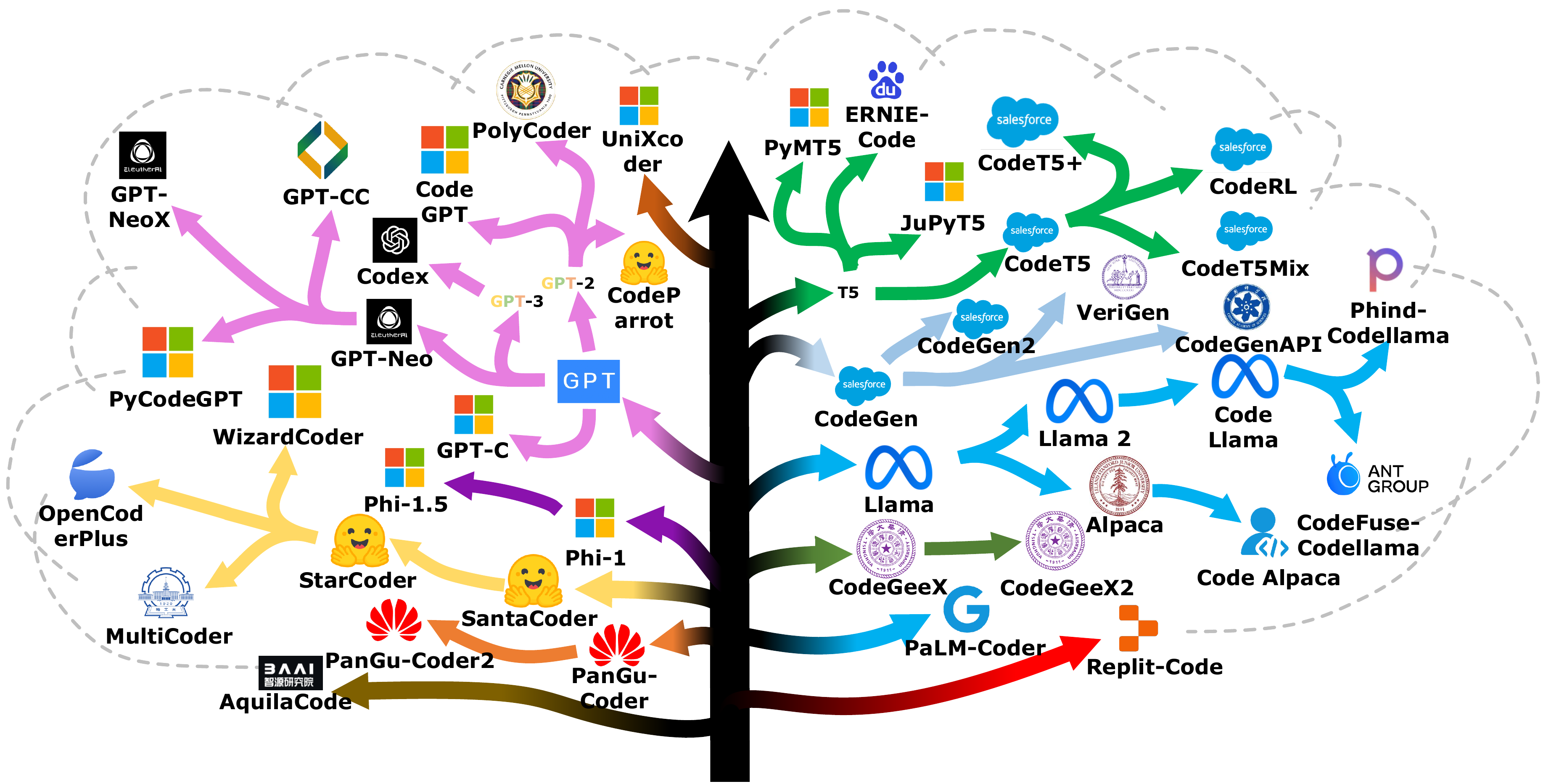}
\caption{The relationships between Code LLMs}
\label{fig:Code LLMs tree}
\end{figure*}

We have also mapped out the relationships between these Code LLMs as shown in Fig.~\ref{fig:Code LLMs tree}. The arrows indicate that the model being pointed to is derived or developed from the model pointing to it. This development can include, but is not limited to, improvements, fine-tuning, or borrowing of techniques.

%% file: RQ2.tex
\section{Do Code LLMs really perform better than general LLMs in  Software Engineering Tasks?}
\label{sec:RQ2}
In this section, we primarily focus on addressing and summarizing \textbf{RQ2}. First, we will filter out the papers from our collection that evaluate Code LLMs, introduce new evaluation benchmarks, or propose new Code LLMs. Then, we will thoroughly examine all the literature and identify papers that compare Code LLMs with general LLMs in their evaluation sections. We will organize this information and focus on presenting the evaluation results.

Indeed, it is important to consider the parameter scale when comparing different LLMs, as variations in parameter size can lead to significant performance differences~\cite{N56}. When comparing LLMs, it is crucial to take into account their specific parameter sizes.

Jiang et al.~\cite{N2} compares the performance of code-davinci-002-175B and text-davinci-002 in three metrics: Pass@1 of HumanEval, CodeBLEU, and AvgPassRatio. Since text-davinci-002is based on code-davinci-002 (175B), there is no significant performance difference between the two. However, compared to text-davinci-003, the performance difference becomes more apparent. After RLHF, text-davinci-003 outperforms code-davinci-002 in all aspects. From the results of N2, it may be difficult to draw useful conclusions. We can conclude that the current code generation capability of text-davinci-003 (175B) is stronger than code-davinci-002.

Maddigan et al.~\cite{N5} evaluate the performance of ChatGPT, Codex, and GPT-3 in the task of natural language generation data visualization. The article evaluates them through five case studies, and in all five case studies, ChatGPT performs better than Codex and GPT-3.

Shirafuji et al.~\cite{N12} evaluate the robustness of source code, Codex, CodeGen, InstructGPT, and ChatGPT, popular LLMs, in solving programming problems by changing prompt formats and modifying problem descriptions. The results of the article show that the new models combined with RLHF techniques have stronger robustness towards the format of problem descriptions. In other words, ChatGPT's ability exceeds the other three models. Li et al.~\cite{N13} mainly compare Codex and ChatGPT in three code generation benchmark metrics, SCoT prompts, and Pass@k of the benchmarks. The results also show that ChatGPT is more powerful than Codex in code generation tasks.

Li et al.~\cite{N14} introduce BIRD, a benchmark test for English large-scale cross-domain text-to-SQL. T5, ChatGPT, and Codex are tested on this benchmark. The experimental results show that ChatGPT and Codex perform better than T5 in both The Valid Efficiency Score and The Execution Accuracy. However, there is not much difference in performance between ChatGPT and Codex, and relatively speaking, ChatGPT is slightly better. However, they still fall far short of human performance.

Yang et al.~\cite{N15} present the performance of CodeParrot-1.5B, CodeParrot-small-110M, PolyCoder-160M, PolyCoder-400M, GPT-Neo-125M, and GPT-Neo-1.3B on OpenAI's HumanEval benchmark. We can see that different parameter sizes of the same model do have an impact on performance but to varying degrees. Among the six models, GPT-Neo-1.3B performs the best, even surpassing the larger parameter size of CodeParrot-1.5B. However, GPT-Neo-125M performs poorly, even worse than the smaller parameter size of CodeParrot-small-110M.

Thakur et al.~\cite{N21} include an experimental section evaluating the ability of LLMs to generate high-quality Verilog code. The experimental results show that the fine-tuned CodeGen-16B model outperforms the GPT-3.5-Turbo model, with an overall improvement of 1.1\%. However, the article also mentions that GPT-4 performs exceptionally well on these tasks, especially in advanced problem-solving. Due to limited API access to GPT-4, the article does not provide a detailed comparative evaluation of GPT-4.

Siddiq et al.~\cite{N26} includes an experiment that quantifies the percentage of compilable code snippets in Java and Python code generation tasks for 11 LLMs. It is observed that CodeGen-2B-multi performs better for Java code, while GPT-3.5 performs relatively better for Python code generation. The article also shows the performance of the 11 LLMs on NDCG@10. Interestingly, GPT-3.5 achieves a higher score in the Java task, while CodeGen-2B-mono obtains a higher score in the Python code generation task. From this article, we can not determine which model performs better between CodeGen-2B and GPT-3.5.

Sun et al.~\cite{N33} presents a performance comparison of several LLMs on the Spider Dev Split test suite, where GPT-4 performs relatively the best.

Sun et al.~\cite{N35} tests the performance of ChatGPT against NCS, CodeBERT, and CodeT5 on the CSN-Python dataset for code summarization. The results show that ChatGPT performs relatively poorly, while CodeT5 achieves the best performance.

Wang et al.~\cite{N36} introduce CodeT5+ and compare it to current major LLMs. Compared to current major Code LLMs, CodeT5+ demonstrates better performance in pass@k(\%) on HumanEval. However, there is still a performance gap between CodeT5+ and models like GPT-4, GPT-3.5, and code-davinci-002.

Li et al.~\cite{N38} compare StarCoder with most mainstream Code LLMs. The results of the article show that StarCoder outperforms general LLMs, PaLM, and various versions of LaMDA in Python program metrics, demonstrating better performance.

Li et al.~\cite{N40} compare the performance of Code LLMs (Codex) and NL LLMs (GPT-3) on the task of information extraction (IE). The article primarily metrics named entity recognition (NER) and relation extraction (RE). The experimental results show that Code LLMs outperform GPT-3 in both tasks.

Zhang et al.~\cite{N42} focuses on the performance of LLMs when executing API operations. The experiments in the article are conducted on three Public library benchmarks and two Private library benchmarks. The results show that CodeGen-2B performs worse than GPT-3.5 in all five metrics. The article also fine-tunes the ToolCoder model based on CodeGen-2B to better handle API operations during code generation. Although ToolCoder achieves results close to GPT-3.5 in some metrics, overall performance still falls short of GPT-3.5.

Siddiq et al.~\cite{N43} validate the performance of LLMs in generating unit test cases. The article conducts experiments with different prompts and compares the results based on compilation rate, test correctness, coverage, and test smells. The experimental results show that CodeGen performs the worst, while Codex performs the best among the LLMs. This could be attributed to CodeGen being a smaller model with only 350 million parameters, much smaller in terms of training data and parameter size compared to Codex and ChatGPT-3.5, which have 12 billion and 175 billion parameters respectively. ChatGPT-3.5's performance falls between the two, but in most cases, it is noticeably inferior to Codex.

Zhuo et al.~\cite{N44} introduce a novel testing framework for code generation tasks and evaluate multiple models on two different aspects (human preferences and execution success) and four programming languages. The results show that the method based on GPT-3.5-turbo performs better. However, there are two caveats to consider regarding the results of this article: (1) The article primarily focuses on example-level Kendall-Tau and Spearman (rs) correlation tests with HumanEval based on executable functional correctness, which may not directly indicate the performance of different language models. (2) The other language models compared to GPT-3.5-turbo in the article, such as CodeBERTScore, are far behind in terms of parameter size and cannot even be considered as LLMs. Therefore, while the work in this article is interesting, we cannot draw useful conclusions from it.

Zheng et al.~\cite{N51-1} introduce CodeGeeX-13B and extensively evaluate its performance on code generation tasks in different programming languages using HumanEval-X. The experimental results of the article show that CodeGeeX-13B and CodeGen-Multi-16B outperform general LLMs, GPT-J-6B, and GPT-NeoX-20B in almost all metrics. The article also provides a comparison between PaLM-540B and PaLMCoder-540B, showing that PaLMCoder-540B performs better than PaLM-540B to some extent on the MBPP dataset.

Rozière et al.~\cite{N52} introduce Code Llama and compare it in detail with SOTA LLMs on the benchmarks of HumanEval, MBPP, and APPS. The article's results demonstrate that Code Llama achieves the best results in almost every evaluation metric. By observing the experimental settings and results, we can draw the following conclusions: (1) The evaluation of GPT-3.5 and GPT-4 in the article is not comprehensive, but based on the evaluations involving GPT-3.5 and GPT-4, they still remain highly competitive and outperform Code LLMs like StarCoder. (2) When comparing Llama 2 with Code Llama and Code Llama with Code Llama Python, Code Llama Python performs better. This also indicates that specialized instruction fine-tuning improves code generation capabilities for the same model. (3) Model size is crucial, as larger parameter models like Code Llama exhibit stronger performance.

Fu et al.~\cite{N55} introduce CodeApex, a bilingual benchmark dataset focused on testing LLMs' programming comprehension and code generation abilities. CodeApex consists of three types of multiple-choice questions: conceptual understanding, commonsense reasoning, and multi-hop reasoning. The results of the article show that GPT-3.5-turbo performs better in most of the metrics.

Zan et al.~\cite{N56} categorize LLMs based on model size and present their performance accordingly. The data results indicate that Code LLMs have better performance when the parameter sizes are similar. At the same time, the impact of parameter size on the model's capabilities may vary among different LLMs. It is observed that the three versions of CodeGen-Mono with different parameter sizes (2.7B, 6.1B, and 16.1B) show minimal differences on the MBPP and HumanEval benchmarks. On the other hand, Codex exhibits relatively significant differences between its two versions: 2.5B and 12B. The marginal effect of model parameter size on performance and the sensitivity of different models to parameter size are important factors worth discussing.

Lu et al.~\cite{N58} introduce CodeGPT, which has the same model architecture and training objective as GPT-2. CodeGPT performs better than GPT-2 in tasks such as code completion and code search. Similarly, Clement et al.~\cite{N63} compare the proposed PyMT5 with GPT-2 and reach the same conclusion. CodeT5 also outperforms GPT-2 with better performance~\cite{N75}.

Xu et al.~\cite{N82} claim that PolyCoder achieves lower complexity than all models, including Codex. However, based on the HumanEval benchmark results provided in the article, PolyCoder falls behind Codex by a considerable margin. It also performs worse than GPT-Neo with similar parameter sizes.

Li et al.~\cite{N83} introduce AlphaCode, which already surpasses GPT-Neo (APPS test) on all difficulty levels with a small 1B parameter model, and it outperforms Codex-12B on interview and contest difficulty levels. The article further analyzes the test results and finds that all encoder-decoder models, including the final AlphaCode model, perform significantly worse in HumanEval compared to decoder-only models. The article attributes this performance difference to the fact that the encoder-decoder models align well with competitive programming settings. In competitive programming settings, there is typically a dataset with clear inputs (programming contest problem descriptions) and outputs (solution code), along with example tests for effective filtering. Therefore, encoder-decoder models can learn effectively and sample efficiently. However, encoder-decoder models are not aligned with the HumanEval setting, where the only training data is GitHub code, which cannot be easily split into meaningful inputs and outputs. Thus, the fact that decoder-only models compute loss for all tokens (instead of only one-third of the tokens) enables them to achieve stronger results in the HumanEval setting. Therefore, more diverse benchmark tests may be needed to quantify the true capabilities of Code LLMs, as a single benchmark may not fully reflect their performance.

Fried et al.~\cite{N84} evaluated the performance of INCODER-6.7B, as proposed in the article, on the HumanEval and MBPP benchmarks. We can see that INCODER outperforms GPT-J-6B, which has a similar number of parameters, by a significant margin, and even performs better than the larger model GPT-NeoX-20B. However, there is still a notable gap between INCODER-6.7B and CodeGen-Mono, Codex, and other Code LLMs.

Wang et al.~\cite{N89} extensively evaluated CodeT5Mix on seven code-related tasks across 20 datasets and demonstrated that it achieves state-of-the-art performance on most tasks. However, CodeT5Mix was not directly compared to general-purpose LLMs in these metrics. In some metrics, the smaller-scale CodeT5Mix outperformed larger models like LaMDA and GPT-2. Unfortunately, the article did not provide a comparison between CodeT5Mix and GPT-3.5/GPT-4. On the other hand, WizardCoder proposed by Luo et al.~\cite{N97} was compared to GPT-3.5 and GPT-4 on HumanEval (pass@1(\%)) and MBPP. The experimental results showed that WizardCoder outperformed GPT-3.5 but was inferior to GPT-4. However, it still performed significantly better than larger models like PaLM-540B and PaLM-Coder-540B.

Li et al.~\cite{N99} also demonstrated impressive performance in leapfrogging, performing well on Common Sense Reasoning Benchmarks and surpassing general LLMs like Llama2-7B. However, it was outperformed by general LLMs on Language Understanding and Knowledge Benchmarks. The article presents results that make it difficult to assess the superiority of Code LLMs over general LLMs. It challenges the popular notion that the ability of LLMs is solely determined by their scale, indicating that data quality plays a more important role than previously believed. The article further demonstrates the feasibility of achieving high-level functionality in smaller LLMs. Perhaps the results presented in the article are far more important than assessing the performance of Code LLMs compared to general LLMs.

Some recent Code LLMs have shown better performance. OpenCoderPlus~\cite{N101} claims to achieve 102.5\% of ChatGPT score on Vicuna GPT-4 evaluation and a 78.7\ win rate on AlpacaEval. CodeFuse-CodeLlama-34B~\cite{N113} also outperformed GPT-4 in HumanEval (Pass@1). Additionally, MFTCoder~\cite{N113} published the performance of WizardCoder-Python-34B-V1.0 in HumanEval (Pass@1), which is slightly better than GPT-4. CodeGeeX2-6B~\cite{N115} also surpassed LLaMA2-70B in HumanEval (Pass@1).

Some evaluation work has focused on analyzing the performance of Code LLMs. Among them, some work has compared the performance differences between Code LLMs and general LLMs. Tang et al.~\cite{N118} evaluate results on BIOCODER showed that ChatGPT has a better understanding of the application of imported toolkits or functions contained in other files. Xia et al.~\cite{N123} explored the vulnerability repair ability of LLMs and found that as the model size increases, there is a scaling effect of performance improvement. Codex exhibited a more prominent performance. Palacio et al.~\cite{N125} evaluated LLMs' features and their corresponding AsC-Eval performance, and in all the test results, CodeGen demonstrated better performance compared to GPT-Neo and CodeParrot. Pan et al.~\cite{N126} evaluated the performance of LLMs on code translation tasks and found that, except for GPT-4 and StarCoder, all other LLMs performed poorly. Among the two, GPT-4 performed better. Kou et al.~\cite{N127} assessed the attention differences between LLMs and programmers in code generation. The article indicated that compared to models like GPT-J-6B and INCODER, CodeGen's attention is closer to that of human programmers. However, the article also found that LLMs pay more attention to different types of keywords, while programmers are more sensitive to adjectives and numbers in task descriptions.

Du et al.~\cite{N128} provided Pass@k with Nucleus Sampling on ClassEval. ClassEval is a class-level code generation benchmark test. The experimental results showed that GPT-4 and GPT-3.5 achieved better performance compared to Code LLMs like WizardCoder, Instruct-StarCoder, and CodeGen.

We organize the results from the aforementioned paper into Table~\ref{tab:RQ2}.

\begin{table*}[]
\centering
\footnotesize
\caption{The performance of Code LLMs and general LLMs in various literatures.}
\renewcommand\arraystretch{1.2}
\label{tab:RQ2}
\begin{tabular}{lllll}
\toprule
\textbf{Work} & \textbf{General LLMs}     & \textbf{Code LLMs}        & \textbf{Unknown}    & \textbf{Task}       \\ \midrule

 Jiang et al.~\cite{N2}             &                           &                           &  \checkmark            & Code Generation \\
  Maddigan et al.~\cite{N5}            &    \checkmark                        &                      &                    & Data visualization       \\
               Shirafuji et al.~\cite{N12}    &   \checkmark                         &                           &                     & Robustness of solving programming problems      \\
              Li et al.~\cite{N13}&      \checkmark                      &                           &                          & Code Generation \\
              Li et al.~\cite{N14}&                           &                           &   \checkmark                  & Text-to-SQL    \\
              Yang et al.~\cite{N15}&                           &                           &    \checkmark& Code Generation     \\
              Thakur et al.~\cite{N21}&                           &\checkmark&                       & Verilog Code Generation    \\
             Siddiq et al.~\cite{N26} &                           &                           &     \checkmark& Code Generation\\
             Sun et al.~\cite{N33} &     \checkmark&                           &                          & Text-to-SQL \\
             Sun et al.~\cite{N35} &                           &\checkmark&                        & Code Summarization   \\
             Wang et al.~\cite{N36}&          \checkmark               &                           &                         & Code Generation  \\
              Li et al.~\cite{N38}&                           &   \checkmark                       &                        & Code Generation   \\
             Li et al.~\cite{N40} &                           &   \checkmark                      &                       & Information extraction     \\
              Zhang et al.~\cite{N42}&       \checkmark                     &                           &                    & API-related Tasks        \\
              Siddiq et al.~\cite{N43} &                           &   \checkmark                       &                 & Unit test Generation          \\
             Zhuo et al.~\cite{N44}&                           &                           &    \checkmark       & Code Generation               \\
              Zheng et al.~\cite{N51-1}&                           &     \checkmark                     &                         & Code Generation  \\
             Rozière et al.~\cite{N52}&                           &   \checkmark                      &                        & Code Generation   \\
              Fu et al.~\cite{N55}&     \checkmark                         &                           &                       & Code Generation\& Understanding    \\    
              Zan et al.~\cite{N56} &                           &\checkmark&                       & Code Generation    \\
             Lu et al.~\cite{N58} &         &\checkmark &                 & Code Completion \& Code Search          \\
             Clement et al.~\cite{N63}  &                           &\checkmark &                      & Code Generation       \\
             Yue et al.~\cite{N75}&                           & \checkmark &                        & Code Generation    \\
             Xu et al.~\cite{N82} & \checkmark &                           &                   & Code Generation        \\
             Li et al.~\cite{N83} &                           & \checkmark &                  & Code Generation         \\
              Fried et al.~\cite{N84} &                           &\checkmark &              & Code Generation             \\
              Wang et al.~\cite{N89} &                           & \checkmark &                 &Seven code-related tasks          \\
               Luo et al.~\cite{N97}&                           &                           & \checkmark   & Code Generation \\
              Li et al.~\cite{N99}&                           &                           & \checkmark & Code Generation \\
             Guan et al.~\cite{N101}&                           & \checkmark &                      & Code Generation     \\
             MFTCoder~\cite{N113} &                           & \checkmark&                      & Code Generation     \\
              THUDM~\cite{N115}&                           & \checkmark &                     & Code Generation      \\
             Tang et al.~\cite{N118} & \checkmark  &                           &                     & Understanding functions      \\
             Xia et al.~\cite{N123} &                           & \checkmark &                        &Vulnerability Remediation   \\
              Palacio et al.~\cite{N125}&                           & \checkmark &             &  Code Syntax Understanding            \\
             Pan et al.~\cite{N126} &\checkmark  &                           &                    &Code Translation       \\
              Kou et al.~\cite{N127}&                           & \checkmark &                   & Attention to Programming Problems        \\
              Du et al.~\cite{N128}& \checkmark  &                           &       & Code Generation                    \\\hline
              \textbf{Total Number}       & \textbf{11}         & \textbf{20}         & \textbf{7}      \\ \bottomrule

\end{tabular}
\end{table*}

From the Table~\ref{tab:RQ2}, we can see that out of the 38 papers, 11 experiments from various works showed that general LLMs perform better on software engineering tasks, while in 20 experiments, Code LLMs demonstrated superior performance. There are 7 papers where we couldn't obtain a clear conclusion regarding whether Code LLMs or general LLMs perform better based on their experimental sections or conclusions.

Based on the above findings, it may be challenging to arrive at a definitive conclusion regarding whether Code LLMs or general LLMs are better. Firstly, the current state-of-the-art general LLMs, such as GPT-4 and GPT-3.5-turbo, are not open-sourced, which limits the ability to compare them in many metrics. Additionally, there are restrictions when accessing the APIs of these models. Secondly, due to the varying publication dates of each work, the choice of Code LLMs and general LLMs differs among them~\cite{N35}, leading to variations in the evaluation results. For example, earlier works often used GPT-2 and GPT-J as baselines for general LLMs~\cite{N127}, while more recent works typically use Llama and GPT-3.5 as baselines~\cite{N26}. Thirdly, the selection of benchmarks and software engineering tasks differs among these studies~\cite{N55,N128}. Lastly, different works may select different versions of the same model with varying parameter sizes, leading to significant performance differences for that model across different studies~\cite{N15}.

Based on the results from the aforementioned literature, we can still draw the following conclusions:
\begin{itemize}
    \item Model parameters have a significant impact on performance, with larger-scale models often exhibiting better performance within the same model architecture~\cite{N123,N56,N52}.
    \item For the same model, fine-tuning the model specifically for software engineering tasks generally leads to improved performance compared to the base model~\cite{N38,N42}.
    \item When parameter sizes are comparable, Code LLMs tend to outperform general LLMs~\cite{N84,N89}.
    \item Currently, the state-of-the-art Code LLMs (such as CodeFuse-CodeLlama-34B and OpenCoderPlus) outperform the current state-of-the-art general LLMs (GPT-4) in code generation tasks~\cite{N101,N103}.
\end{itemize}

Furthermore, we can derive an interesting finding from the literature:
\begin{itemize}
    \item The impact of parameter size on model performance may vary among different LLMs. In Zan et al.~\cite{N56}, it is observed that different versions of CodeGen-Mono with parameter sizes of 2.7B, 6.1B, and 16.1B exhibit relatively small differences in performance on the MBPP and HumanEval benchmarks. On the other hand, the variations between the two versions of Codex, 2.5B and 12B, are relatively larger. Therefore, for Code LLMs, the marginal effect of model parameter size on performance and the sensitivity of different models to parameter size could be a topic worth discussing.
    \item All encoder-decoder models perform significantly worse in HumanEval compared to decoder-only models. Li et al.~\cite{N83} suggest that this discrepancy arises from the fact that encoder-decoder models are not well-aligned with the HumanEval setup, whereas they align well with competitive programming settings. In competitive programming, there is typically a well-defined input (programming competition problem description) and output (solution code). Encoder-decoder models may struggle to split the input and output into meaningful segments for HumanEval, thus calculating the loss for all tokens in the decoder-only models enables them to achieve stronger results in the HumanEval setting. Therefore, a more diverse set of evaluation benchmarks may be needed to accurately quantify the true capabilities of Code LLMs, as a single benchmark may not fully capture their performance.
    \item The abilities of LLMs may not be solely determined by their model size~\cite{N99}, and data quality can play a crucial role. This is exemplified by phi-1.5-1.3B, which has been able to outperform LLMs with parameter sizes five times larger.
\end{itemize}

%% file: RQ3.tex
\section{Which software engineering tasks are different LLMs particularly good at?}
\label{sec:RQ3}
In this section, we primarily focus on addressing and summarizing \textbf{RQ3}. First, we will gather the relevant papers and extract the sections that evaluate LLMs in software engineering tasks. Next, we will classify these works according to different software engineering tasks and further categorize them based on the evaluation benchmarks used. Our goal is to compile a relatively comprehensive list for each benchmark.

\subsection{Code Generation}


Code generation is currently one of the most prominent software engineering tasks of interest for Code LLMs. Consequently, there is a rich variety of benchmarks available to evaluate LLMs in code generation tasks. Some commonly used benchmarks include HumanEval~\cite{N150}, DS-1000~\cite{N151}, and MBPP~\cite{N152}. Additionally, there have been efforts to propose new benchmarks from different perspectives to assess the code generation capabilities of LLMs. For instance, Muennighoff et al.\cite{N18} introduced HUMANEVALPACK to evaluate the multilingual code generation ability of LLMs, and Du et al.\cite{N128} proposed ClassEval, a class-level code generation benchmark.

However, to the best of our knowledge, there is currently no organization or individual actively maintaining these benchmarks. Therefore, it is challenging to find a comprehensive list that showcases the performance of LLMs on a specific benchmark. In this section, we will organize LLMs according to different benchmarks to facilitate a better comparison of their performance.

\textbf{HumanEval: }HumanEval consists of 164 handwritten Python programming problems. Each problem provides a prompt that includes a description of the function to be generated, the function signature, and example test cases in the form of assertions. The models are required to complete the functionality based on the prompt, aiming to pass all the provided test cases, thereby measuring their performance in terms of functional correctness. HumanEval is currently the most commonly used benchmark to test the code generation capabilities of LLMs. We have organized the results from the collected papers on HumanEval and obtained Table~\ref{tab:RQ3-humaneval}. In the case of *, this result is sourced from Shen et al.~\cite{N23}.

We have marked the top five scores from each evaluation data using underlines and bold formatting. From the table, we can see that LATS( GPT-4 based )~\cite{N152} performs the best in Pass@1. In Pass@10, the best performer is Unnatural-Code-LLaMA-34B~\cite{N52}, and it remains the best in Pass@100 as well. Considering the results from all three criteria, Unnatural-Code-LLaMA-34B is currently the top-performing LLM (Language Model) according to HumanEval. To a certain extent, GPT-4 and Unnatural-Code-LLaMA-34B is also the best-performing LLM for code generation tasks at the moment. The next best performers are Code-LLaMA-Python-34B and Code-LLaMA-Python-13B.

\begin{table*}[]
\centering
\scriptsize
\caption{Performance of LLMs in HumanEval Benchmark.}
\setlength{\tabcolsep}{0.5mm}
\renewcommand\arraystretch{1}
\label{tab:RQ3-humaneval}
\begin{tabular}{llll|llll}
\toprule
\multirow{2}{*}{\textbf{LLMs}}         & \multicolumn{3}{l|}{\textbf{HumanEval}} & \multirow{2}{*}{\textbf{LLMs}} & \textbf{HumanEval} &         &          \\ \cline{2-4} \cline{6-8} 
                              & \textbf{Pass@1}  & \textbf{Pass@10}  & \textbf{Pass@100}  &                       & \textbf{Pass@1}    & \textbf{Pass@10} & \textbf{Pass@100} \\ \midrule
LATS(GPT-4 based)             & \underline{\textbf{94.4}}    & -        & -         & LLaMA2-13B            & 20.1      & 34.8    & 61.2     \\
Reflexion(GPT-4 based)        & \underline{\textbf{91}}      & -        & -         & CodeGen-multi-16.1B   & 19.22     & 34.64   & 55.17    \\
LATS(GPT-3.5 based)           & \underline{\textbf{86.9}}    & -        & -         & CodeGen-16B-multi     & 19.2      & 34.6    & 55.2     \\
Parsel                        & 85.1    & -        & -         & CodeGen2-7B           & 18.83     & 31.78   & 50.41    \\
GPT-4(*)                      & 82      & -        & -         & CodeGen-multi-16B     & 18.3      & -       & -        \\
MetaGPT                       & 81.7    & -        & -         & CodeGen-multi-6.1B    & 18.16     & 27.81   & 44.85    \\
CodeFuse-CodeLlama-34B        & 74.4    & -        & -         & SantaCoder-1.1B       & 18        & 29      & 49       \\
Phind-CodeLlama-34B-v2        & 73.8    & -        & -         & AlphaCode-1.1B        & 17.1      & 28.2    & 45.3     \\
WizardCoder-Python-34B        & 73.2    & -        & -         & PanGu-Coder-317M      & 17.07     & 24.05   & 34.55    \\
Phind-CodeLlama-Python-34B-v1 & 69.5    & -        & -         & Codex-679M            & 16.22     & 25.70   & 40.95    \\
GPT-3.5(*)                    & 68.9    & -        & -         & PaLM-62B              & 15.9      & -       & 46.3     \\
Phind-CodeLlama-34B-v1        & 67.6    & -        & -         & LLaMA-13B             & 15.8      & -       & 52.5     \\
GPT-4(OpenAI)                 & 67      & -        & -         & BLOOM-176B            & 15.52     & 32.20   & 55.45    \\
Unnatural-Code-LLaMA-34B      & 62.2    & \underline{\textbf{85.2}}     & \underline{\textbf{95.4}}      & CodeT5+-770M          & 15.5      & 27.2    & 42.7     \\
PanGu-Coder2-15B              & 61.64   & \underline{\textbf{79.55}}    & 91.75    & GPT-NeoX-20B          & 15.4      & 25.6    & 41.2     \\
WizardCoder-15B               & 57.3    & 73.2     & 90.46     & InCoder-6.7B          & 15.2      & 27.8    & 47.0     \\
Code-LLaMA-Python-34B         & 53.7    & \underline{\textbf{82.8}}     & \underline{\textbf{94.7}}      & InCoder-multi-6.7B    & 15.2      & 27.8    & 47.0     \\
Phi-1-1.3B                    & 50.6    & -        & -         & InCoder-6B            & 15.2      & 27.8    & 47.0     \\
Code-LLaMA-34B                & 48.8    & 76.8     & \underline{\textbf{93.0}}      & CodeGen-multi-2.7B    & 14.51     & 24.67   & 38.56    \\
GPT-3.5(OpenAI)               & 48.1    & -        & -         & Codegen-NL-16.1B      & 14.24     & 23.46   & 38.33    \\
code-davinci-002              & 47.0    & 74.9     & 92.1      & AlphaCode(dec)-685M   & 14.2      & 24.4    & 38.8     \\
OctoCoder                     & 46.2    & -        & -         & LaMDA-137B            & 14.0      & -       & 47.3     \\
Code-LLaMA-Python-13B         & 43.3    & 77.4     & \underline{\textbf{94.1}}     & Codex-300M            & 13.17     & 20.37   & 36.27    \\
Code-LLaMA-Instruct-13B       & 42.7    & 71.6     & 91.6      & CodeGen-Mono-350M     & 12.76     & 23.11   & 35.19    \\
Code-LLaMA-Instruct-34B       & 41.5    & 77.2     & \underline{\textbf{93.5 }}     & CodeGen-mono-350M     & 12.76     & 23.11   & 35.19    \\
StarCoder-Prompted-15B        & 40.8    & -        & -         & LLaMA2-7B             & 12.2      & 25.2    & 44.4     \\
code-davinci-001              & 39      & 60.6     & 84.1      & CodeT5-770M           & 12.09     & 19.24   & 30.93    \\
Code-LLaMA-Python-7B          & 38.4    & 70.3     & 90.6      & CodeT5+-220M          & 12.0      & 20.7    & 31.6     \\
PaLM-2-S                      & 37.6    & -        & 88.4      & GPT-J-6B              & 11.62     & 15.74   & 27.74    \\
PaLM-Coder-540B               & 36      & -        & 88.4      & AlphaCode-302M        & 11.6      & 18.8    & 31.8     \\
Code-LLaMA-13B                & 36.0    & 69.4     & 89.8      & LLaMA-7B              & 10.5      & -       & 36.5     \\
CodeGeeX2-6B                  & 35.9    & 62.6     & 88.3      & CODEGEN-NL-6.1B       & 10.43     & 18.36   & 29.85.   \\
InstructCodeT5+-16B           & 35.0    & 54.5     & 77.9      & InCoder-1.3B          & 8.9       & 16.7    & 25.6     \\
Code-LLaMA-Instruct-7B        & 34.8    & 64.3     & 88.1      & PyCodeGPT-110M        & 8.33      & 13.36   & 19.13    \\
CodeGen-16.1B                 & 34.6    & -        & -         & Codex-85M             & 8.22      & 12.81   & 22.40    \\
StarCoder-Python-15B          & 33.6    & -        & -         & BLOOM-7.1B            & 7.73      & 17.38   & 29.47    \\
StarCoder-15B                 & 33.60   & 45.78    & 79.82     & CODEGEN-NL-2.7B       & 6.7       & 14.15   & 22.84    \\
code-cushman-001              & 33.5    & 54.3     & 77.4      & CODEGEN-MULTI-350M    & 6.67      & 10.61   & 16.84    \\
Code-LLaMA-7B                 & 33.5    & 59.6     & 85.9      & BLOOM-3B              & 6.48      & 11.35   & 20.43    \\
CodeGen2.5-7B-mono            & 33.4    & 58.4     & 82.7      & GPT-NEO-2.7B          & 6.41      & 11.27   & 21.37    \\
CodeT5+-16B-mono              & 30.9    & 51.6     & 76.7      & PolyCoder-2.7B        & 5.59      & 9.84    & 17.68    \\
MIM-2.7B                      & 30.7    & 48.22    & 69.6      & JuPyT5-300M           & 5.4       & 15.46   & 25.60    \\
Replit-Finetuned-2.7B         & 30.5    & -        & -         & Codex-42M             & 5.06      & 8.8     & 15.55    \\
LLaMA2-70B                    & 30.5    & 59.4     & 87.0      & GPT-Neo-1.3B(1.5B)    & 4.79      & 7.47    & 16.30    \\
StarCoderBase-15B             & 30.4    & -        & -         & GPT-Neo-1.3B          & 4.79      & 7.47    & 16.3     \\
CodeGen-mono-16B(16.1B)       & 29.28   & 49.86    & 75.00     & AlphaCode(dec)-89M    & 4.3       & 12.2    & 20.0     \\
Codex-12B                     & 28.81   & 46.81    & 72.31     & AlphaCode(dec)-55M    & 4.2       & 8.2     & 16.9     \\
CodeGen2.5-7B                 & 28.36   & 47.46    & 75.15     & BLOOM-1.7B            & 4.03      & 7.45    & 12.75    \\
CodeT5+-6B                    & 28.0    & 47.2     & 69.8      & CodeParrot-multi-1.5B & 4.0       & 8.7     & 17.9     \\
PaLM-540B                     & 26.2    & -        & 76.2      & CodeParrot-1.5B       & 3.99      & 8.69    & 17.88    \\
CodeGen-Mono-6.1B             & 26.13   & 42.29    & 65.82     & CodeParrot-110M       & 3.8       & 6.57    & 12.78    \\
CodeGen-mono-6B               & 26.1    & 42.3     & 65.8      & CodeParrot-1.5B       & 3.8       & 6.57    & 12.78    \\
CodeT5+-2B                    & 24.2    & 38.2     & 57.8      & PaLM-8B               & 3.6       & -       & 18.7     \\
PanGu-Coder-2.6B              & 23.78   & 35.36    & 51.24     & CodeParrot-small-110M & 3.58      & 8.03    & 14.96    \\
CodeGen-Mono-2.7B             & 23.7    & 36.64    & 57.01     & AlphaCode(dec)-29M    & 3.4       & 5.8     & 11.2     \\
LLaMA-65B                     & 23.7    & -        & 79.3      & Codex-25M             & 3.21      & 7.1     & 12.89    \\
CodeGen-mono-2B               & 23.7    & 36.6     & 57        & PolyCoder-400M        & 2.96      & 5.29    & 11.59    \\
CodeGeeX-13B                  & 22.89   & 39.57    & 60.92     & BLOOM-1.1B            & 2.48      & 5.93    & 9.62     \\
LLaMA2-34B                    & 22.6    & 47.0     & 79.5      & PolyCoder-160M        & 2.13      & 3.35    & 4.88     \\
MIM-1.3B                      & 22.4    & 41.7     & 53.8      & CODEGEN-NL-350M       & 2.12      & 4.10    & 7.38     \\
Replit-3B                     & 21.9    & -        & -         & Codex-12M             & 2         & 3.62    & 8.58     \\
Replit-2.7B                   & 21.9    & -        & -         & AlphaCode-13M    & 1.5       & 3.6     & 8.6      \\
LLaMA-33B                     & 21.7    & -        & 70.7      & GPT-NEO-350M          & 0.85      & 2.55    & 5.95     \\
Codex-2.5B                    & 21.36   & 35.42    & 59.50     & BLOOM-560M            & 0.82      & 3.02    & 5.91     \\
CodeGen2-16B                  & 20.46   & 36.5     & 56.71     & GPT-Neo-125M          & 0.75      & 1.88    & 2.97     \\ \bottomrule
\end{tabular}
\end{table*}

It is important to note that there may be variations in the data reported for the same model across different papers. We have selected data from more recent publications. Additionally, for cases where there are differences in the reported data, we have included them in Table~\ref{tab:RQ3-differ} as well.

\begin{table}[]
\centering
\footnotesize
\caption{Performance inconsistency reported in different works on HumanEval.}
\setlength{\tabcolsep}{9mm}
\renewcommand\arraystretch{1}
\label{tab:RQ3-differ}
\begin{tabular}{llll}
\toprule \multirow{2}{*}{ \textbf{LLMs} } & \multicolumn{3}{l}{ \textbf{HumanEval} } \\
\cline { 2 - 4 } & \textbf{Pass@1} & \textbf{Pass@10} & \textbf{Pass@100} \\
 \midrule
 InCoder-1.3B~\cite{N36} & 8.9 & 16.7 & 25.6 \\

 InCoder-1.3B~\cite{N56} & 11.09 & 16.14 & 24.20 \\
 CodeGen-multi-16.1B~\cite{N51-1} & 19.22 & 34.64 & 55.17 \\
 CodeGen-multi-16.1B~\cite{N76} & 18.32 & 32.07 & 50.80 \\
\bottomrule
\end{tabular}
\end{table}

There are also some benchmarks such as APPS~\cite{N153} and CodeXGLUE~\cite{N58} where the available data is limited, making it difficult to showcase performance differences among different LLMs. Therefore, we won't be presenting them here. Additionally, there are more complex benchmarks that require different settings for different tasks, such as HumanEvalX. As a result, each paper provides different experimental setups and results. It is also challenging to observe the performance of these LLMs in the context of HumanEvalX.

There have been various studies that have explored the performance of LLMs in code generation from different perspectives. Zan et al.~\cite{N22}, focused on evaluating LLMs' performance in generating code tailored to private libraries. The study proposed four benchmark metrics for private libraries, including TorchDataEval, TorchDataComplexEval, MonkeyEval, and BeatNumEval. Based on the data provided in the paper, we can observe that the proposed CodeGenAPI-6B performs the best in this task. CodeGenAPI-6B is developed based on CodeGen. If we exclude CodeGenAPI, then the best performer is code-davinci-002 (Codex), followed by CodeGen-6B.

Siddiq et al.~\cite{N26} introduced a lightweight framework for recommending more secure source code derived from LLMs. The article also provides performance results for several LLMs, as shown in Tables~\ref{tab:RQ3-N26-1}and Tables~\ref{tab:RQ3-N26-2}. We can see that GPT-3.5 performs better than other LLMs. However, it's important to note that the evaluated LLMs in the article may not be state-of-the-art LLMs, so the obtained results have limited value as a comparison to current top-performing models. However, from the tables, we can also observe that the performance of the same LLM can vary significantly across different programming languages.

\begin{table*}[]
\centering
\tiny
\caption{Percentage of compilable suggestions (code snippets) in~\cite{N26}.}
\setlength{\tabcolsep}{1mm}
\renewcommand\arraystretch{1}
\label{tab:RQ3-N26-1}
\begin{tabular}{p{25pt} p{27pt}p{27pt}p{27pt}p{27pt}p{27pt}p{27pt}p{27pt}p{27pt}p{27pt}p{27pt}p{27pt}}
\toprule  & \textbf{Code Parrot small} & \textbf{CodeParrot regular} & \textbf{InCoder 1B} & \textbf{CodeGen 350M mono}  &  \textbf{CodeGen 350M multi}  &  \textbf{CodeGen 2B mono} &  \textbf{CodeGen 2B multi} &  \textbf{PolyCoder 160M} & \textbf{PolyCoder 0.4B} & \textbf{PolyCoder 2.7B} & \textbf{GPT 3.5} \\
\midrule  
 Java     & - & $0.15 \%$ & - & $5.86 \%$ & - & $7.43 \%$ & $0.13 \%$ & $0.29 \%$ & $0.19 \%$ & $0.89 \%$ \\
Python  & $22.22 \%$ & $25.98 \%$ & & $34.56 \%$ & $37.06 \%$ & $34.29 \%$ & $40.49 \%$ & $19.73 \%$ & $21.33 \%$ & $22.72 \%$ & $57.75 \%$ \\
\bottomrule
\end{tabular}
\end{table*}

\begin{table*}[]
\centering
\tiny
\caption{NDCG@10 scores for the original model ranking in~\cite{N26}.}
\setlength{\tabcolsep}{1mm}
\renewcommand\arraystretch{1}
\label{tab:RQ3-N26-2}
\begin{tabular}{p{25pt} p{27pt}p{27pt}p{27pt}p{27pt}p{27pt}p{27pt}p{27pt}p{27pt}p{27pt}p{27pt}p{27pt}}
\toprule  & \textbf{Code Parrot small} & \textbf{CodeParrot regular} & \textbf{InCoder 1B} & \textbf{CodeGen 350M mono}  &  \textbf{CodeGen 350M multi}  &  \textbf{CodeGen 2B mono} &  \textbf{CodeGen 2B multi} &  \textbf{PolyCoder 160M} & \textbf{PolyCoder 0.4B} & \textbf{PolyCoder 2.7B} & \textbf{GPT 3.5} \\
\midrule  
 Java     & - & - & 0.0979 & 0.2330 & - & - & 0.3019 & 0.1385 & 0.1525 & 0.2264 & 0.5775 \\
Python  & 0.3297 & 0.4021 & 0.2012 & 0.3944 & 0.3640 & 0.4745 & 0.4738 & 0.3740 & 0.4200 & 0.4541 & 0.3742 \\
\bottomrule
\end{tabular}
\end{table*}

Zhang et al.~\cite{N42} also evaluated the performance of several LLMs on PublicLibrary Benchmark and Private Library Benchmark, as shown in Tables~\ref{tab:RQ3-N42} and Tables~\ref{tab:RQ3-N42-2}. Although the primary focus of Zhang et al~\cite{N42} was to introduce the proposed ToolCoder, we can also observe the differences in LLMs' ability to call APIs during code generation from Tables A and B. Based on the results provided in the article, ToolCoder is undoubtedly the best performer, followed by GPT-3.5. However, the performance of CodeGenAPI, which is CodeGen fine-tuned for this task, did not surpass that of CodeGen.

\begin{table*}[]
\centering
\footnotesize
\caption{Pass rates of models on public library benchmarks in~\cite{N42}.}
\setlength{\tabcolsep}{1.2mm}
\renewcommand\arraystretch{1}
\label{tab:RQ3-N42}
\begin{tabular}{p{130pt} p{35pt} p{35pt} p{35pt} p{35pt} p{35pt} p{35pt} p{35pt}}
\toprule 
\multirow{2}{*}{ \textbf{Model} }  & \multicolumn{2}{l}{ \textbf{NumpyEval} } & \multicolumn{2}{l}{ \textbf{PandasEval} } & \multicolumn{2}{l}{ \textbf{TorchDataEval} } \\
\cline { 2 - 7 }  & \textbf{pass@1} & \textbf{pass@10} & \textbf{pass@1} & \textbf{pass@10} & \textbf{pass@1} & \textbf{pass@10} \\
\midrule CodeT5-220M & 0 & 0.1 & 0 & 0 & 0 & 0 \\
 PyCodeGPT-11M & 18.04 & 38.61 & 12.75 & 37.62 & 3.80 & 14.00 \\
 CodeGen-350M & 18.51 & 43.56 & 16.73 & 29.70 & 4.60 & 14.00 \\
 CodeGen2B-2B & 29.10 & 53.46 & 30.69 & 42.57 & 7.00 & 18.00 \\
 GPT-3.5  & 58.41 & 66.21 & 30.09 & 33.16 & 6.00 & 24.00 \\
 CERT-numpy-220M & 31.47 & 46.42 & 16.03 & 27.72 & 2.20 & 14.00 \\
 CERT-pandas-220M & 18.81 & 33.66 & 28.42 & 48.04 & 2.80 & 6.00 \\
 CodeGenAPI-350M & 16.55 & 29.48 & 13.58 & 34.95 & 7.19 & 16.93 \\
 CodeGenAPI-retrieval-475M & 12.67 & 27.32 & 11.25 & 28.61 & 10.41 & 23.50 \\
 CodeGen-retrieval-475M & 18.30 & 35.12 & 9.54 & 29.02 & 7.52 & 16.36 \\
 ToolCoder-OnlineTool-350M & 35.64 & 50.50 & 22.77 & 37.62 & 7.40 & 20.00 \\
 ToolCoder-OnlineTool-2B & 41.58 & 55.44 & 31.68 & 47.52 & 11.80 & 24.00 \\
\bottomrule
\end{tabular}
\end{table*}

\begin{table}[]
\centering
\footnotesize
\caption{Pass rates of models on private library benchmarks in~\cite{N42}.}
\setlength{\tabcolsep}{5.8mm}
\renewcommand\arraystretch{1}
\label{tab:RQ3-N42-2}
\begin{tabular}{lllll}
\toprule \multirow{2}{*}{ \textbf{Model} }  & \multicolumn{2}{l}{ \textbf{MonkeyEval} } & \multicolumn{2}{l}{ \textbf{BeatNumEval} } \\
\cline { 2 - 5 }  & \textbf{pass@1} & \textbf{pass@10} & \textbf{pass@1} & \textbf{pass@10} \\
\midrule 
CodeT5-220M & 0 & 0 & 0 & 0 \\
CodeGen-350M  & 0.95 & 4.90 & 5.15 & 11.96 \\
CodeGen-2B & 1.59 & 5.94 & 5.94 & 11.88 \\
GPT-3.5  & 2.47 & 8.91 & 6.68 & 17.82 \\
CodeGenAPI-350M & 1.19 & 4.68 & 4.44 & 8.24 \\
CodeGenAPI-retrieval-475M & 3.41 & 8.33 & 5.90 & 11.79 \\
CodeGen-retrieval-475M & 2.46 & 6.35 & 6.65 & 13.68 \\
ToolCoder-DocTool-350M & 2.98 & 5.94 & 6.73 & 12.87 \\
ToolCoder-DocTool-2B & 3.02 & 7.92 & 6.93 & 13.86 \\
\bottomrule
\end{tabular}
\end{table}

Zhong et al.~\cite{N124} also evaluated the issue of API misuse in code generation by LLMs. Table~\ref{tab:RQ3-N124} in the article presents the test results for GPT-3.5/GPT-4/Llama/Vicunad on the ROBUSTAPI benchmark. The misuse rate refers to the proportion of misuse cases among executable cases, exec. The sample represents the proportion of executable cases among all questions, and the overall misuse percentage is the proportion of misuse cases among all questions. However, from Table~\ref{tab:RQ3-N124}, it is difficult to determine which LLM performs better. We can only conclude that LLMs commonly exhibit API misuse issues, even when generating code that is executable and aligns with the user's intent.

\begin{table*}[]
\centering
\footnotesize
\caption{Performance of LLMs on ROBUSTAPI in~\cite{N124}.}
\setlength{\tabcolsep}{1.4mm}
\renewcommand\arraystretch{1}
\label{tab:RQ3-N124}
\begin{tabular}{p{30pt} p{30pt}p{30pt}p{30pt}p{30pt}p{30pt}p{30pt}p{30pt}p{30pt}p{30pt}}
\toprule \textbf{LLMs} & \multicolumn{3}{l}{ \textbf{Zero-shot} } & \multicolumn{3}{l}{ \textbf{One-shot-irrelevant} } & \multicolumn{3}{l}{ \textbf{One-shot-relevant} } \\
\cline { 2 - 10 } & $\begin{array}{l}\textbf { Misuse } \\
\textbf { Rate } \end{array}$ & $\begin{array}{l}\textbf { Exec. } \\
\textbf { Sample  } \end{array}$ & $\begin{array}{l}\textbf { Overall } \\
\textbf { Misuse } \end{array}$ & $\begin{array}{l}\textbf { Misuse } \\
\textbf { Rate } \end{array}$ & $\begin{array}{l}\textbf { Exec. } \\
\textbf { Sample } \end{array}$ & $\begin{array}{l}\textbf { Overall } \\
\textbf { Misuse } \end{array}$ & $\begin{array}{l}\textbf { Misuse } \\
\textbf { Rate } \end{array}$ & $\begin{array}{l}\textbf { Exec. } \\
\textbf { Sample  } \end{array}$ & $\begin{array}{l}\textbf { Overall } \\
\textbf { Misuse  } \end{array}$ \\
\midrule GPT-3.5 & $62.97 \%$ & $79.14 \%$ & $49.83 \%$ & $68.09 \%$ & $91.06 \%$ & $62.00 \%$ & $38.56 \%$ & $80.71 \%$ & $31.13 \%$ \\
GPT-4 & $68.81 \%$ & $90.23 \%$ & $62.09 \%$ & $70.38 \%$ & $91.39 \%$ & $64.32 \%$ & $54.40 \%$ & $90.40 \%$ & $49.17 \%$ \\
Llama-2 & $7.34 \%$ & $9.02 \%$ & $0.66 \%$ & $61.36 \%$ & $80.13 \%$ & $49.17 \%$ & $64.47 \%$ & $72.93 \%$ & $47.02 \%$ \\
Vicuna-1.5 & $45.66 \%$ & $37.17 \%$ & $16.97 \%$ & $57.85 \%$ & $83.86 \%$ & $48.51 \%$ & $42.53 \%$ & $64.24 \%$ & $27.32 \%$ \\
\bottomrule
\end{tabular}
\end{table*}

Tu et al.~\cite{N31} introduce and define the problem of error code completion, where given a problem statement and partially coded program with potential errors, the task is to complete the coding program. The article selects InCoder and CodeGen as the compared Code LLMs in the experiments. According to the experimental data provided in Table~\ref{tab:RQ3-N31}, CodeGen-2B consistently achieves better results across various metrics. However, it's important to note that the article only compares InCoder and CodeGen, which limits the scope of the results presented in the article.

\begin{table}[]
\centering
\footnotesize
\caption{Pass@1 of completion methods on buggy-HumanEval and buggy-FixEval datasets in~\cite{N31}.}
\setlength{\tabcolsep}{3.5mm}
\renewcommand\arraystretch{1}
\label{tab:RQ3-N31}
\begin{tabular}{llllllllll}
\toprule \multirow{2}{*}{ \textbf{Prefix} } & \multirow{2}{*}{ \textbf{Method} } & \multicolumn{3}{l}{\textbf{ buggy-HumanEval} } & \multicolumn{4}{l}{\textbf{ buggy-FixEval} } \\
\cline { 3 - 10 } & & \multicolumn{2}{l}{ \textbf{CodeGen} } & \multicolumn{2}{l}{ \textbf{InCoder} } & \multicolumn{2}{l}{ \textbf{CodeGen} } & \multicolumn{2}{l}{ \textbf{InCoder} } \\
\cline { 3 - 10 } & & $350 \mathrm{M}$ & $2 \mathrm{~B}$ & $1 \mathrm{~B}$ & $6 \mathrm{~B}$ & $350 \mathrm{M}$ & $2 \mathrm{~B}$ & $1 \mathrm{~B}$ & $6 \mathrm{~B}$ \\
\midrule  Clean  & completion & 43.0 & 54.9 & 41.1 & 50.6 & 27.6 & 37.8 & 24.1 & 32.3 \\
  Buggy  & completion & 0.7 & 3.1 & 0.5 & 1.0 & 2.4 & 4.3 & 1.2 & 1.8 \\
\bottomrule
\end{tabular}
\end{table}

Fu et al.~\cite{N55} introduce CodeApex, a benchmark focused on evaluating LLMs' programming understanding and code generation capabilities. In terms of programming understanding, GPT-3.5-turbo consistently ranks first in almost all tasks, followed by InternLM-Chat-7B. The experimental conclusions for code generation are similar, with GPT-3.5-turbo performing the best. Additionally, WizardCoder-15B also achieves very good results. It's worth noting that the article's evaluation of LLMs is limited, and there is relatively little work that has tested LLMs using CodeApex.

Du et al.~\cite{N128} attempted to evaluate LLMs' class-level code generation capabilities and introduced a class-level code generation benchmark called ClassEval. The article also presents the performance of 11 state-of-the-art LLMs on ClassEval, as shown in Table~\ref{tab:RQ3-N128}. We can observe that GPT-4 achieves the best performance in almost all metrics, followed by GPT-3.5. Among the Code LLMs, WizardCoder achieves the best scores.
\begin{table}[]
\centering
\footnotesize
\caption{Pass@k with Nucleus Sampling on ClassEval in~\cite{N128}.}
\setlength{\tabcolsep}{4mm}
\renewcommand\arraystretch{1}
\label{tab:RQ3-N128}
\begin{tabular}{lllllll}
\toprule \multirow{2}{*}{\textbf{Model} } & \multicolumn{3}{l}{ \textbf{Class-level} } & \multicolumn{3}{l}{ \textbf{Method-level} } \\
\cline { 2 - 7 } & \textbf{Pass@1} & \textbf{Pass@3} & \textbf{Pass@5} & \textbf{Pass@1} & \textbf{Pass@3} & \textbf{Pass@5} \\
\midrule GPT-4 & $\mathbf{3 7 . 6 \%}$ & $\mathbf{4 1 . 3 \%}$ & $\mathbf{4 2 . 0 \%}$ & $\mathbf{6 2 . 8 \%}$ & $\mathbf{6 7 . 4 \%}$ & $\mathbf{6 8 . 5} \%$ \\
 GPT-3.5 & $29.6 \%$ & $34.9 \%$ & $36.0 \%$ & $50.4 \%$ & $59.0 \%$ & $61.1 \%$ \\
 WizardCoder & $12.2 \%$ & $20.0 \%$ & $23.0 \%$ & $35.2 \%$ & $47.1 \%$ & $51.1 \%$ \\
 Instruct-StarCoder & $10.2 \%$ & $12.7 \%$ & $14.0 \%$ & $23.1 \%$ & $26.5 \%$ & $27.7 \%$ \\
 SantaCoder & $8.6 \%$ & $9.9 \%$ & $10.0 \%$ & $27.7 \%$ & $33.0 \%$ & $34.9 \%$ \\
 Instruct-CodeGen & $8.2 \%$ & $12.3 \%$ & $13.0 \%$ & $24.9 \%$ & $34.3 \%$ & $37.1 \%$ \\
 CodeGeeX & $7.2 \%$ & $9.4 \%$ & $10.0 \%$ & $21.2 \%$ & $27.1 \%$ & $29.5 \%$ \\
 InCoder & $6.2 \%$ & $7.6 \%$ & $8.0 \%$ & $21.1 \%$ & $26.5 \%$ & $29.1 \%$ \\
 Vicuna & $3.0 \%$ & $3.6 \%$ & $4.0 \%$ & $11.0 \%$ & $15.8 \%$ & $18.4 \%$ \\
 ChatGLM & $1.4 \%$ & $2.6 \%$ & $3.0 \%$ & $8.2 \%$ & $11.2 \%$ & $12.4 \%$ \\
 PolyCoder & $1.4 \%$ & $2.2 \%$ & $3.0 \%$ & $13.2 \%$ & $17.5 \%$ & $19.6 \%$ \\
\bottomrule
\end{tabular}
\end{table}

Yu et al.~\cite{N130} introduce a benchmark called CoderEval to evaluate the performance of models in generating practical code. Compared to the HumanEval benchmark, CoderEval includes programming tasks from various open-source projects and provides full coverage testing to assess models' performance in practical code generation. The article does not conduct large-scale evaluation experiments on existing LLMs but compares the performance of CodeGen, Codex, and PanGu-Coder on CoderEval. The results indicate that Codex performs better in various testing scenarios.

Based on the information provided, we can see that the Code-LLaMA series of LLMs performs well in several commonly used code generation benchmarks. Among them, Unnatural-Code-LLaMA-34B stands out with outstanding performance. For API-related code generation tasks, ToolCoder performs better. Additionally, GPT-4 and GPT-3.5 (GPT-3.5-turbo) also exhibit good performance.

\subsection{Test Case Generation}

In Schäfer et al.~\cite{N3}, GPT-3.5-turbo, StarCoder, and code-cushman-002 were tested for the number of generated tests and the percentage of passed generated tests. The experimental results in the article showed that the code-Cushman-002 model had a test coverage rate comparable to that of GPT-3.5-turbo, with the latter having slightly higher median statement and branch coverage. StarCoder exhibited relatively poorer performance in comparison.

Shirafuji et al.~\cite{N12} aims to demonstrate the high capability of LLMs in solving a wide range of programming problems, specifically their robustness in solving programming problems. Therefore, the article assigns each LLM to generate 100 programs for each of the 40 questions and tests their performance, as shown in Table~\ref{tab:RQ3-N12}. On average, we can observe that Codex performs three times better than CodeGen. Codex's successor, InstructGPT, improves the average success rate by twofold, while ChatGPT exhibits even greater improvement. The article suggests that these significant performance differences also reflect the impact of fine-tuning on programming problems, as Codex is fine-tuned while CodeGen is not.

\begin{table}[]
\centering
\footnotesize
\caption{Average solved rates $(\%)$ for each type of problem formatting in~\cite{N12}. }
\setlength{\tabcolsep}{6mm}
\renewcommand\arraystretch{1}
\label{tab:RQ3-N12}
\begin{tabular}{lllll}
\toprule \textbf{Formatting} & \textbf{CodeGen} & \textbf{Codex} & \textbf{InstructGPT} & \textbf{ChatGPT} \\
\midrule Raw HTML & 10.9 & 30.9 & $\mathbf{7 5 . 9}$ & $\mathbf{9 0 . 1}$ \\
Parsed plain & 9.2 & 34.7 & 73.5 & 89.0 \\
AlphaCode-inspired & 9.7 & 35.2 & 72.0 & 88.7 \\
APPS-inspired & $\mathbf{1 1 . 7}$ & 39.3 & 73.7 & 88.6 \\
Fully Markdown-formatted & 9.9 & $\mathbf{3 9 . 9}$ & 74.5 & 89.0 \\ Average & 10.3 & 36.0 & 73.9 & 89.1 \\
Variance & 1.01 & 13.61 & 2.04 & 0.36 \\
\bottomrule
\end{tabular}
\end{table}

According to Sun et al.~\cite{N33}, a performance comparison of GPT-4, GPT-3.5, and Codex was conducted on the Spider Dev Split test suite to evaluate their accuracy. The results are presented in Table~\ref{tab:RQ3-N33}. It can be observed that on the Spider Dev Split, GPT-4 outperforms Codex, while Codex's performance is better than that of GPT-3.5.

\begin{table}[]
\centering
\footnotesize
\caption{Performance comparison on Test Suite accuracy on Spider Dev Split in~\cite{N12}. }
\setlength{\tabcolsep}{7mm}
\renewcommand\arraystretch{1}
\label{tab:RQ3-N33}
\begin{tabular}{llll}
\toprule  \textbf{LLMs}  & \textbf{Execution Accuracy}  & \textbf{Test-suite Accuracy} \\
\midrule 
 GPT-3 ada (0-shot) & 2.3 & 0.3 \\
 GPT-3 babbage (0-shot) & 5.7 & 3.9 \\
 GPT-3 curie (0-shot)  & 12.6 & 8.3 \\
 GPT-3 davinci (0-shot) & 26.3 & 21.7 \\
 CodeX cushman (0-shot) & 63.7 & 53.0 \\
 CodeX davinci (0-shot) & 67.0 & 55.1 \\
 CodeX davinci (few-shot)  & 71.0 & 61.5 \\
 ChatGPT (w/ OpenAI-default Prompt) & 70.1 & 60.1 \\
 GPT-4 (Zero-shot) & 72.9 & 64.9 \\
 GPT-4 (Few-shot) & 76.8 & 67.4 \\
\bottomrule
\end{tabular}
\end{table}

In Siddiq et al.~\cite{N43}, ChatGPT, codegen-350M-multi, and Codex (2K and 4K) were evaluated for their ability to generate unit tests. Although the article provides a detailed test analysis for these three LLMs, it does not provide definitive results. Even when considering the compilation success rate, it is not possible to draw deterministic conclusions, as shown in Table~\ref{tab:RQ3-N43}.

\begin{table}[]
\centering
\footnotesize
\caption{Compilation status of the generated unit tests in~\cite{N34}.}
\setlength{\tabcolsep}{4mm}
\renewcommand\arraystretch{1}
\label{tab:RQ3-N43}
\begin{tabular}{llllll}
\toprule \textbf{Benchmark} & \textbf{LLM} &  \textbf{Compilable} & $\begin{array}{c}\textbf {  Compilable } \\
\textbf { (after fix) }\end{array}$ & $\begin{array}{c}\textbf {  Test } \\
\textbf { Methods }\end{array}$ & $\begin{array}{c}\textbf {  Test } \\
\textbf { Files }\end{array}$ \\
\midrule 
HumanEval &ChatGPT & $43.1 \%$ & $81.3 \%$ & 1,117 & 130 \\
HumanEval &CodeGen & $23.8 \%$ & $33.1 \%$ & 844 & 529 \\
HumanEval &Codex (2K) & $37.5 \%$ & $100 \%$ & 697 & 160 \\
HumanEval &Codex (4K) & $44.4 \%$ & $99.4 \%$ & 774 & 159 \\
SF110 &ChatGPT & $9.7 \%$ & $85.9 \%$ & 194 & 87 \\
SF110 &CodeGen & $21.0 \%$ & $58.5 \%$ & 83 & 139 \\
SF110 &Codex (2K) & $2.7 \%$ & $74.5 \%$ & 1,406 & 222 \\
SF110 &Codex (4K) & $3.4 \%$ & $83.5 \%$ & 1,039 & 152 \\
\bottomrule 
\end{tabular}
\end{table}

Based on the available information, it can be concluded that in the task of test case generation, GPT-4 and GPT-3.5 (GPT-3.5-turbo) show better performance.

\subsection{Code Summarization}

Sun et al.~\cite{N35} present the overall performance of ChatGPT (one sentence), NCS, CodeBERT, and CodeT5 on the CSN-Python dataset. In Table~\ref{tab:RQ3-N35}, We can see that CodeT5 performs the best among the four models in the code summarization task, outperforming ChatGPT. However, it is important to note that the article's evaluation includes a limited selection of LLMs, which still imposes limitations on the presented results.

\begin{table}[]
\centering
\footnotesize
\caption{Performance comparison on Test Suite accuracy on Spider Dev Split in~\cite{N35}.}
\setlength{\tabcolsep}{9.5mm}
\renewcommand\arraystretch{1}
\label{tab:RQ3-N35}
\begin{tabular}{llll}
\toprule \multirow{2}{*}{ \textbf{Method} } & \multicolumn{3}{l}{ \textbf{CSN-Python} } \\
\cline { 2 - 4 } & \textbf{BLEU} & \textbf{METEOR} & \textbf{ROUGE-L} \\
\midrule NCS & 15.8 & 10.6 & 31.3 \\
CodeBERT & 18.7 & 12.4 & 34.8 \\
CodeT5 & 20.0 & 14.7 & 37.7 \\
ChatGPT (one sentence) & 10.28 & 14.40 & 20.81 \\
\bottomrule
\end{tabular}
\end{table}

Wang et al.~\cite{N36} also presented the performance of several LLMs on CSN and released CodeT5+. The details of their performance can be found in Table~\ref{tab:RQ3-N36}. Furthermore, Wanget al.~\cite{N89} demonstrates the performance of CodeT5Mix on smoothed BLEU-4. CodeT5Mix is an improved version of CodeT5. However, based on the results, CodeT5Mix performs almost on par with CodeT5 and does not show significant performance improvement.

\begin{table}[]
\centering
\footnotesize
\caption{Performance (smoothed BLEU-4) on code summarization on CodeSearchNet in~\cite{N36}. }
\setlength{\tabcolsep}{3.8mm}
\renewcommand\arraystretch{1}
\label{tab:RQ3-N36}
\begin{tabular}{llllllll}
\toprule \textbf{LLMs} & \textbf{Ruby} & \textbf{JS} & \textbf{Go} & \textbf{Python} & \textbf{Java} & \textbf{PHP} & \textbf{Overall} \\
\midrule RoBERTa-125M & 11.17 & 11.90 & 17.72 & 18.14 & 16.47 & 24.02 & 16.57 \\
CodeBERT-125M & 12.16 & 14.90 & 18.07 & 19.06 & 17.65 & 25.16 & 17.83 \\
UniXcoder-125M & 14.87 & 15.85 & 19.07 & 19.13 & 20.31 & 26.54 & 19.30 \\
CodeGen-multi-350M & 13.48 & 16.54 & 18.09 & 18.31 & 19.41 & 24.41 & 18.37 \\
PLBART-140M & 14.11 & 15.56 & 18.91 & 19.30 & 18.45 & 23.58 & 18.32 \\
CodeT5-220M & 15.24 & 16.16 & 19.56 & 20.01 & 20.31 & 26.03 & 19.55 \\ CodeT5+-220M & 15.51 & 16.27 & 19.60 & 20.16 & 20.53 & 26.78 & 19.81 \\
CodeT5+-770M & 15.63 & 17.93 & 19.64 & 20.47 & 20.83 & 26.39 & 20.15 \\
\bottomrule
\end{tabular}
\end{table}

Based on the available information, it can be concluded that in the task of code summarization, CodeT5+ demonstrates better performance compared to GPT-3.5 (GPT-3.5-turbo).

\subsection{Code Translation}
As shown in Tables~\ref{tab:RQ3-N51}, the performance of CodeGeeX on code translation tasks and compared with two other models., it can be observed that CodeGeeX-13B-FT exhibits relatively better performance, while CodeGen-Multi-16B also performs exceptionally well. However, their strengths lie in different multilingual scenarios. Nonetheless, the non-fine-tuned CodeGeeX-13B does not perform as well as CodeGen-Multi-16B in code translation. On the XLCoST benchmark, CodeGeeX outperforms CodeT5, although the difference in scores between the two is not significant.

\begin{table*}[]
\centering
\footnotesize
\caption{Performance of code translation task in HumanEval-X in~\cite{N51-1}. }
\setlength{\tabcolsep}{0.4mm}
\renewcommand\arraystretch{1}
\label{tab:RQ3-N51}
\begin{tabular}{lllllllllllllllll}
\toprule & \multirow{3}{*}{ \textbf{Model} } & \multicolumn{15}{l}{ \textbf{Target Language} } \\
 & & \multicolumn{3}{l}{ \textbf{Python} } & \multicolumn{3}{l}{\textbf{C++}} & \multicolumn{3}{l}{ \textbf{Java} } & \multicolumn{3}{l}{ \textbf{JavaScript} } & \multicolumn{3}{l}{ \textbf{Go} } \\
\cline { 3 - 17 } & & \textbf{@1} & \textbf{@10} & \textbf{@100} & \textbf{@1} & \textbf{@10} & \textbf{@100} & \textbf{@1} & \textbf{@10} & \textbf{@100} & \textbf{@1} & \textbf{@10} & \textbf{@100} & \textbf{@1} & \textbf{@10} & \textbf{@100} \\
\midrule \multirow{4}{*}{ Py } & InCoder-6.7B & - & - & - & 26.11 & 41.00 & 54.25 & 26.74 & 42.66 & 61.20 & 37.05 & 58.85 & 78.91 & 15.69 & 27.57 & 43.67 \\
 & CodeGen-Multi-16B & - & - & - & 35.94 & 47.81 & 59.37 & 29.27 & 45.70 & 64.45 & 43.40 & 66.26 & 82.55 & 28.87 & 41.01 & $\mathbf{5 7 . 7 2}$ \\
 & CodeGeeX-13B & - & - & - & 26.54 & 43.56 & 56.48 & 25.84 & 41.52 & 59.72 & 23.22 & 47.33 & 65.87 & 9.56 & 23.83 & 33.56 \\
 & CodeGeeX-13B-FT & - & - & - & 34.16 & 46.86 & 61.22 & 41.98 & 58.17 & 72.78 & 34.81 & 53.05 & 66.08 & 16.41 & 30.76 & 46.37 \\
\midrule \multirow{4}{*}{$\mathrm{C}++$} & InCoder-6.7B & 34.37 & 58.41 & 78.57 & - & - & - & 34.04 & 57.02 & 68.70 & 37.05 & 65.05 & 79.61 & 25.54 & 39.11 & 58.02 \\
 & CodeGen-Multi-16B & 33.83 & 55.37 & 76.64 & - & - & - & 43.20 & 69.84 & 88.82 & 54.51 & 71.50 & 83.14 & 27.94 & 49.73 & 68.32 \\
 & CodeGeeX-13B & 27.18 & 49.02 & 67.69 & - & - & - & 22.56 & 40.91 & 64.08 & 30.23 & 55.68 & 75.58 & 8.64 & 18.79 & 31.76 \\
 & CodeGeeX-13B-FT & 62.79 & 80.39 & 87.10 & - & - & - & 71.68 & 81.62 & 85.84 & 50.83 & 64.55 & 74.57 & 16.71 & 34.18 & 52.98 \\
\midrule \multirow{4}{*}{ Java } & InCoder-6.7B & 42.76 & 65.55 & 80.43 & 40.01 & 55.17 & 70.39 & - & - & - & 43.20 & 68.24 & 84.39 & 21.58 & 35.20 & 54.97 \\
 & CodeGen-Multi-16B & 52.73 & 69.30 & 82.74 & 41.42 & 54.68 & 65.50 & - & - & - & 57.65 & 67.90 & 79.22 & 34.00 & 48.49 & 67.94 \\
 & CodeGeeX-13B & 43.41 & 68.46 & 84.03 & 39.33 & 58.48 & 72.36 & - & - & - & 44.19 & 64.22 & 82.89 & 17.17 & 32.74 & 47.71 \\
 & CodeGeeX-13B-FT & 75.03 & 87.71 & 95.13 & 49.67 & 65.65 & 75.40 & - & - & - & 49.95 & 62.82 & 79.64 & 18.85 & 32.92 & 48.93 \\
\midrule \multirow{4}{*}{ JS } & InCoder-6.7B & 23.18 & 50.47 & 67.26 & 35.47 & 54.48 & 70.71 & 30.67 & 50.90 & 71.03 & - & - & - & 25.79 & 42.96 & 61.47 \\
 & CodeGen-Multi-16B & 35.52 & 52.23 & 69.78 & 35.41 & 53.12 & 64.47 & 33.79 & 56.06 & 74.00 & - & - & - & 33.38 & 49.08 & 64.14 \\
 & CodeGeeX-13B & 31.15 & 54.02 & 72.36 & 30.32 & 51.63 & 69.37 & 24.68 & 48.35 & 69.03 & - & - & - & 11.91 & 26.39 & 39.81 \\
 & CodeGeeX-13B-FT & 67.63 & 81.88 & 89.30 & 46.87 & 60.82 & 73.18 & 56.55 & 70.27 & 80.71 & - & - & - & 16.46 & 32.99 & 50.29 \\
\midrule \multirow{4}{*}{ Go } & InCoder-6.7B & 34.14 & 54.52 & 70.88 & 30.45 & 48.47 & 62.81 & 34.52 & 53.95 & 69.92 & 39.37 & 63.63 & 80.75 & - & - & - \\
 & CodeGen-Multi-16B & 38.32 & 50.57 & 68.65 & 32.95 & 45.88 & 59.56 & 36.55 & 59.12 & 78.70 & 38.93 & 56.68 & 70.68 & - & - & - \\
 & CodeGeeX-13B & 35.92 & 56.02 & 77.32 & 29.83 & 41.98 & 58.15 & 22.89 & 41.04 & 61.46 & 25.24 & 46.50 & 69.93 & - & - & - \\
 & CodeGeeX-13B-FT & 57.98 & 79.04 & 93.57 & 38.97 & 53.05 & 63.92 & 54.22 & 69.03 & 79.40 & 43.07 & 59.78 & 74.04 & - & - & - \\
\bottomrule
\end{tabular}
\end{table*}

Pan et al.~\cite{N126} also provide the performance of seven LLMs, including GPT-4 and StarCoder, on code translation tasks across seven datasets. From the test results presented in the article (as shown in Table~\ref{tab:RQ3-N126}, for detailed experimental settings please refer to Pan et al.~\cite{N126}), it can be observed that, except for GPT-4 and StarCoder, the other models perform poorly. The article also points out a strong correlation between the average number of test attempts per translation sample and unsuccessful translations. Additionally, unsuccessful translations do not exhibit consistent patterns between the source and target languages, but due to stricter GO syntax constraints, code translations related to GO perform poorly.

\begin{table*}[]
\centering
\footnotesize
\caption{Performance of subject LLMs in translating code. }
\setlength{\tabcolsep}{1.4mm}
\renewcommand\arraystretch{1}
\label{tab:RQ3-N126}
\begin{tabular}{lllllllll}
\toprule \multirow{2}{*}{ \textbf{Dataset} }  & \multicolumn{7}{l}{ \textbf{\% Unsuccessful Translations} } \\
\cline { 2-8 }  & \textbf{CodeGen} & \textbf{CodeGeeX} & \textbf{StarCoder} & \textbf{GPT-4} & \textbf{Llama 2} & \textbf{TB-Airoboros} & \textbf{TB-Vicuna} \\
\midrule \multirow{5}{*}{ CodeNet  } & $76.6 \%$ & $85.1 \%$ & $58.0 \%$ & $17.0 \%$ & $85.1 \%$ & $81.2 \%$ & $95.6 \%$ \\
  & $86.0 \%$ & $96.4 \%$ & $60.9 \%$ & $20.0 \%$ & $90.5 \%$ & $91.7 \%$ & $96.6 \%$ \\
 & $85.7 \%$ & $94.1 \%$ & $58.0 \%$ & $14.5 \%$ & $83.1 \%$ & $93.4 \%$ & $99.1 \%$ \\
 & $78.7 \%$ & $89.7 \%$ & $69.7 \%$ & $18.7 \%$ & $86.1 \%$ & $93.5 \%$ & $99.9 \%$ \\
 & $82.5 \%$ & $92.7 \%$ & $66.7 \%$ & $20.1 \%$ & $89.0 \%$ & $93.5 \%$ & $99.0 \%$ \\
\midrule Total/Average (CodeNet)  & $81.9 \%$ & $91.6 \%$ & $62.7 \%$ & $18.0 \%$ & $86.8 \%$ & $90.7 \%$ & $98.0 \%$ \\
\midrule \multirow{2}{*}{ AVATAR  }  & $91.9 \%$ & $98.2 \%$ & $88.1 \%$ & $29.2 \%$ & $98.2 \%$ & $94.9 \%$ & $100 \%$ \\
 & $96.2 \%$ & $98.4 \%$ & $85.8 \%$ & $47.8 \%$ & $95.3 \%$ & $99.1 \%$ & $99.1 \%$ \\
\midrule Total/Average (AvATAR)  & $94.1 \%$ & $98.3 \%$ & $87.0 \%$ & $38.5 \%$ & $96.8 \%$ & $97.0 \%$ & $99.6 \%$ \\
\midrule EvalPlus   & $83.5 \%$ & $96.3 \%$ & $78.0 \%$ & $20.7 \%$ & $98.8 \%$ & $86.0 \%$ & $92.1 \%$ \\
\midrule Commons CLI   & $100 \%$ & $100 \%$ & $100 \%$ & $86.4 \%$ & $100 \%$ & $100 \%$ & $100 \%$ \\
\midrule Click   & $100 \%$ & $100 \%$ & $100 \%$ & $100 \%$ & $100 \%$ & $100 \%$ & $100 \%$ \\
\midrule Total/Average (All)  & $91.9 \%$ & $97.2 \%$ & $85.5 \%$ & $52.7 \%$ & $96.5 \%$ & $94.7 \%$ & $97.9 \%$ \\
\bottomrule
\end{tabular}
\end{table*}

In the task of code translation, GPT-4 performs better. This is supported by Pan et al.~\cite{N126}, who found that GPT-4 outperforms CodeGeeX significantly in terms of performance.

\subsection{Vulnerability Repair}

\textbf{MBPP: }MBPP is also one of the important benchmarks for evaluating the code generation capabilities of LLMs. MBPP, which stands for Massively Bugs and Performance Problems, is a benchmark that consists of a large number of code snippets with defects and performance issues. The models are required to generate the correct repair code that is relevant to the given problem. The benchmark aims to assess the models' ability to identify and resolve software errors and performance problems. By using the MBPP benchmark, the practicality and robustness of the models in real-world software engineering scenarios can be evaluated. We have organized the results from the collected papers on MBPP and obtained Table~\ref{tab:RQ3-mbpp}.

\begin{table*}[]
\centering
\scriptsize
\caption{Performance of LLMs in MBPP benchmark.}
\setlength{\tabcolsep}{0.15mm}
\renewcommand\arraystretch{1}
\label{tab:RQ3-mbpp}
\begin{tabular}{lllll|lllll}
\toprule
\multirow{2}{*}{\textbf{LLMs}}    & \multicolumn{4}{l|}{\textbf{MBPP}}                                  & \multirow{2}{*}{\textbf{LLMs}}   & \multicolumn{4}{l}{\textbf{MBPP}}              \\ \cline{2-5} \cline{7-10} 
                         & \textbf{Pass@1} & \textbf{Pass@10} & \textbf{Pass@80} & \multicolumn{1}{l|}{\textbf{Pass@100}} &                         & \textbf{Pass@1} & \textbf{Pass@10} & \textbf{Pass@80} & \textbf{Pass@100} \\ \midrule
WizardCoder-16B          & 51.8   & -       & -       & -                             & CodeParrot-110M         & 0.48   & 3.89    & -       & 15.93    \\
Unnatural-Code-LLaMA-34B & \underline{\textbf{61.2}}   & \underline{\textbf{76.6}}    & -       & \underline{\textbf{86.7}}                          & CodeParro-1.5B          & 1.29   & 8.66    & -       & 27.17    \\
StarCoder-Python-15B     & \underline{\textbf{52.7}}   & -       & -       & -                             & Code-LLaMA-Python-7B    & 47.6   & 70.3    & -       & 84.8     \\
StarCoder-Prompted-15.5B & 49.5   & -       & -       & -                             & Code-LLaMA-Python-34B   & \underline{\textbf{56.2}}   & \underline{\textbf{76.4}}    & -       & \underline{\textbf{88.2}}     \\
StarCoderBase-15B        & 49.0   & -       & -       & -                             & Code-LLaMA-Python-13B   & 49     & \underline{\textbf{74}}      & -       & \underline{\textbf{87.6}}     \\
StarCoder-5.5B           & \underline{\textbf{52.7}}   & -       & -       & -                             & Code-LLaMA-Instruct-7B  & 44.4   & 65.4    & -       & 76.8     \\
SantaCoder-1.1B          & 3.65   & 21.33   & -       & 41.92                         & Code-LLaMA-Instruct-34B & \underline{\textbf{57}}     & \underline{\textbf{74.6}}    & -       & 85.4     \\
SantaCoder-1.1B          & 35     & -       & -       & -                             & Code-LLaMA-Instruct-13B & 49.4   & 71.2    & -       & 84.1     \\
PyCodeGPT-110M           & 9.39   & 28.37   & -       & 48.71                         & Code-LLaMA-7B           & 41.4   & 66.7    & -       & 82.5     \\
PolyCoder-400M           & 1.31   & 7.98    & -       & 21.55                         & Code-LLaMA-34B          & \underline{\textbf{55}}     & \underline{\textbf{76.2}}    & -       & \underline{\textbf{86.6}}     \\
PolyCoder-2.7B           & 4.39   & 17.99   & -       & 38.17                         & Code-LLaMA-13B          & 47     & 71.7    & -       & \underline{\textbf{87.1}}     \\
PolyCoder-160M           & 1.08   & 6.67    & -       & 18.97                         & CodeGen-NL 6.1B         & 8.15   & 31.21   & -       & 55.27    \\
phi-1-1.3B               & 55.5   & -       & -       & -                             & CodeGen-NL 350M         & 0.96   & 6.37    & -       & 19.91    \\
PaLM-Coder-540B          & 47     & -       & 80.8    & -                             & CodeGen-NL 2.7B         & 5.34   & 24.63   & -       & 48.95    \\
PaLMCoder-540B           & 47     & -       & 80.8    & -                             & CodeGen-NL 16.1B        & 10.92  & 38.43   & -       & 62.76    \\
PaLM-540B                & 36.8   & -       & 75      & -                             & CodeGen-Multi-16B       & 20.9   & -       & -       & -        \\
PaLM-2-S                 & 50     & -       & -       & -                             & CodeGen-Multi-6.1B      & 18.35  & 47.27   & -       & 67.92    \\
LLaMA-7B                 & 17.7   & -       & -       & -                             & CodeGen-Multi-350M      & 7.46   & 24.18   & -       & 46.37    \\
LLaMA-65B                & 37.7   & -       & -       & -                             & CodeGen-Multi-2.7B      & 18.06  & 45.80   & -       & 65.34    \\
LLaMA-33B                & 30.2   & -       & -       & -                             & CodeGen-Multi-16.1B     & 20.94  & 51.61   & -       & 70.02    \\
LLaMA2-7B                & 20.8   & 41.8    & -       & 65.5                          & CodeGen-Mono-6.1B       & 33.70  & 62.70   & -       & 70.25    \\
LLaMA2-70B               & 45.4   & 66.2    & -       & 83.1                          & CodeGen-Mono-350M       & 15.44  & 42.50   & -       & 64.40    \\
LLaMA2-34B               & 33.8   & 56.9    & -       & 77.6                          & CodeGen-Mono-350M       & -      & -       & -       & -        \\
LLaMA2-13B               & 27.6   & 48.1    & -       & 69.5                          & CodeGen-Mono-2.7B       & 28.80  & 60.73   & -       & 75.41    \\
LLaMA-13B                & 22     & -       & -       & -                             & CodeGen-Mono-16B        & 35.3   & -       & -       & -        \\
LaMDA-137B               & 14.8   & -       & 62.4    & -                             & CodeGen-Mono-16.1B      & 35.28  & 67.32   & -       & 80.09    \\
JuPuT5-300M              & -      & -       & 52.2    & -                             & CodeGen-Mono-16.1B      & 35.3   & -       & -       & -        \\
InstructCodeT5+-16B      & -      & -       & -       & -                             & CODEGEN-Mono-6.1B       & 32.48  & 64.20   & -       & 76.81    \\
InCoder-6.7B             & 19.4   & -       & -       & -                             & CODEGEN-Mono-350M       & 14.59  & 41.49   & -       & 63.00    \\
InCoder-6.7B             & 21.3   & 46.5    & -       & 66.2                          & CODEGEN-Mono-2.7B       & 27.31  & 59.19   & -       & 74.24    \\
InCoder-1.3B             & 10.00  & 34.02   & -       & 55.50                         & CODEGEN-Mono-16.1B      & 35.28  & 67.32   & -       & 80.09    \\
InCoder-6B               & 21.30  & 46.50   & -       & 66.20                         & CodeGen2-1B             & -      & -       & -       & -        \\
GPT-Neo-2.7B             & 5.89   & 23.09   & -       & 44.26                         & CodeGeeX-13B            & 24.4   & 48      & 68.5    & -        \\
GPT-Neo-125M             & 0.26   & 2.15    & -       & 7.96                          & code-davinci-002        & 58.10  & 76.70   & -       & 84.50    \\
GPT-Neo-1.3B             & 3.77   & 16.26   & -       & 29.51                         & code-davinci-001        & 51.80  & 72.80   & -       & 84.10    \\
GPT-J-6B                 & 11.30  & 35.62   & -       & 53.63                         & code-cushman-001        & 45.90  & 66.90   & -       & 79.90    \\
GPT-4                    & -      & -       & -       & -                             & CodcGen2-7B             & -      & -       & -       & -        \\
GPT-3.5                    & -      & -       & -       & -                             & BLOOM-7.1B              & 1.01   & 7.91    & -       & 24.12    \\
GPT-3.5-turbo                 & 52.2   & -       & -       & -                             & BLOOM-560M              & 0.26   & 2.04    & -       & 8.90     \\
CodeT5-770M              & 15.78  & 38.63   &         & 50.35                         & BLOOM-3B                & 2.25   & 13.58   & -       & 32.08    \\
CodeT5+-2B               & -      & -       & -       & -                             & BLOOM-1.7B              & 3.16   & 14.23   & -       & 31.38    \\
CodeT5+-16B              & -      & -       & -       & -                             & BLOOM-1.1B              & 1.90   & 9.20    & -       & 23.42   \\
\bottomrule
\end{tabular}
\end{table*}

Similarly, while organizing the data for MBPP, we have also noticed variations in the reported data across different literature sources. We have compiled and included these variations in Table~\ref{tab:RQ3-differ-2} as well. We have followed the principle of selecting the most recent literature to obtain the data.

\begin{table}[]
\centering
\footnotesize
\caption{Data in the literature that differ for MBPP.}
\setlength{\tabcolsep}{6mm}
\renewcommand\arraystretch{1}
\label{tab:RQ3-differ-2}
\begin{tabular}{lllll}
\toprule \multirow{2}{*}{ \textbf{LLMs} } & \multicolumn{4}{l}{ \textbf{MBPP} } \\
\cline { 2 - 5 } & \textbf{Pass@1} & \textbf{Pass@10} & \textbf{Pass@80} & \textbf{Pass@100} \\
 \midrule
InCoder-6.7B~\cite{N15}                            & 19.4                                & -                                    & -                                    & -                                     \\
InCoder-6.7B~\cite{N76}                            & 21.3                                & 46.5                                 & -                                    & 66.2                                  \\
CodeGen-Mono-16.1B~\cite{N76}                     & 35.28                               & 67.32                                & -                                    & 80.09                                 \\
CodeGen-Mono-16.1B~\cite{N38}                    & 35.3                                & -                                    & -                                    & -                                     \\ \bottomrule
\end{tabular}
\end{table}

We have also marked the top five performing data with underlines and bold for each metric (except for pass@80). We can see that the Code-LLaMA series continues to exhibit strong performance, carrying forward its excellent performance on HumanEval. Among them, Unnatural-Code-LLaMA shows the best overall performance, followed by Code-LLaMA-Python-34B. However, it should be noted that, unlike HumanEval, it is challenging to find scores for many LLMs on MBPP. We attempted to search using a snowballing approach but did not make significant progress. Based on the currently available data, Unnatural-Code-LLaMA-34B is the top-performing LLM on MBPP. To some extent, Unnatural-Code-LLaMA-34B is also the best-performing LLM for code generation tasks currently available.

Pearce et al.~\cite{N119} explore the ability of LLMs to fix software vulnerabilities in a zero-shot setting. The experimental section of the article mainly utilizes the following LLMs: code-cushman-001, code-davinci-001, code-davinci-002, j1-large, j1-jumbo, and polycoder. Overall, based on the combined evaluation datasets, code-davinci-002 performs relatively well. However, the article does not explicitly provide performance differences between the LLMs. The study finds that LLMs can generate fix programs for security vulnerabilities when provided with carefully constructed prompts. However, the evaluation of LLM performance indicates that the current state of the technology is not sufficient to deliver true value in the context of program repair frameworks.

Xia et al.~\cite{N123} evaluates LLMs used for direct program repair. The article reveals the scaling effects of increasing model size on various crucial factors in APR, such as the number of fixed bugs, patch generation speed, and compilation rate. LLMs are also tested on widely used APR benchmarks, resulting in Table~\ref{tab:RQ3-N123-1} and Table~\ref{tab:RQ3-N123-2} (Columns CF, CI, SL refer to complete function, correct infilling and single line generation, respectively). It can be observed that many models achieve similar (or even better) performance through carefully designed APR tools. Additionally, in the Defects4J 2.0, QuixBugs-Java, and QuixBugs-Python benchmarks, all nine LLMs outperform TBar (state-of-the-art template-based APR tool). Among these nine LLMs, Codex demonstrates the best performance, followed by InCoder-6.7B and GPT-NeoX. 

\begin{table}[]
\centering
\footnotesize
\caption{Number of samples generated per minute for different PLMs on Defects4J 1.2 and QuixBugs with the 3 repair generation settings in~\cite{N123}.}
\setlength{\tabcolsep}{4.3mm}
\renewcommand\arraystretch{1}
\label{tab:RQ3-N123-1}
\begin{tabular}{lllll}
\toprule
\textbf{Tools / Models} & \textbf{Single func (255 bugs)}  & \textbf{Patch func} & \textbf{Correct hunk} & \textbf{Single line} \\ \midrule
 AlphaRepair & 67 & -&- &- \\
RewardRepair & 48 &-& -& -\\
Recoder & 61 &- & -& -\\
TBar & 54 &- &- &- \\
CURE & 52 &- &- & -\\
GPT-Neo 125M & 9 & 6 & - & 5 \\
GPT-Neo 1.3B & 18 & 7 & - & 12 \\
GPT-Neo 2.7B & 20 & 10 & - & 13 \\
GPT-J & 28 & 14 & - & 16 \\
GPT-NeoX & 34 & 18 & - & 21 \\
CodeT5 & 6 & - & 6 & - \\
INCODER 1.3B & 32 & - & 32 & - \\
INCODER 6.7B & 37 & - & 37 & - \\
Codex & 99 & 63 & 62 & 32 \\ \midrule
Total & 109 & 69 & 74 & 40 \\
\bottomrule
\end{tabular}
\end{table}

\begin{table}[]
\centering
\footnotesize
\caption{Performance on Defects4J 2.0, QuixBugs-Java and -Python in~\cite{N123}}
\setlength{\tabcolsep}{2.8mm}
\renewcommand\arraystretch{1}
\label{tab:RQ3-N123-2}
\begin{tabular}{llll}
\toprule \textbf{Tools / Models} & \textbf{Defects4J 2.0 (78 bugs)}& \textbf{QuixBugs Java (40 bugs)} & \textbf{QuixBugs Python (40 bugs)} \\
\midrule AlphaRepair & 35 & 28 & 27 \\
RewardRepair & 25 & 20 & - \\
DeepDebug & - & - & 21 \\
Recoder & 11 & 17 & - \\
CURE & - & 21 & - \\
TBar & 8 & - & - \\
CoCoNuT & - & 13 & 19 \\
\hline GPT-Neo 125M & 10 & 8 & 9 \\
GPT-Neo 1.3B & 11 & 20 & 17 \\
GPT-Neo 2.7B & 19 & 18 & 24 \\
GPT-J & 16 & 22 & 29 \\
GPT-NeoX & 24 & 21 & 31 \\
CodeT5 & 9 & 10 & 7 \\
INCODER 1.3B & 15 & 21 & 25 \\
INCODER 6.7B & 21 & 26 & 27 \\
Codex & 45 & 38 & 40 \\
\midrule Total & 52 & 38 & 40 \\
\bottomrule
\end{tabular}
\end{table}

In summary, based on the limited available results, Codex demonstrates better performance in the task of vulnerability repair. However, the information available for this task is still limited, and there is a lack of performance comparison between state-of-the-art (SOTA) Code LLMs and GPT-4.

\subsection{Other Evaluation or New Benchmarks}
Yuan et al.~\cite{N19} provide a detailed evaluation of 10 open-source guided LLMs on four representative code understanding and generation tasks: defect detection, clone detection, assertion generation, and code summarization. While the main focus of the article is on the impact of instruction fine-tuning on Code LLMs, it also provides valuable insights. Table~\ref{tab:RQ3-N19-1} showcases the performance of instruction-tuned LLMs on software engineering tasks under zero-shot and one-shot settings. We can observe that WizardCoder-15B performs relatively well, particularly in the assertion generation task. The paper also presents several interesting findings: (1) For zero-shot settings, guided LLMs sometimes outperform small-scale SOTA models fine-tuned specifically for each downstream task in code understanding and generation tasks. (2) For few-shot settings, the addition of demonstration examples can significantly improve the performance of guided LLMs on most code understanding and generation tasks. (3) For fine-tuning settings, further performance enhancement on downstream code understanding and generation tasks can be achieved through fine-tuning.

\begin{table*}[]
\centering
\footnotesize
\caption{Average solved rate $(\%)$ for each type of problem in~\cite{N19}.}
\setlength{\tabcolsep}{1mm}
\renewcommand\arraystretch{1}
\label{tab:RQ3-N19-1}
\begin{tabular}{ccccccccc}
\toprule \multirow{2}{*}{ \textbf{LLM} } & \multicolumn{2}{c}{ \textbf{Defect Detection  $(\boldsymbol{\%})$}} & \multicolumn{2}{c}{ \textbf{Clone Detection (\%) }} & \multicolumn{2}{c}{ \textbf{Assert Generation (\%) } }& \multicolumn{2}{c}{ \textbf{Code Summarization  (\%) }} \\
\cline { 2 - 9 } & Zero-shot & One-shot & Zero-shot & One-shot & Zero-shot & One-shot & Zero-shot & One-shot \\
\midrule CodeGen-6B & 0.3 & 43.6 & 1.4 & 23.4 & 0.0 & 56.2 & 0.0 & 13.0 \\
 ChatGLM-6B & 7.1 & 54.2 & 17.5 & 12.8 & 1.7 & 46.2 & 45.0 & $\mathbf{5 4 . 0}$ \\
 Vicuna-7B & 54.0 & 54.1 & 13.2 & - & 10.1 & 31.2 & 48.0 & 37.0 \\
 Alpaca-7B & 45.8 & $\mathbf{5 5 . 4}$ & 22.1 & - & 5.3 & 41.4 & 32.0 & 6.0 \\
 Dolly-7B & 33.1 & 49.9 & 21.3 & $\mathbf{2 3 . 5}$ & 1.9 & 51.0 & 12.0 & 14.0 \\
 StableLM-7B & 44.3 & 43.4 & $\mathbf{2 4 . 3}$ & - & 1.1 & 44.4 & 30.0 & 19.0 \\
 CodeAlpaca-7B & 51.9 & 50.3 & 1.4 & 10.3 & 4.4 & 35.1 & 9.0 & 34.0 \\
 Dolly-12B & 33.8 & 52.7 & 23.5 & 22.6 & 1.0 & 51.7 & 5.0 & 8.0 \\
 Vicuna-13B & 49.8 & 53.0 & 14.1 & 6.5 & 12.0 & 44.0 & 63.0 & 24.0 \\
 WizardCoder-15B & $\mathbf{5 4 . 4}$ & 53.8 & 23.8 & 7.3 & $\mathbf{1 9 . 4}$ & $\mathbf{6 3 . 3}$ & $\mathbf { 7 1 . 0 }$ & 50.0 \\
 Instruct-CodeGen-16B & 47.8 & 54.6 & 14.2 & 20.7 & 8.4 & 55.0 & 9.0 & 41.0 \\
\bottomrule
\end{tabular}
\end{table*}

Zan et al.~\cite{N56} conducted a comprehensive investigation of Code LLMs on 27 existing LLMs and reviewed widely used benchmarks and metrics, as shown in Table~\ref{tab:RQ3-N56}. In the table, P.NL represents the Problem description's Natural Language, S.PL denotes the code Solution's Programming Language, and T.N. denotes the average Number of Test cases. P.C. and P.L. (S.C. and S.L.) stand for the average number of Characters and Lines in the Problem description (code Solution), respectively. 

\begin{table*}[]
\centering
\footnotesize
\caption{Average solved rate $(\%)$ for each type of problem formatting in~\cite{N56}. The asterisk (*) indicates the number of instances per programming language.}
\setlength{\tabcolsep}{1.2mm}
\renewcommand\arraystretch{1}
\label{tab:RQ3-N56}
\begin{tabular}{llllllllll}
\toprule \multirow{2}{*}{ \textbf{Benchmark} } & \multirow{2}{*}{ \textbf{Number of instances }} & \multirow{2}{*}{ \textbf{P.NL} } & \multirow{2}{*}{ \textbf{S.PL} } & \multicolumn{5}{c}{ \textbf{Data Statistics} } & \multirow{2}{*}{ \textbf{Scenario} } \\
\cline { 5 - 9 }
 & & & & T.N. & P.C. & P.L. & S.C. & S.L. & \\
\midrule HumanEval& 164 & English & Python & 7.8 & 450.6 & 13.7 & 180.9 & 6.8 & Code Exercise \\
 MBPP & 974 & English & Python & 3.1 & 78.6 & 1.0 & 181.1 & 6.7 & Code Exercise \\
 APPS  & 5,000 & English & Python & 21.0 & 1743.4 & 41.6 & 473.8 & 21.4 & Competitions \\
 CodeContests & 165 & English & Multi. & 203.7 & 1989.2 & 66.4 & 2239.3 & 92.1 & Competitions \\
 DS-1000  & 1,000 & English & Python & 1.6 & 879.1 & 31.6 & 137.4 & 5.0 & Data Science \\
 DSP & 1,119 & English & Python & 2.1 & 756.9 & 17.8 & 226.3 & 7.6 & Data Science \\
 MBXP & $974^*$ & English & Multi. & 3.1 & 419.9 & 14.8 & - & - & Multilingual \\
 MBXP-HumanEval & $164^*$ & English & Multi. & 7.8 & 825.6 & 30.0 & - & - & Multilingual \\
 HumanEval-X & $164^*$ & English & Multi. & 7.8 & 468.4 & 15.5 & 264.6 & 12.1 & Multilingual \\
 MultiPL-HumanEval & $164^*$ & English & Multi. & 7.8 & 453.9 & 13.0 & - & - & Multilingual \\
 MultiPL-MBPP & $974^*$ & English & Multi. & 3.1 & 181.2 & 5.4 & - & - & Multilingual \\
 PandasEval & 101 & English & Python & 6.5 & 244.5 & 7.2 & 46.2 & 1.3 & Public Library \\
 NumpyEval & 101 & English & Python & 3.5 & 222.9 & 7.0 & 29.9 & 1.1 & Public Library \\
 TorchDataEval & 50 & English & Python & 1.1 & 329.0 & 8.6 & 50.7 & 1.3 & Private Library \\
 MTPB & 115 & English & Python & - & 72.7 & 1.0 & - & - & Multi-Turn \\
 ODEX 2022 c & 945 & Multi. & Python & 1.8 & 26.6 & 2.0 & 50.4 & 1.9 & Open-Domain \\
 BIG-Bench & 32 & English & Python & 4.7 & 341.8 & 3.0 & - & - & Code Exercise \\
\bottomrule
\end{tabular}
\end{table*}

Athiwaratkun et al.~\cite{N122} introduce new benchmarks for evaluating code generation models: MBXP, Multilingual HumanEval, and MathQA-X. The article provides a detailed comparison of the performance of the CodeGen, OPT, and BLOOM models in multilingual code generation and code translation tasks. In each test, CodeGen-Mono-16B achieves higher scores. The article also highlights some important findings:
(1) Given the same model size, multilingual models generally outperform the best monolingual models trained with equivalent training resources, especially when the model is large enough. This observation suggests that training a single model on all programming languages is beneficial, and as long as the model has sufficient capacity, its performance will surpass the best monolingual models. (2) LLMs have the potential to learn from their uncurated programming languages through unit tests. (3) Few-shot prompts can effectively help the model acquire knowledge of new languages not seen during training, significantly improving out-of-domain code generation capabilities. Through error analysis, it is observed that fewer prompts help reduce compilation or parsing errors, which are the main source of errors when dealing with programming languages the model is unfamiliar with. (4) Language models possess zero-shot code translation capabilities, and this translation ability extends to monolingual models. (5) Multilingual models are more robust to prompt perturbations and can better summarize code.

Tang et al.~\cite{N118} introduce BIOCODER, a benchmark for evaluating LLMs' ability to generate bioinformatics code. The article presents the performance of InCoder, CodeGen, CodeGen2, SantaCoder, StarCoder, StarCoder+, InstructCodeT5+, and ChatGPT on BIOCODER, as shown in Table~\ref{tab:RQ3-N118}. Notably, StarCoder+ is the result of fine-tuning StarCoder on Java for 2000 steps, while all others are zero-shot results. It can be observed that ChatGPT performs the best on all tasks. The article also highlights an interesting phenomenon: despite InstructCodeT5+, CodeGen, and CodeGen2 having larger parameter sizes than InCoder and SantaCoder, their performance is significantly worse. The authors attribute this to InstructCodeT5+, CodeGen, and CodeGen2 being trained on single-line completions rather than function completions. Additionally, InstructCodeT5+, CodeGen, and CodeGen2 have relatively smaller context constraints. The authors further note that context constraints have a significant impact on how different models perform under different prompts.
\begin{table*}[]
\centering
\footnotesize
\caption{Performance with five prompt versions of BIOCODER~\cite{N118}. }
\setlength{\tabcolsep}{0.7mm}
\renewcommand\arraystretch{1}
\label{tab:RQ3-N118}
\begin{tabular}{llllllllll}
\toprule \multirow{2}{*}{ \textbf{Model} } & \multirow{2}{*}{ \textbf{Prompt} } & \multicolumn{4}{c}{ \textbf{Java }} & \multicolumn{4}{c}{ \textbf{Python} } \\
\cline { 3 - 10 } & & Pass@1 & Pass@5 & 5ass@10 & Pass@20 & Pass@1 & Pass@5 & Pass@10 & Pass@20 \\
\midrule \multirow{5}{*}{ InCoder-6B } & Summary at Top & 0 & 0 & 0 & 0 & 0.828 & 2.016 & 3.006 & 4.459 \\
 & Uncommented & 0 & 0 & 0 & 0 & 0.032 & 0.159 & 0.318 & 0.637 \\
 & Summary Only & 0 & 0 & 0 & 0 & 1.688 & 5.320 & 8.332 & 12.006 \\ & Summary at Bottom & - & - & - & - & 0.610 & 2.587 & 4.303 & 6.274 \\
 & Necessary Only & 0 & 0 & 0 & 0 & 0.032 & 0.159 & 0.318 & 0.637 \\
\hline \multirow{5}{*}{ SantaCoder-1.1B } & Summary at Top & 0 & 0 & 0 & 0 & 0.637 & 1.338 & 1.844 & 2.548 \\
 & Uncommented & 0 & 0 & 0 & 0 & 0.287 & 0.764 & 0.955 & 1.274 \\
 & Summary Only & 0 & 0 & 0 & 0 & 2.965 & 9.848 & 14.227 & 18.181 \\
 & Summary at Bottom & - & - & - & - & 0.510 & 1.949 & 3.013 & 4.459 \\
 & Necessary Only & 0 & 0 & 0 & 0 & 0.032 & 0.159 & 0.318 & 0.637 \\
\hline \multirow{5}{*}{ StarCoder-15.5B } & Summary at Top & 0 & 0 & 0 & 0 & 3.694 & 13.197 & 19.359 & 24.554 \\
 & Uncommented & 0 & 0 & 0 & 0 & 0.318 & 1.062 & 1.591 & 2.548 \\
 & Summary Only & 0 & 0 & 0 & 0 & 4.682 & 15.225 & 21.200 & 27.166 \\
 & Summary at Bottom & - & - & - & - & 6.465 & 13.824 & 16.746 & 19.076 \\
 & Necessary Only & 0 & 0 & 0 & 0 & 0.127 & 0.603 & 1.123 & 1.911 \\
\hline \multirow{5}{*}{$\begin{array}{l}\text { StarCoder-15.5B } \\
\quad \text { (finetuned) }\end{array}$} & Summary at top & 0 & 0 & 0 & 0 & - & - & - & - \\
 & Uncommented & 0 & 0 & 0 & 0 & - & -& - & - \\
 & Summary Only & 0.200 & 1.000 & 2.000 & 4.000 & - & - & - & - \\
 & Summary at bottom & - & - & - & - & - & - & - & - \\
 & Necessary Only & 3.300 & 12.097 & 19.545 & 30.000 & - & - & - & - \\
\hline \multirow{5}{*}{ StarCoder+ } & Summary at Top & 0 & 0 & 0 & 0 & 2.675 & 9.133 & 14.019 & 19.650 \\
 & Uncommented & 0 & 0 & 0 & 0 & 0.510 & 0.955 & 1.274 & 1.911 \\
 & Summary Only & 1.300 & 5.031 & 8.042 & 12.000 & 2.548 & 8.279 & 12.864 & 18.057 \\
 & Summary at Bottom & - & - & - & - & 4.172 & 11.772 & 14.933 & 17.197 \\
 & Necessary Only & 0 & 0 & 0 & 0 & 0.127 & 0.457 & 0.609 & 0.637 \\
\hline InstructCodeT5+ & All prompt types & 0 & 0 & 0 & 0 & 0 & 0 & 0 & 0 \\
\hline \multirow{5}{*}{ CodeGen-6B-mono } & Summary at Top & 0 & 0 & 0 & 0 & 0.637 & 0.637 & 0.637 & 0.637 \\
 & Uncommented & 0 & 0 & 0 & 0 & 0 & 0 & 0 & 0 \\
 & Summary Only & 0 & 0 & 0 & 0 & 0.637 & 0.637 & 0.637 & 0.637 \\
 & Summary at Bottom & - & - & - & - & 2.070 & 4.535 & 5.896 & 7.006 \\
 & Necessary Only & 0 & 0 & 0 & 0 & 0 & 0 & 0 & 0 \\
\hline \multirow{5}{*}{ CodeGen-16B-mono } & Summary at Top & 0 & 0 & 0 & 0 & 0.637 & 0.637 & 0.637 & 0.637 \\
 & Uncommented & 0 & 0 & 0 & 0 & 0 & 0 & 0 & 0 \\
 & Summary Only & 0 & 0 & 0 & 0 & 0.637 & 0.637 & 0.637 & 0.637 \\
 & Summary at Bottom & - & - & - & - & 2.166 & 5.137 & 6.022 & 6.369 \\
 & Necessary Only & 0 & 0 & 0 & 0 & 0 & 0 & 0 & 0 \\
\hline \multirow{5}{*}{ CodeGen2-7B } & Summary at Top & 0 & 0 & 0 & 0 & 0.637 & 0.637 & 0.637 & 0.637 \\
 & Uncommented & 0 & 0 & 0 & 0 & 0.510 & 0.637 & 0.637 & 0.637 \\
 & Summary Only & 0 & 0 & 0 & 0 & 0.860 & 2.494 & 3.962 & 6.242 \\
 & Summary at Bottom & - & - & - & - & 0.510 & 1.019 & 1.207 & 1.274 \\
 & Necessary Only & 0 & 0 & 0 & 0 & 0 & 0 & 0 & 0 \\
\hline \multirow{5}{*}{ GPT-3.5-Turbo } & Summary at Top & 4.100 & 7.235 & 8.989 & 11.600 & 22.771 & 33.461 & 36.551 & 39.490 \\
 & Uncommented & 6.300 & 11.563 & 14.436 & 18.000 & 11.019 & 19.075 & 21.680 & 24.204 \\
 & Summary Only & 17.400 & 33.199 & 37.878 & 42.000 & 24.682 & 33.997 & 37.132 & 40.127 \\
 & Summary at Bottom & - & - & - & - & 13.439 & 20.040 & 22.460 & 25.478 \\
 & Necessary Only & 43.500 & 52.582 & 53.995 & 55.400 & 28.758 & 39.529 & 44.029 & 47.771 \\
\bottomrule
\end{tabular}
\end{table*}

Kou et al.~\cite{N127} evaluated the differences between LLMs and human model attention in programming based on keyword coverage and Cohen's Kappa consistency level. The article shows that among the five models evaluated, CodeGen exhibited the highest consistency with human programmers. However, the results presented in the article are limited, and there was no testing conducted on state-of-the-art (SOTA) models.

In conclusion, we can summarize the following points:
\begin{itemize}
    \item The current evaluation of LLMs focuses more on code generation tasks, with less emphasis on evaluating or researching other tasks such as vulnerability repair. There is a lack of relevant evaluation work, and when new models are released, there is limited attention given to tasks like vulnerability repair.
    \item Code generation tasks have well-known benchmarks like HumanEval. However, other tasks lack such widely recognized benchmarks.
    \item In code generation tasks, the Code-LLaMA series of LLMs perform the best, especially with Unnatural-Code-LLaMA-34B showing outstanding performance. In API-related code generation tasks, ToolCoder performs better. GPT-4 and GPT-3.5 (GPT-3.5-turbo) also exhibit good performance in code generation.
    \item For test case generation tasks, GPT-4 and GPT-3.5 (GPT-3.5-turbo) demonstrate better performance.
    \item In code summarization tasks, CodeT5+ outperforms GPT-3.5 (GPT-3.5-turbo).
    \item In code translation tasks, GPT-4 performs better.
    \item For vulnerability repair tasks, based on limited results, Codex shows better performance.
\end{itemize}

However it should be noted that, except for the relatively accurate results in code generation tasks, the results in other tasks are not precise enough. For example, in tasks such as code translation and code summarization, there is a lack of comparative evaluation work between SOTA Code LLMs such as Unnatural-Code-LLaMA-34B and GPT-4. Furthermore, apart from code generation, we also lack relevant benchmarks to measure the differences in capabilities among various models in other software engineering tasks. An interesting thing to note is that different literature records vary in terms of LLMs' performance in HumanEval, especially for ChatGPT and GPT-3.5 models, and these inconsistencies are not simply due to errors in quoting. The reasons for this occurrence are not yet clear, and it is worth discussing.

%% file: RelatedWork.tex
\section{Related Work}
\label{sec:RW}

Artetxe et al.~\cite{N85} conducted a comprehensive analysis and summary of 27 code-based large models released before December 2022. These 27 models were tested and evaluated on the HumanEval benchmark using a zero-shot approach to provide intuitive comparative results. The article identifies the key factors for the success of code-based large models as model parameters, data quality, and expert tuning. However, there are still many open challenges in code testing benchmarks for large language models (LLMs). For example, most of these benchmarks only have problem descriptions in English and Python code solutions, which cannot cover multiple natural languages and programming languages. The article suggests that current code-based large models still face challenges in terms of comprehension ability, inference ability, explanatory ability, adaptive learning ability, and multitasking ability.

Zheng et al.~\cite{N154} discuss the applications of LLMs in the field of software engineering. The article organizes and categorizes 123 selected works and literature on the intersection of software engineering and LLMs. It classifies them according to software engineering tasks, revealing the research focuses and potential directions for combining various software engineering tasks with LLMs. Additionally, the article reveals the performance of LLMs in these tasks, along with their limitations, and provides directions for future research and optimization.

Some efforts aim to enhance the capabilities of existing LLMs in software engineering tasks. Gong et al.~\cite{N9} introduces CODETF, an open-source library based on transformers for code LLMs and code intelligence. CODETF is designed with a unified interface to enable fast access and development across different types of models, datasets, and tasks. Strictly speaking, CODETF is not a conventional LLM in the traditional sense but more of a methodology. Lu et al.~\cite{N54} addresses the phenomenon of training code LLMs on large, uncleaned source code corpora scraped from the internet. The paper discusses the security, privacy, and licensing issues associated with generated LLMs and provides four feasible recommendations to address these concerns. Le et al.~\cite{N77} presents a Cross-language Code representation with a large-scale pre-training (XCode) method, which utilizes several abstract syntax trees and ELMo-enhanced variational autoencoders to train multiple pre-trained source code language models on approximately 1.5 million code snippets. Maddigan et al.~\cite{N5} enhances the robustness of existing pre-trained models by designing nine PL-NL enhancement operators to group semantically equivalent variants.

Some works have utilized the capabilities of LLMs to design automated tools or frameworks. Gao et al.~\cite{N48} introduces CollabCoder, a system that supports users in completing qualitative coding through multiple stages using LLMs. Tanaka et al.~\cite{N17} proposes a new learning approach called Inductive Bias Learning (IBL), which combines Inductive Concept Learning (ICL) and code generation techniques. The article suggests that the prediction models generated by IBL have the potential to replace existing machine learning models in terms of interpretability and inference speed. Das et al.~\cite{N27} addresses the cumbersome and time-consuming process of generating and integrating code views for each programming language and presents a tool called COMEX. COMEX allows researchers and developers to create and harvest multiple code views that can be used by LLMs for various software engineering tasks.

 
Karmakar et al.~\cite{N10} evaluated the code synthesis ability of the Codex model using a set of 115 Python problem statements from HackerRank and found clear evidence of Codex's ability to generate code from memory. Ding et al.~\cite{N32} proposed a static evaluation framework for code completion generated by large language models and performed error analysis on the CodeGen model using a large-scale real-world Python evaluation set. Martínez et al.~\cite{N25} explored the application of law graduates in code detection and evaluated the code detection capability of GPT-3.5 using matrix multiplication. N41 conducted an empirical study to assess ChatGPT's unit test generation capability and proposed a method called CHATTESTER, which utilizes ChatGPT itself to improve the quality of generated tests. The results showed that GPT-3.5 achieved an accuracy rate close to 100\% in the evaluation. Furthermore, there are several related works that have developed new benchmarks for testing the code generation capabilities of LLMs, such as CODETASKCL~\cite{N28}, CodeBLEU~\cite{N103}, CodeSearchNet~\cite{N64}, and Galeras~\cite{N53}.

%% file: Conclusion.tex
\section{Conclusion}
\label{sec:conclusion1}


After a comprehensive review, this paper explores the performance and value of specialized LLMs in the field of software engineering. Firstly, we collected and screened 134 works related to Code LLMs. Next, we organized Code LLMs based on the types of institutions to which their main developers belong, revealing the relationships between Code LLMs, general LLMs, and among Code LLMs themselves. Furthermore, we conducted a comprehensive analysis and compilation of the performance of general LLMs and Code LLMs in software engineering tasks. We provided statistical results and analyzed interesting phenomena. Lastly, we maintained the scores of 126 Code LLMs on major benchmarks and conducted a detailed analysis of their performance across different software engineering tasks. The contribution of this paper lies in the comprehensive overview of Code LLMs and their performance. This work can assist Code LLM developers in making informed decisions regarding base models and fine-tuning approaches, and it also provides key improvement directions for Code LLMs.

%% file: main-manuscript.bbl

\begin{thebibliography}{125}


\ifx \showCODEN    \undefined \def \showCODEN     #1{\unskip}     \fi
\ifx \showDOI      \undefined \def \showDOI       #1{#1}\fi
\ifx \showISBNx    \undefined \def \showISBNx     #1{\unskip}     \fi
\ifx \showISBNxiii \undefined \def \showISBNxiii  #1{\unskip}     \fi
\ifx \showISSN     \undefined \def \showISSN      #1{\unskip}     \fi
\ifx \showLCCN     \undefined \def \showLCCN      #1{\unskip}     \fi
\ifx \shownote     \undefined \def \shownote      #1{#1}          \fi
\ifx \showarticletitle \undefined \def \showarticletitle #1{#1}   \fi
\ifx \showURL      \undefined \def \showURL       {\relax}        \fi
\providecommand\bibfield[2]{#2}
\providecommand\bibinfo[2]{#2}
\providecommand\natexlab[1]{#1}
\providecommand\showeprint[2][]{arXiv:#2}

\bibitem[Ahmad et~al\mbox{.}(2021)]%
        {N65}
\bibfield{author}{\bibinfo{person}{Wasi~Uddin Ahmad}, \bibinfo{person}{Saikat Chakraborty}, \bibinfo{person}{Baishakhi Ray}, {and} \bibinfo{person}{Kai{-}Wei Chang}.} \bibinfo{year}{2021}\natexlab{}.
\newblock \showarticletitle{Unified Pre-training for Program Understanding and Generation}. In \bibinfo{booktitle}{\emph{Proceedings of the 2021 Conference of the North American Chapter of the Association for Computational Linguistics: Human Language Technologies, {NAACL-HLT} 2021, Online, June 6-11, 2021}}. \bibinfo{publisher}{Association for Computational Linguistics}, \bibinfo{pages}{2655--2668}.
\newblock
\urldef\tempurl%
\url{https://doi.org/10.18653/v1/2021.naacl-main.211}
\showDOI{\tempurl}


\bibitem[Ahmed and Devanbu(2022)]%
        {N140}
\bibfield{author}{\bibinfo{person}{Toufique Ahmed} {and} \bibinfo{person}{Premkumar~T. Devanbu}.} \bibinfo{year}{2022}\natexlab{}.
\newblock \showarticletitle{Few-shot training LLMs for project-specific code-summarization}. In \bibinfo{booktitle}{\emph{37th {IEEE/ACM} International Conference on Automated Software Engineering, {ASE} 2022, Rochester, MI, USA, October 10-14, 2022}}. \bibinfo{publisher}{{ACM}}.
\newblock


\bibitem[Allal et~al\mbox{.}(2023)]%
        {N7}
\bibfield{author}{\bibinfo{person}{Loubna~Ben Allal}, \bibinfo{person}{Raymond Li}, \bibinfo{person}{Denis Kocetkov}, \bibinfo{person}{Chenghao Mou}, \bibinfo{person}{Christopher Akiki}, \bibinfo{person}{Carlos~Mu{\~{n}}oz Ferrandis}, \bibinfo{person}{Niklas Muennighoff}, \bibinfo{person}{Mayank Mishra}, \bibinfo{person}{Alex Gu}, \bibinfo{person}{Manan Dey}, \bibinfo{person}{Logesh~Kumar Umapathi}, \bibinfo{person}{Carolyn~Jane Anderson}, \bibinfo{person}{Yangtian Zi}, \bibinfo{person}{Joel Lamy{-}Poirier}, \bibinfo{person}{Hailey Schoelkopf}, \bibinfo{person}{Sergey Troshin}, \bibinfo{person}{Dmitry Abulkhanov}, \bibinfo{person}{Manuel Romero}, \bibinfo{person}{Michael Lappert}, \bibinfo{person}{Francesco~De Toni}, \bibinfo{person}{Bernardo~Garc{\'{\i}}a del R{\'{\i}}o}, \bibinfo{person}{Qian Liu}, \bibinfo{person}{Shamik Bose}, \bibinfo{person}{Urvashi Bhattacharyya}, \bibinfo{person}{Terry~Yue Zhuo}, \bibinfo{person}{Ian Yu}, \bibinfo{person}{Paulo Villegas}, \bibinfo{person}{Marco Zocca},
  \bibinfo{person}{Sourab Mangrulkar}, \bibinfo{person}{David Lansky}, \bibinfo{person}{Huu Nguyen}, \bibinfo{person}{Danish Contractor}, \bibinfo{person}{Luis Villa}, \bibinfo{person}{Jia Li}, \bibinfo{person}{Dzmitry Bahdanau}, \bibinfo{person}{Yacine Jernite}, \bibinfo{person}{Sean Hughes}, \bibinfo{person}{Daniel Fried}, \bibinfo{person}{Arjun Guha}, \bibinfo{person}{Harm de Vries}, {and} \bibinfo{person}{Leandro von Werra}.} \bibinfo{year}{2023}\natexlab{}.
\newblock \showarticletitle{SantaCoder: don't reach for the stars!}
\newblock \bibinfo{journal}{\emph{CoRR}}  \bibinfo{volume}{abs/2301.03988} (\bibinfo{year}{2023}).
\newblock
\urldef\tempurl%
\url{https://doi.org/10.48550/arXiv.2301.03988}
\showDOI{\tempurl}
\showeprint[arXiv]{2301.03988}


\bibitem[Allamanis and Sutton(2013)]%
        {N70}
\bibfield{author}{\bibinfo{person}{Miltiadis Allamanis} {and} \bibinfo{person}{Charles Sutton}.} \bibinfo{year}{2013}\natexlab{}.
\newblock \showarticletitle{Mining source code repositories at massive scale using language modeling}. In \bibinfo{booktitle}{\emph{Proceedings of the 10th Working Conference on Mining Software Repositories, {MSR} '13, San Francisco, CA, USA, May 18-19, 2013}}, \bibfield{editor}{\bibinfo{person}{Thomas Zimmermann}, \bibinfo{person}{Massimiliano~Di Penta}, {and} \bibinfo{person}{Sunghun Kim}} (Eds.). \bibinfo{publisher}{{IEEE} Computer Society}, \bibinfo{pages}{207--216}.
\newblock
\urldef\tempurl%
\url{https://doi.org/10.1109/MSR.2013.6624029}
\showDOI{\tempurl}


\bibitem[Artetxe et~al\mbox{.}(2022)]%
        {N85}
\bibfield{author}{\bibinfo{person}{Mikel Artetxe}, \bibinfo{person}{Shruti Bhosale}, \bibinfo{person}{Naman Goyal}, \bibinfo{person}{Todor Mihaylov}, \bibinfo{person}{Myle Ott}, {et~al\mbox{.}}} \bibinfo{year}{2022}\natexlab{}.
\newblock \showarticletitle{Efficient Large Scale Language Modeling with Mixtures of Experts}. In \bibinfo{booktitle}{\emph{Proceedings of the 2022 Conference on Empirical Methods in Natural Language Processing, {EMNLP} 2022, Abu Dhabi, United Arab Emirates, December 7-11, 2022}}. \bibinfo{publisher}{Association for Computational Linguistics}, \bibinfo{pages}{11699--11732}.
\newblock
\urldef\tempurl%
\url{https://aclanthology.org/2022.emnlp-main.804}
\showURL{%
\tempurl}


\bibitem[Athiwaratkun et~al\mbox{.}(2023)]%
        {N122}
\bibfield{author}{\bibinfo{person}{Ben Athiwaratkun}, \bibinfo{person}{Sanjay~Krishna Gouda}, \bibinfo{person}{Zijian Wang}, \bibinfo{person}{Xiaopeng Li}, {et~al\mbox{.}}} \bibinfo{year}{2023}\natexlab{}.
\newblock \bibinfo{title}{Multi-lingual Evaluation of Code Generation Models}.
\newblock
\newblock
\showeprint[arxiv]{2210.14868}~[cs.LG]


\bibitem[Bavarian et~al\mbox{.}(2022)]%
        {N74}
\bibfield{author}{\bibinfo{person}{Mohammad Bavarian}, \bibinfo{person}{Heewoo Jun}, \bibinfo{person}{Nikolas Tezak}, \bibinfo{person}{John Schulman}, \bibinfo{person}{Christine McLeavey}, \bibinfo{person}{Jerry Tworek}, {and} \bibinfo{person}{Mark Chen}.} \bibinfo{year}{2022}\natexlab{}.
\newblock \showarticletitle{Efficient Training of Language Models to Fill in the Middle}.
\newblock \bibinfo{journal}{\emph{CoRR}}  \bibinfo{volume}{abs/2207.14255} (\bibinfo{year}{2022}).
\newblock
\urldef\tempurl%
\url{https://doi.org/10.48550/arXiv.2207.14255}
\showDOI{\tempurl}


\bibitem[Black et~al\mbox{.}(2022)]%
        {N132}
\bibfield{author}{\bibinfo{person}{Sid Black}, \bibinfo{person}{Stella Biderman}, \bibinfo{person}{Eric Hallahan}, \bibinfo{person}{Quentin Anthony}, \bibinfo{person}{Leo Gao}, {et~al\mbox{.}}} \bibinfo{year}{2022}\natexlab{}.
\newblock \showarticletitle{{GPT-NeoX-20B}: An Open-Source Autoregressive Language Model}. In \bibinfo{booktitle}{\emph{Proceedings of the ACL Workshop on Challenges \& Perspectives in Creating Large Language Models}}.
\newblock
\urldef\tempurl%
\url{https://arxiv.org/abs/2204.06745}
\showURL{%
\tempurl}


\bibitem[Bui et~al\mbox{.}(2023)]%
        {N34}
\bibfield{author}{\bibinfo{person}{Nghi D.~Q. Bui}, \bibinfo{person}{Hung Le}, \bibinfo{person}{Yue Wang}, \bibinfo{person}{Junnan Li}, \bibinfo{person}{Akhilesh~Deepak Gotmare}, {and} \bibinfo{person}{Steven C.~H. Hoi}.} \bibinfo{year}{2023}\natexlab{}.
\newblock \bibinfo{title}{CodeTF: One-stop Transformer Library for State-of-the-art Code LLM}.
\newblock
\newblock
\showeprint[arxiv]{2306.00029}~[cs.SE]


\bibitem[Cassano et~al\mbox{.}(2023)]%
        {N16}
\bibfield{author}{\bibinfo{person}{Federico Cassano}, \bibinfo{person}{John Gouwar}, \bibinfo{person}{Francesca Lucchetti}, \bibinfo{person}{Claire Schlesinger}, \bibinfo{person}{Carolyn~Jane Anderson}, \bibinfo{person}{Michael Greenberg}, \bibinfo{person}{Abhinav Jangda}, {and} \bibinfo{person}{Arjun Guha}.} \bibinfo{year}{2023}\natexlab{}.
\newblock \bibinfo{title}{Knowledge Transfer from High-Resource to Low-Resource Programming Languages for Code LLMs}.
\newblock
\newblock
\showeprint[arxiv]{2308.09895}~[cs.PL]


\bibitem[Chan et~al\mbox{.}(2023)]%
        {N149}
\bibfield{author}{\bibinfo{person}{Aaron Chan}, \bibinfo{person}{Anant Kharkar}, \bibinfo{person}{Roshanak~Zilouchian Moghaddam}, \bibinfo{person}{Yevhen Mohylevskyy}, \bibinfo{person}{Alec Helyar}, \bibinfo{person}{Eslam Kamal}, \bibinfo{person}{Mohamed Elkamhawy}, {and} \bibinfo{person}{Neel Sundaresan}.} \bibinfo{year}{2023}\natexlab{}.
\newblock \bibinfo{title}{Transformer-based Vulnerability Detection in Code at EditTime: Zero-shot, Few-shot, or Fine-tuning?}
\newblock
\newblock
\showeprint[arxiv]{2306.01754}~[cs.CR]


\bibitem[Chandel et~al\mbox{.}(2022)]%
        {N73}
\bibfield{author}{\bibinfo{person}{Shubham Chandel}, \bibinfo{person}{Colin~B. Clement}, \bibinfo{person}{Guillermo Serrato}, {and} \bibinfo{person}{Neel Sundaresan}.} \bibinfo{year}{2022}\natexlab{}.
\newblock \showarticletitle{Training and Evaluating a Jupyter Notebook Data Science Assistant}.
\newblock \bibinfo{journal}{\emph{CoRR}}  \bibinfo{volume}{abs/2201.12901} (\bibinfo{year}{2022}).
\newblock
\showeprint[arXiv]{2201.12901}
\urldef\tempurl%
\url{https://arxiv.org/abs/2201.12901}
\showURL{%
\tempurl}


\bibitem[Chaudhary(2023)]%
        {N92}
\bibfield{author}{\bibinfo{person}{Sahil Chaudhary}.} \bibinfo{year}{2023}\natexlab{}.
\newblock \bibinfo{title}{Code Alpaca: An Instruction-following LLaMA model for code generation}.
\newblock \bibinfo{howpublished}{\url{https://github.com/sahil280114/codealpaca}}.
\newblock


\bibitem[Chen et~al\mbox{.}(2021d)]%
        {N137}
\bibfield{author}{\bibinfo{person}{Jiachi Chen}, \bibinfo{person}{Xin Xia}, \bibinfo{person}{David Lo}, \bibinfo{person}{John Grundy}, {and} \bibinfo{person}{Xiaohu Yang}.} \bibinfo{year}{2021}\natexlab{d}.
\newblock \showarticletitle{Maintenance-related concerns for post-deployed Ethereum smart contract development: issues, techniques, and future challenges}.
\newblock \bibinfo{journal}{\emph{Empirical Software Engineering}} \bibinfo{volume}{26}, \bibinfo{number}{6} (\bibinfo{year}{2021}), \bibinfo{pages}{1--44}.
\newblock


\bibitem[Chen et~al\mbox{.}(2021a)]%
        {N142}
\bibfield{author}{\bibinfo{person}{Mark Chen}, \bibinfo{person}{Jerry Tworek}, \bibinfo{person}{Heewoo Jun}, \bibinfo{person}{Qiming Yuan}, \bibinfo{person}{Henrique Ponde de~Oliveira Pinto}, \bibinfo{person}{Jared Kaplan}, \bibinfo{person}{Harri Edwards}, \bibinfo{person}{Yuri Burda}, \bibinfo{person}{Nicholas Joseph}, \bibinfo{person}{Greg Brockman}, {et~al\mbox{.}}} \bibinfo{year}{2021}\natexlab{a}.
\newblock \showarticletitle{Evaluating large language models trained on code}.
\newblock \bibinfo{journal}{\emph{arXiv preprint arXiv:2107.03374}} (\bibinfo{year}{2021}).
\newblock


\bibitem[Chen et~al\mbox{.}(2021b)]%
        {N150}
\bibfield{author}{\bibinfo{person}{Mark Chen}, \bibinfo{person}{Jerry Tworek}, \bibinfo{person}{Heewoo Jun}, \bibinfo{person}{Qiming Yuan}, \bibinfo{person}{Henrique Ponde de~Oliveira Pinto}, \bibinfo{person}{Jared Kaplan}, \bibinfo{person}{Harri Edwards}, \bibinfo{person}{Yuri Burda}, \bibinfo{person}{Nicholas Joseph}, \bibinfo{person}{Greg Brockman}, {et~al\mbox{.}}} \bibinfo{year}{2021}\natexlab{b}.
\newblock \showarticletitle{Evaluating large language models trained on code}.
\newblock \bibinfo{journal}{\emph{arXiv preprint arXiv:2107.03374}} (\bibinfo{year}{2021}).
\newblock


\bibitem[Chen et~al\mbox{.}(2021c)]%
        {N61}
\bibfield{author}{\bibinfo{person}{Mark Chen}, \bibinfo{person}{Jerry Tworek}, \bibinfo{person}{Heewoo Jun}, \bibinfo{person}{Qiming Yuan}, \bibinfo{person}{Henrique Ponde de~Oliveira Pinto}, \bibinfo{person}{Jared Kaplan}, \bibinfo{person}{Harri Edwards}, \bibinfo{person}{Yuri Burda}, \bibinfo{person}{Nicholas Joseph}, \bibinfo{person}{Greg Brockman}, {et~al\mbox{.}}} \bibinfo{year}{2021}\natexlab{c}.
\newblock \showarticletitle{Evaluating large language models trained on code}.
\newblock \bibinfo{journal}{\emph{arXiv preprint arXiv:2107.03374}} (\bibinfo{year}{2021}).
\newblock


\bibitem[Chowdhery et~al\mbox{.}(2022)]%
        {N81}
\bibfield{author}{\bibinfo{person}{Aakanksha Chowdhery}, \bibinfo{person}{Sharan Narang}, \bibinfo{person}{Jacob Devlin}, \bibinfo{person}{Maarten Bosma}, {et~al\mbox{.}}} \bibinfo{year}{2022}\natexlab{}.
\newblock \showarticletitle{PaLM: Scaling Language Modeling with Pathways}.
\newblock \bibinfo{journal}{\emph{CoRR}}  \bibinfo{volume}{abs/2204.02311} (\bibinfo{year}{2022}).
\newblock
\urldef\tempurl%
\url{https://doi.org/10.48550/arXiv.2204.02311}
\showDOI{\tempurl}
\showeprint[arXiv]{2204.02311}


\bibitem[Christopoulou et~al\mbox{.}(2022)]%
        {N86}
\bibfield{author}{\bibinfo{person}{Fenia Christopoulou}, \bibinfo{person}{Gerasimos Lampouras}, \bibinfo{person}{Milan Gritta}, \bibinfo{person}{Guchun Zhang}, \bibinfo{person}{Yinpeng Guo}, \bibinfo{person}{Zhongqi Li}, \bibinfo{person}{Qi Zhang}, \bibinfo{person}{Meng Xiao}, \bibinfo{person}{Bo Shen}, \bibinfo{person}{Lin Li}, {et~al\mbox{.}}} \bibinfo{year}{2022}\natexlab{}.
\newblock \showarticletitle{Pangu-coder: Program synthesis with function-level language modeling}.
\newblock \bibinfo{journal}{\emph{arXiv preprint arXiv:2207.11280}} (\bibinfo{year}{2022}).
\newblock


\bibitem[Clement et~al\mbox{.}(2020)]%
        {N63}
\bibfield{author}{\bibinfo{person}{Colin~B. Clement}, \bibinfo{person}{Dawn Drain}, \bibinfo{person}{Jonathan Timcheck}, \bibinfo{person}{Alexey Svyatkovskiy}, {and} \bibinfo{person}{Neel Sundaresan}.} \bibinfo{year}{2020}\natexlab{}.
\newblock \showarticletitle{PyMT5: multi-mode translation of natural language and Python code with transformers}. In \bibinfo{booktitle}{\emph{Proceedings of the 2020 Conference on Empirical Methods in Natural Language Processing, {EMNLP} 2020, Online, November 16-20, 2020}}, \bibfield{editor}{\bibinfo{person}{Bonnie Webber}, \bibinfo{person}{Trevor Cohn}, \bibinfo{person}{Yulan He}, {and} \bibinfo{person}{Yang Liu}} (Eds.). \bibinfo{publisher}{Association for Computational Linguistics}, \bibinfo{pages}{9052--9065}.
\newblock
\urldef\tempurl%
\url{https://doi.org/10.18653/v1/2020.emnlp-main.728}
\showDOI{\tempurl}


\bibitem[CodedotAl(2021)]%
        {N131}
\bibfield{author}{\bibinfo{person}{CodedotAl}.} \bibinfo{year}{2021}\natexlab{}.
\newblock \bibinfo{booktitle}{\emph{{GPT-Code-Clippy}}}.
\newblock
\urldef\tempurl%
\url{https://github.com/CodedotAl/gpt-code-clippy}
\showURL{%
\tempurl}


\bibitem[Das et~al\mbox{.}(2023)]%
        {N27}
\bibfield{author}{\bibinfo{person}{Debeshee Das}, \bibinfo{person}{Noble~Saji Mathews}, \bibinfo{person}{Alex Mathai}, \bibinfo{person}{Srikanth Tamilselvam}, \bibinfo{person}{Kranthi Sedamaki}, \bibinfo{person}{Sridhar Chimalakonda}, {and} \bibinfo{person}{Atul Kumar}.} \bibinfo{year}{2023}\natexlab{}.
\newblock \bibinfo{title}{COMEX: A Tool for Generating Customized Source Code Representations}.
\newblock
\newblock
\showeprint[arxiv]{2307.04693}~[cs.SE]


\bibitem[Ding et~al\mbox{.}(2023)]%
        {N32}
\bibfield{author}{\bibinfo{person}{Hantian Ding}, \bibinfo{person}{Varun Kumar}, \bibinfo{person}{Yuchen Tian}, \bibinfo{person}{Zijian Wang}, \bibinfo{person}{Rob Kwiatkowski}, \bibinfo{person}{Xiaopeng Li}, \bibinfo{person}{Murali~Krishna Ramanathan}, \bibinfo{person}{Baishakhi Ray}, \bibinfo{person}{Parminder Bhatia}, \bibinfo{person}{Sudipta Sengupta}, \bibinfo{person}{Dan Roth}, {and} \bibinfo{person}{Bing Xiang}.} \bibinfo{year}{2023}\natexlab{}.
\newblock \bibinfo{title}{A Static Evaluation of Code Completion by Large Language Models}.
\newblock
\newblock
\showeprint[arxiv]{2306.03203}~[cs.CL]


\bibitem[Dinh et~al\mbox{.}(2023)]%
        {N31}
\bibfield{author}{\bibinfo{person}{Tuan Dinh}, \bibinfo{person}{Jinman Zhao}, \bibinfo{person}{Samson Tan}, \bibinfo{person}{Renato Negrinho}, \bibinfo{person}{Leonard Lausen}, \bibinfo{person}{Sheng Zha}, {and} \bibinfo{person}{George Karypis}.} \bibinfo{year}{2023}\natexlab{}.
\newblock \bibinfo{title}{Large Language Models of Code Fail at Completing Code with Potential Bugs}.
\newblock
\newblock
\showeprint[arxiv]{2306.03438}~[cs.LG]


\bibitem[Du et~al\mbox{.}(2023)]%
        {N128}
\bibfield{author}{\bibinfo{person}{Xueying Du}, \bibinfo{person}{Mingwei Liu}, \bibinfo{person}{Kaixin Wang}, \bibinfo{person}{Hanlin Wang}, \bibinfo{person}{Junwei Liu}, \bibinfo{person}{Yixuan Chen}, \bibinfo{person}{Jiayi Feng}, \bibinfo{person}{Chaofeng Sha}, \bibinfo{person}{Xin Peng}, {and} \bibinfo{person}{Yiling Lou}.} \bibinfo{year}{2023}\natexlab{}.
\newblock \bibinfo{title}{ClassEval: A Manually-Crafted Benchmark for Evaluating LLMs on Class-level Code Generation}.
\newblock
\newblock
\showeprint[arxiv]{2308.01861}~[cs.CL]


\bibitem[Fried et~al\mbox{.}(2023)]%
        {N84}
\bibfield{author}{\bibinfo{person}{Daniel Fried}, \bibinfo{person}{Armen Aghajanyan}, \bibinfo{person}{Jessy Lin}, \bibinfo{person}{Sida Wang}, \bibinfo{person}{Eric Wallace}, \bibinfo{person}{Freda Shi}, \bibinfo{person}{Ruiqi Zhong}, \bibinfo{person}{Scott Yih}, \bibinfo{person}{Luke Zettlemoyer}, {and} \bibinfo{person}{Mike Lewis}.} \bibinfo{year}{2023}\natexlab{}.
\newblock \showarticletitle{InCoder: {A} Generative Model for Code Infilling and Synthesis}. In \bibinfo{booktitle}{\emph{The Eleventh International Conference on Learning Representations, {ICLR} 2023, Kigali, Rwanda, May 1-5, 2023}}. \bibinfo{publisher}{OpenReview.net}.
\newblock
\urldef\tempurl%
\url{https://openreview.net/pdf?id=hQwb-lbM6EL}
\showURL{%
\tempurl}


\bibitem[Fu et~al\mbox{.}(2023)]%
        {N55}
\bibfield{author}{\bibinfo{person}{Lingyue Fu}, \bibinfo{person}{Huacan Chai}, \bibinfo{person}{Shuang Luo}, \bibinfo{person}{Kounianhua Du}, \bibinfo{person}{Weiming Zhang}, \bibinfo{person}{Longteng Fan}, \bibinfo{person}{Jiayi Lei}, \bibinfo{person}{Renting Rui}, \bibinfo{person}{Jianghao Lin}, \bibinfo{person}{Yuchen Fang}, \bibinfo{person}{Yifan Liu}, \bibinfo{person}{Jingkuan Wang}, \bibinfo{person}{Siyuan Qi}, \bibinfo{person}{Kangning Zhang}, \bibinfo{person}{Weinan Zhang}, {and} \bibinfo{person}{Yong Yu}.} \bibinfo{year}{2023}\natexlab{}.
\newblock \bibinfo{title}{CodeApex: A Bilingual Programming Evaluation Benchmark for Large Language Models}.
\newblock
\newblock
\showeprint[arxiv]{2309.01940}~[cs.CL]


\bibitem[Gao et~al\mbox{.}(2023)]%
        {N48}
\bibfield{author}{\bibinfo{person}{Jie Gao}, \bibinfo{person}{Yuchen Guo}, \bibinfo{person}{Gionnieve Lim}, \bibinfo{person}{Tianqin Zhang}, \bibinfo{person}{Zheng Zhang}, \bibinfo{person}{Toby Jia-Jun Li}, {and} \bibinfo{person}{Simon~Tangi Perrault}.} \bibinfo{year}{2023}\natexlab{}.
\newblock \bibinfo{title}{CollabCoder: A GPT-Powered Workflow for Collaborative Qualitative Analysis}.
\newblock
\newblock
\showeprint[arxiv]{2304.07366}~[cs.HC]


\bibitem[Gong et~al\mbox{.}(2022)]%
        {N9}
\bibfield{author}{\bibinfo{person}{Zi Gong}, \bibinfo{person}{Yinpeng Guo}, \bibinfo{person}{Pingyi Zhou}, \bibinfo{person}{Cuiyun Gao}, \bibinfo{person}{Yasheng Wang}, {and} \bibinfo{person}{Zenglin Xu}.} \bibinfo{year}{2022}\natexlab{}.
\newblock \bibinfo{title}{MultiCoder: Multi-Programming-Lingual Pre-Training for Low-Resource Code Completion}.
\newblock
\newblock
\showeprint[arxiv]{2212.09666}~[cs.CL]


\bibitem[Gu et~al\mbox{.}(2020)]%
        {N112}
\bibfield{author}{\bibinfo{person}{Albert Gu}, \bibinfo{person}{Tri Dao}, \bibinfo{person}{Stefano Ermon}, \bibinfo{person}{Atri Rudra}, {and} \bibinfo{person}{Christopher R{\'{e}}}.} \bibinfo{year}{2020}\natexlab{}.
\newblock \showarticletitle{HiPPO: Recurrent Memory with Optimal Polynomial Projections}. In \bibinfo{booktitle}{\emph{Advances in Neural Information Processing Systems 33: Annual Conference on Neural Information Processing Systems 2020, NeurIPS 2020, December 6-12, 2020, virtual}}, \bibfield{editor}{\bibinfo{person}{Hugo Larochelle}, \bibinfo{person}{Marc'Aurelio Ranzato}, \bibinfo{person}{Raia Hadsell}, \bibinfo{person}{Maria{-}Florina Balcan}, {and} \bibinfo{person}{Hsuan{-}Tien Lin}} (Eds.).
\newblock


\bibitem[Gunasekar et~al\mbox{.}(2023)]%
        {N98}
\bibfield{author}{\bibinfo{person}{Suriya Gunasekar}, \bibinfo{person}{Yi Zhang}, \bibinfo{person}{Jyoti Aneja}, \bibinfo{person}{Caio C{\'{e}}sar~Teodoro Mendes}, \bibinfo{person}{Allie~Del Giorno}, \bibinfo{person}{Sivakanth Gopi}, {et~al\mbox{.}}} \bibinfo{year}{2023}\natexlab{}.
\newblock \showarticletitle{Textbooks Are All You Need}.
\newblock \bibinfo{journal}{\emph{CoRR}}  \bibinfo{volume}{abs/2306.11644} (\bibinfo{year}{2023}).
\newblock
\urldef\tempurl%
\url{https://doi.org/10.48550/arXiv.2306.11644}
\showDOI{\tempurl}
\showeprint[arXiv]{2306.11644}


\bibitem[Hendrycks et~al\mbox{.}(2021)]%
        {N153}
\bibfield{author}{\bibinfo{person}{Dan Hendrycks}, \bibinfo{person}{Steven Basart}, \bibinfo{person}{Saurav Kadavath}, \bibinfo{person}{Mantas Mazeika}, \bibinfo{person}{Akul Arora}, \bibinfo{person}{Ethan Guo}, \bibinfo{person}{Collin Burns}, \bibinfo{person}{Samir Puranik}, \bibinfo{person}{Horace He}, \bibinfo{person}{Dawn Song}, {and} \bibinfo{person}{Jacob Steinhardt}.} \bibinfo{year}{2021}\natexlab{}.
\newblock \showarticletitle{Measuring Coding Challenge Competence With APPS}.
\newblock \bibinfo{journal}{\emph{NeurIPS}} (\bibinfo{year}{2021}).
\newblock


\bibitem[Hochreiter and Schmidhuber(1997)]%
        {N66}
\bibfield{author}{\bibinfo{person}{Sepp Hochreiter} {and} \bibinfo{person}{J{\"{u}}rgen Schmidhuber}.} \bibinfo{year}{1997}\natexlab{}.
\newblock \showarticletitle{Long Short-Term Memory}.
\newblock \bibinfo{journal}{\emph{Neural Comput.}} \bibinfo{volume}{9}, \bibinfo{number}{8} (\bibinfo{year}{1997}), \bibinfo{pages}{1735--1780}.
\newblock
\urldef\tempurl%
\url{https://doi.org/10.1162/neco.1997.9.8.1735}
\showDOI{\tempurl}


\bibitem[Hou et~al\mbox{.}(2023)]%
        {N138}
\bibfield{author}{\bibinfo{person}{Xinyi Hou}, \bibinfo{person}{Yanjie Zhao}, \bibinfo{person}{Yue Liu}, \bibinfo{person}{Zhou Yang}, \bibinfo{person}{Kailong Wang}, \bibinfo{person}{Li Li}, \bibinfo{person}{Xiapu Luo}, \bibinfo{person}{David Lo}, \bibinfo{person}{John Grundy}, {and} \bibinfo{person}{Haoyu Wang}.} \bibinfo{year}{2023}\natexlab{}.
\newblock \showarticletitle{Large Language Models for Software Engineering: A Systematic Literature Review}.
\newblock \bibinfo{journal}{\emph{arXiv preprint arXiv:2308.10620}} (\bibinfo{year}{2023}).
\newblock


\bibitem[Huaggingface(2021)]%
        {N79}
\bibfield{author}{\bibinfo{person}{Huaggingface}.} \bibinfo{year}{2021}\natexlab{}.
\newblock \showarticletitle{Training CodeParrot from Scratch}.
\newblock
\urldef\tempurl%
\url{https://huggingface.co/blog/codeparrot.}
\showURL{%
\tempurl}


\bibitem[Husain et~al\mbox{.}(2019)]%
        {N64}
\bibfield{author}{\bibinfo{person}{Hamel Husain}, \bibinfo{person}{Ho{-}Hsiang Wu}, \bibinfo{person}{Tiferet Gazit}, \bibinfo{person}{Miltiadis Allamanis}, {and} \bibinfo{person}{Marc Brockschmidt}.} \bibinfo{year}{2019}\natexlab{}.
\newblock \showarticletitle{CodeSearchNet Challenge: Evaluating the State of Semantic Code Search}.
\newblock \bibinfo{journal}{\emph{CoRR}}  \bibinfo{volume}{abs/1909.09436} (\bibinfo{year}{2019}).
\newblock
\showeprint[arXiv]{1909.09436}
\urldef\tempurl%
\url{http://arxiv.org/abs/1909.09436}
\showURL{%
\tempurl}


\bibitem[Iyer et~al\mbox{.}(2016)]%
        {N68}
\bibfield{author}{\bibinfo{person}{Srinivasan Iyer}, \bibinfo{person}{Ioannis Konstas}, \bibinfo{person}{Alvin Cheung}, {and} \bibinfo{person}{Luke Zettlemoyer}.} \bibinfo{year}{2016}\natexlab{}.
\newblock \showarticletitle{Summarizing Source Code using a Neural Attention Model}. In \bibinfo{booktitle}{\emph{Proceedings of the 54th Annual Meeting of the Association for Computational Linguistics, {ACL} 2016, August 7-12, 2016, Berlin, Germany, Volume 1: Long Papers}}. \bibinfo{publisher}{The Association for Computer Linguistics}.
\newblock
\urldef\tempurl%
\url{https://doi.org/10.18653/v1/p16-1195}
\showDOI{\tempurl}


\bibitem[Jiang et~al\mbox{.}(2023)]%
        {N2}
\bibfield{author}{\bibinfo{person}{Xue Jiang}, \bibinfo{person}{Yihong Dong}, \bibinfo{person}{Lecheng Wang}, \bibinfo{person}{Zheng Fang}, \bibinfo{person}{Qiwei Shang}, \bibinfo{person}{Ge Li}, \bibinfo{person}{Zhi Jin}, {and} \bibinfo{person}{Wenpin Jiao}.} \bibinfo{year}{2023}\natexlab{}.
\newblock \bibinfo{title}{Self-planning Code Generation with Large Language Models}.
\newblock
\newblock
\showeprint[arxiv]{2303.06689}~[cs.SE]


\bibitem[Karmakar et~al\mbox{.}(2022)]%
        {N10}
\bibfield{author}{\bibinfo{person}{Anjan Karmakar}, \bibinfo{person}{Julian~Aron Prenner}, \bibinfo{person}{Marco D'Ambros}, {and} \bibinfo{person}{Romain Robbes}.} \bibinfo{year}{2022}\natexlab{}.
\newblock \bibinfo{title}{Codex Hacks HackerRank: Memorization Issues and a Framework for Code Synthesis Evaluation}.
\newblock
\newblock
\showeprint[arxiv]{2212.02684}~[cs.SE]


\bibitem[Khan et~al\mbox{.}(2023)]%
        {N120}
\bibfield{author}{\bibinfo{person}{Mohammad Abdullah~Matin Khan}, \bibinfo{person}{M~Saiful Bari}, \bibinfo{person}{Xuan~Long Do}, \bibinfo{person}{Weishi Wang}, \bibinfo{person}{Md~Rizwan Parvez}, {and} \bibinfo{person}{Shafiq Joty}.} \bibinfo{year}{2023}\natexlab{}.
\newblock \bibinfo{title}{xCodeEval: A Large Scale Multilingual Multitask Benchmark for Code Understanding, Generation, Translation and Retrieval}.
\newblock
\newblock
\showeprint[arxiv]{2303.03004}~[cs.CL]


\bibitem[Kitchenham(2007)]%
        {N135}
\bibfield{author}{\bibinfo{person}{B.~A. Kitchenham}.} \bibinfo{year}{2007}\natexlab{}.
\newblock \showarticletitle{Kitchenham, B.: Guidelines for performing Systematic Literature Reviews in software engineering. EBSE Technical Report EBSE-2007-01}.
\newblock \bibinfo{journal}{\emph{IEEE Computer Society}} (\bibinfo{year}{2007}).
\newblock


\bibitem[Kocetkov et~al\mbox{.}(2022)]%
        {N100}
\bibfield{author}{\bibinfo{person}{Denis Kocetkov}, \bibinfo{person}{Raymond Li}, \bibinfo{person}{Loubna~Ben Allal}, \bibinfo{person}{Jia Li}, \bibinfo{person}{Chenghao Mou}, {et~al\mbox{.}}} \bibinfo{year}{2022}\natexlab{}.
\newblock \showarticletitle{The Stack: 3 {TB} of permissively licensed source code}.
\newblock \bibinfo{journal}{\emph{CoRR}}  \bibinfo{volume}{abs/2211.15533} (\bibinfo{year}{2022}).
\newblock
\urldef\tempurl%
\url{https://doi.org/10.48550/arXiv.2211.15533}
\showDOI{\tempurl}
\showeprint[arXiv]{2211.15533}


\bibitem[Kou et~al\mbox{.}(2023)]%
        {N127}
\bibfield{author}{\bibinfo{person}{Bonan Kou}, \bibinfo{person}{Shengmai Chen}, \bibinfo{person}{Zhijie Wang}, \bibinfo{person}{Lei Ma}, {and} \bibinfo{person}{Tianyi Zhang}.} \bibinfo{year}{2023}\natexlab{}.
\newblock \bibinfo{title}{Is Model Attention Aligned with Human Attention? An Empirical Study on Large Language Models for Code Generation}.
\newblock
\newblock
\showeprint[arxiv]{2306.01220}~[cs.SE]


\bibitem[Lai et~al\mbox{.}(2022)]%
        {N151}
\bibfield{author}{\bibinfo{person}{Yuhang Lai}, \bibinfo{person}{Chengxi Li}, \bibinfo{person}{Yiming Wang}, \bibinfo{person}{Tianyi Zhang}, \bibinfo{person}{Ruiqi Zhong}, \bibinfo{person}{Luke Zettlemoyer}, \bibinfo{person}{Scott~Wen tau Yih}, \bibinfo{person}{Daniel Fried}, \bibinfo{person}{Sida Wang}, {and} \bibinfo{person}{Tao Yu}.} \bibinfo{year}{2022}\natexlab{}.
\newblock \bibinfo{title}{DS-1000: A Natural and Reliable Benchmark for Data Science Code Generation}.
\newblock
\newblock
\showeprint[arxiv]{2211.11501}~[cs.SE]


\bibitem[Le et~al\mbox{.}(2022)]%
        {N77}
\bibfield{author}{\bibinfo{person}{Hung Le}, \bibinfo{person}{Yue Wang}, \bibinfo{person}{Akhilesh~Deepak Gotmare}, \bibinfo{person}{Silvio Savarese}, {and} \bibinfo{person}{Steven~Chu{-}Hong Hoi}.} \bibinfo{year}{2022}\natexlab{}.
\newblock \showarticletitle{CodeRL: Mastering Code Generation through Pretrained Models and Deep Reinforcement Learning}. In \bibinfo{booktitle}{\emph{NeurIPS}}.
\newblock
\urldef\tempurl%
\url{http://papers.nips.cc}
\showURL{%
\tempurl}


\bibitem[Lhoest et~al\mbox{.}(2021)]%
        {N111}
\bibfield{author}{\bibinfo{person}{Quentin Lhoest}, \bibinfo{person}{Albert~Villanova del Moral}, \bibinfo{person}{Yacine Jernite}, \bibinfo{person}{Abhishek Thakur}, \bibinfo{person}{Patrick von Platen}, \bibinfo{person}{Suraj Patil}, {et~al\mbox{.}}} \bibinfo{year}{2021}\natexlab{}.
\newblock \showarticletitle{Datasets: {A} Community Library for Natural Language Processing}. In \bibinfo{booktitle}{\emph{Proceedings of the 2021 Conference on Empirical Methods in Natural Language Processing: System Demonstrations, {EMNLP} 2021, Online and Punta Cana, Dominican Republic, 7-11 November, 2021}}, \bibfield{editor}{\bibinfo{person}{Heike Adel} {and} \bibinfo{person}{Shuming Shi}} (Eds.). \bibinfo{publisher}{Association for Computational Linguistics}, \bibinfo{pages}{175--184}.
\newblock
\urldef\tempurl%
\url{https://doi.org/10.18653/v1/2021.emnlp-demo.21}
\showDOI{\tempurl}


\bibitem[Li et~al\mbox{.}(2023c)]%
        {N14}
\bibfield{author}{\bibinfo{person}{Jinyang Li}, \bibinfo{person}{Binyuan Hui}, \bibinfo{person}{Ge Qu}, \bibinfo{person}{Binhua Li}, \bibinfo{person}{Jiaxi Yang}, \bibinfo{person}{Bowen Li}, \bibinfo{person}{Bailin Wang}, \bibinfo{person}{Bowen Qin}, \bibinfo{person}{Rongyu Cao}, \bibinfo{person}{Ruiying Geng}, \bibinfo{person}{Nan Huo}, \bibinfo{person}{Xuanhe Zhou}, \bibinfo{person}{Chenhao Ma}, \bibinfo{person}{Guoliang Li}, \bibinfo{person}{Kevin C.~C. Chang}, \bibinfo{person}{Fei Huang}, \bibinfo{person}{Reynold Cheng}, {and} \bibinfo{person}{Yongbin Li}.} \bibinfo{year}{2023}\natexlab{c}.
\newblock \bibinfo{title}{Can LLM Already Serve as A Database Interface? A BIg Bench for Large-Scale Database Grounded Text-to-SQLs}.
\newblock
\newblock
\showeprint[arxiv]{2305.03111}~[cs.CL]


\bibitem[Li et~al\mbox{.}(2023d)]%
        {N13}
\bibfield{author}{\bibinfo{person}{Jia Li}, \bibinfo{person}{Ge Li}, \bibinfo{person}{Yongmin Li}, {and} \bibinfo{person}{Zhi Jin}.} \bibinfo{year}{2023}\natexlab{d}.
\newblock \bibinfo{title}{Structured Chain-of-Thought Prompting for Code Generation}.
\newblock
\newblock
\showeprint[arxiv]{2305.06599}~[cs.SE]


\bibitem[Li et~al\mbox{.}(2023e)]%
        {N40}
\bibfield{author}{\bibinfo{person}{Peng Li}, \bibinfo{person}{Tianxiang Sun}, \bibinfo{person}{Qiong Tang}, \bibinfo{person}{Hang Yan}, \bibinfo{person}{Yuanbin Wu}, \bibinfo{person}{Xuanjing Huang}, {and} \bibinfo{person}{Xipeng Qiu}.} \bibinfo{year}{2023}\natexlab{e}.
\newblock \bibinfo{title}{CodeIE: Large Code Generation Models are Better Few-Shot Information Extractors}.
\newblock
\newblock
\showeprint[arxiv]{2305.05711}~[cs.CL]


\bibitem[Li et~al\mbox{.}(2023a)]%
        {N38}
\bibfield{author}{\bibinfo{person}{Raymond Li}, \bibinfo{person}{Loubna~Ben Allal}, \bibinfo{person}{Yangtian Zi}, \bibinfo{person}{Niklas Muennighoff}, \bibinfo{person}{Denis Kocetkov}, \bibinfo{person}{Chenghao Mou}, \bibinfo{person}{Marc Marone}, \bibinfo{person}{Christopher Akiki}, \bibinfo{person}{Jia Li}, \bibinfo{person}{Jenny Chim}, \bibinfo{person}{Qian Liu}, \bibinfo{person}{Evgenii Zheltonozhskii}, \bibinfo{person}{Terry~Yue Zhuo}, \bibinfo{person}{Thomas Wang}, \bibinfo{person}{Olivier Dehaene}, \bibinfo{person}{Mishig Davaadorj}, \bibinfo{person}{Joel Lamy-Poirier}, \bibinfo{person}{João Monteiro}, \bibinfo{person}{Oleh Shliazhko}, \bibinfo{person}{Nicolas Gontier}, \bibinfo{person}{Nicholas Meade}, \bibinfo{person}{Armel Zebaze}, \bibinfo{person}{Ming-Ho Yee}, \bibinfo{person}{Logesh~Kumar Umapathi}, \bibinfo{person}{Jian Zhu}, \bibinfo{person}{Benjamin Lipkin}, \bibinfo{person}{Muhtasham Oblokulov}, \bibinfo{person}{Zhiruo Wang}, \bibinfo{person}{Rudra Murthy}, \bibinfo{person}{Jason
  Stillerman}, \bibinfo{person}{Siva~Sankalp Patel}, \bibinfo{person}{Dmitry Abulkhanov}, \bibinfo{person}{Marco Zocca}, \bibinfo{person}{Manan Dey}, \bibinfo{person}{Zhihan Zhang}, \bibinfo{person}{Nour Fahmy}, \bibinfo{person}{Urvashi Bhattacharyya}, \bibinfo{person}{Wenhao Yu}, \bibinfo{person}{Swayam Singh}, \bibinfo{person}{Sasha Luccioni}, \bibinfo{person}{Paulo Villegas}, \bibinfo{person}{Maxim Kunakov}, \bibinfo{person}{Fedor Zhdanov}, \bibinfo{person}{Manuel Romero}, \bibinfo{person}{Tony Lee}, \bibinfo{person}{Nadav Timor}, \bibinfo{person}{Jennifer Ding}, \bibinfo{person}{Claire Schlesinger}, \bibinfo{person}{Hailey Schoelkopf}, \bibinfo{person}{Jan Ebert}, \bibinfo{person}{Tri Dao}, \bibinfo{person}{Mayank Mishra}, \bibinfo{person}{Alex Gu}, \bibinfo{person}{Jennifer Robinson}, \bibinfo{person}{Carolyn~Jane Anderson}, \bibinfo{person}{Brendan Dolan-Gavitt}, \bibinfo{person}{Danish Contractor}, \bibinfo{person}{Siva Reddy}, \bibinfo{person}{Daniel Fried}, \bibinfo{person}{Dzmitry Bahdanau},
  \bibinfo{person}{Yacine Jernite}, \bibinfo{person}{Carlos~Muñoz Ferrandis}, \bibinfo{person}{Sean Hughes}, \bibinfo{person}{Thomas Wolf}, \bibinfo{person}{Arjun Guha}, \bibinfo{person}{Leandro von Werra}, {and} \bibinfo{person}{Harm de Vries}.} \bibinfo{year}{2023}\natexlab{a}.
\newblock \bibinfo{title}{StarCoder: may the source be with you!}
\newblock
\newblock
\showeprint[arxiv]{2305.06161}~[cs.CL]


\bibitem[Li et~al\mbox{.}(2023f)]%
        {N144}
\bibfield{author}{\bibinfo{person}{Tsz-On Li}, \bibinfo{person}{Wenxi Zong}, \bibinfo{person}{Yibo Wang}, \bibinfo{person}{Haoye Tian}, \bibinfo{person}{Ying Wang}, \bibinfo{person}{Shing-Chi Cheung}, {and} \bibinfo{person}{Jeff Kramer}.} \bibinfo{year}{2023}\natexlab{f}.
\newblock \bibinfo{title}{Finding Failure-Inducing Test Cases with ChatGPT}.
\newblock
\newblock
\showeprint[arxiv]{2304.11686}~[cs.SE]


\bibitem[Li et~al\mbox{.}(2023b)]%
        {N99}
\bibfield{author}{\bibinfo{person}{Yuanzhi Li}, \bibinfo{person}{Sébastien Bubeck}, \bibinfo{person}{Ronen Eldan}, \bibinfo{person}{Allie~Del Giorno}, \bibinfo{person}{Suriya Gunasekar}, {and} \bibinfo{person}{Yin~Tat Lee}.} \bibinfo{year}{2023}\natexlab{b}.
\newblock \bibinfo{title}{Textbooks Are All You Need II: phi-1.5 technical report}.
\newblock
\newblock
\showeprint[arxiv]{2309.05463}~[cs.CL]


\bibitem[Li et~al\mbox{.}(2022)]%
        {N83}
\bibfield{author}{\bibinfo{person}{Yujia Li}, \bibinfo{person}{David Choi}, \bibinfo{person}{Junyoung Chung}, \bibinfo{person}{Nate Kushman}, \bibinfo{person}{Julian Schrittwieser}, {et~al\mbox{.}}} \bibinfo{year}{2022}\natexlab{}.
\newblock \showarticletitle{Competition-level code generation with AlphaCode}.
\newblock \bibinfo{journal}{\emph{Science}} \bibinfo{volume}{378}, \bibinfo{number}{6624} (\bibinfo{year}{2022}), \bibinfo{pages}{1092--1097}.
\newblock
\urldef\tempurl%
\url{https://doi.org/10.1126/science.abq1158}
\showDOI{\tempurl}
\showeprint{https://www.science.org/doi/pdf/10.1126/science.abq1158}


\bibitem[Liu et~al\mbox{.}(2019)]%
        {N104}
\bibfield{author}{\bibinfo{person}{Yinhan Liu}, \bibinfo{person}{Myle Ott}, \bibinfo{person}{Naman Goyal}, \bibinfo{person}{Jingfei Du}, \bibinfo{person}{Mandar Joshi}, \bibinfo{person}{Danqi Chen}, \bibinfo{person}{Omer Levy}, \bibinfo{person}{Mike Lewis}, \bibinfo{person}{Luke Zettlemoyer}, {and} \bibinfo{person}{Veselin Stoyanov}.} \bibinfo{year}{2019}\natexlab{}.
\newblock \showarticletitle{RoBERTa: {A} Robustly Optimized {BERT} Pretraining Approach}.
\newblock \bibinfo{journal}{\emph{CoRR}}  \bibinfo{volume}{abs/1907.11692} (\bibinfo{year}{2019}).
\newblock
\showeprint[arXiv]{1907.11692}
\urldef\tempurl%
\url{http://arxiv.org/abs/1907.11692}
\showURL{%
\tempurl}


\bibitem[Lu et~al\mbox{.}(2023)]%
        {N54}
\bibfield{author}{\bibinfo{person}{Junyi Lu}, \bibinfo{person}{Lei Yu}, \bibinfo{person}{Xiaojia Li}, \bibinfo{person}{Li Yang}, {and} \bibinfo{person}{Chun Zuo}.} \bibinfo{year}{2023}\natexlab{}.
\newblock \bibinfo{title}{LLaMA-Reviewer: Advancing Code Review Automation with Large Language Models through Parameter-Efficient Fine-Tuning}.
\newblock
\newblock
\showeprint[arxiv]{2308.11148}


\bibitem[Lu et~al\mbox{.}(2021)]%
        {N58}
\bibfield{author}{\bibinfo{person}{Shuai Lu}, \bibinfo{person}{Daya Guo}, \bibinfo{person}{Shuo Ren}, \bibinfo{person}{Junjie Huang}, {et~al\mbox{.}}} \bibinfo{year}{2021}\natexlab{}.
\newblock \showarticletitle{CodeXGLUE: {A} Machine Learning Benchmark Dataset for Code Understanding and Generation}. In \bibinfo{booktitle}{\emph{Proceedings of the Neural Information Processing Systems Track on Datasets and Benchmarks 1, NeurIPS Datasets and Benchmarks 2021, December 2021, virtual}}, \bibfield{editor}{\bibinfo{person}{Joaquin Vanschoren} {and} \bibinfo{person}{Sai{-}Kit Yeung}} (Eds.).
\newblock
\urldef\tempurl%
\url{https://datasets-benchmarks-proceedings.neurips.cc}
\showURL{%
\tempurl}


\bibitem[Luo et~al\mbox{.}(2023)]%
        {N97}
\bibfield{author}{\bibinfo{person}{Ziyang Luo}, \bibinfo{person}{Can Xu}, \bibinfo{person}{Pu Zhao}, \bibinfo{person}{Qingfeng Sun}, \bibinfo{person}{Xiubo Geng}, \bibinfo{person}{Wenxiang Hu}, \bibinfo{person}{Chongyang Tao}, \bibinfo{person}{Jing Ma}, \bibinfo{person}{Qingwei Lin}, {and} \bibinfo{person}{Daxin Jiang}.} \bibinfo{year}{2023}\natexlab{}.
\newblock \showarticletitle{WizardCoder: Empowering Code Large Language Models with Evol-Instruct}.
\newblock \bibinfo{journal}{\emph{CoRR}}  \bibinfo{volume}{abs/2306.08568} (\bibinfo{year}{2023}).
\newblock
\urldef\tempurl%
\url{https://doi.org/10.48550/arXiv.2306.08568}
\showDOI{\tempurl}
\showeprint[arXiv]{2306.08568}


\bibitem[Luong et~al\mbox{.}(2015)]%
        {N105}
\bibfield{author}{\bibinfo{person}{Thang Luong}, \bibinfo{person}{Hieu Pham}, {and} \bibinfo{person}{Christopher~D. Manning}.} \bibinfo{year}{2015}\natexlab{}.
\newblock \showarticletitle{Effective Approaches to Attention-based Neural Machine Translation}. In \bibinfo{booktitle}{\emph{Proceedings of the 2015 Conference on Empirical Methods in Natural Language Processing}}. \bibinfo{publisher}{Association for Computational Linguistics}, \bibinfo{address}{Lisbon, Portugal}, \bibinfo{pages}{1412--1421}.
\newblock
\urldef\tempurl%
\url{https://doi.org/10.18653/v1/D15-1166}
\showDOI{\tempurl}


\bibitem[Ma et~al\mbox{.}(2023)]%
        {N147}
\bibfield{author}{\bibinfo{person}{Wei Ma}, \bibinfo{person}{Shangqing Liu}, \bibinfo{person}{Wenhan Wang}, \bibinfo{person}{Qiang Hu}, \bibinfo{person}{Ye Liu}, \bibinfo{person}{Cen Zhang}, \bibinfo{person}{Liming Nie}, {and} \bibinfo{person}{Yang Liu}.} \bibinfo{year}{2023}\natexlab{}.
\newblock \bibinfo{title}{The Scope of ChatGPT in Software Engineering: A Thorough Investigation}.
\newblock
\newblock
\showeprint[arxiv]{2305.12138}~[cs.SE]


\bibitem[Maddigan and Susnjak(2023)]%
        {N5}
\bibfield{author}{\bibinfo{person}{Paula Maddigan} {and} \bibinfo{person}{Teo Susnjak}.} \bibinfo{year}{2023}\natexlab{}.
\newblock \bibinfo{title}{Chat2VIS: Generating Data Visualisations via Natural Language using ChatGPT, Codex and GPT-3 Large Language Models}.
\newblock
\newblock
\showeprint[arxiv]{2302.02094}~[cs.HC]


\bibitem[Martínez et~al\mbox{.}(2023)]%
        {N25}
\bibfield{author}{\bibinfo{person}{Pablo~Antonio Martínez}, \bibinfo{person}{Gregorio Bernabé}, {and} \bibinfo{person}{José~Manuel García}.} \bibinfo{year}{2023}\natexlab{}.
\newblock \bibinfo{title}{Code Detection for Hardware Acceleration Using Large Language Models}.
\newblock
\newblock
\showeprint[arxiv]{2307.10348}~[cs.SE]


\bibitem[MFTCoder(2023)]%
        {N113}
\bibfield{author}{\bibinfo{person}{MFTCoder}.} \bibinfo{year}{2023}\natexlab{}.
\newblock \bibinfo{booktitle}{\emph{{CodeFuse-MFTCoder: Multitask Fine-Tuned Code LLMs}}}.
\newblock
\urldef\tempurl%
\url{https://github.com/codefuse-ai/MFTCoder}
\showURL{%
\tempurl}


\bibitem[MOI et~al\mbox{.}(2022)]%
        {N109}
\bibfield{author}{\bibinfo{person}{Anthony MOI}, \bibinfo{person}{Nicolas Patry}, \bibinfo{person}{Pierric Cistac}, \bibinfo{person}{Pete}, {et~al\mbox{.}}} \bibinfo{year}{2022}\natexlab{}.
\newblock \bibinfo{booktitle}{\emph{huggingface/tokenizers: Rust 0.13.2}}.
\newblock
\urldef\tempurl%
\url{https://doi.org/10.5281/zenodo.7298413}
\showDOI{\tempurl}


\bibitem[Muennighoff et~al\mbox{.}(2023)]%
        {N18}
\bibfield{author}{\bibinfo{person}{Niklas Muennighoff}, \bibinfo{person}{Qian Liu}, \bibinfo{person}{Armel Zebaze}, \bibinfo{person}{Qinkai Zheng}, \bibinfo{person}{Binyuan Hui}, \bibinfo{person}{Terry~Yue Zhuo}, \bibinfo{person}{Swayam Singh}, \bibinfo{person}{Xiangru Tang}, \bibinfo{person}{Leandro von Werra}, {and} \bibinfo{person}{Shayne Longpre}.} \bibinfo{year}{2023}\natexlab{}.
\newblock \bibinfo{title}{OctoPack: Instruction Tuning Code Large Language Models}.
\newblock
\newblock
\showeprint[arxiv]{2308.07124}~[cs.CL]


\bibitem[Nathan~Coooper et~al\mbox{.}(2022)]%
        {N78}
\bibfield{author}{\bibinfo{person}{Artashes~Arutiunian Nathan~Coooper} {et~al\mbox{.}}} \bibinfo{year}{2022}\natexlab{}.
\newblock \bibinfo{title}{{Code Clippy Data: A large dataset of code data from Github for research into code language models}}.
\newblock
\newblock
\urldef\tempurl%
\url{https://github.com/ncoop57/gpt-code-clippy}
\showURL{%
\tempurl}


\bibitem[Nijkamp et~al\mbox{.}(2023a)]%
        {N133}
\bibfield{author}{\bibinfo{person}{Erik Nijkamp}, \bibinfo{person}{Hiroaki Hayashi}, \bibinfo{person}{Caiming Xiong}, \bibinfo{person}{Silvio Savarese}, {and} \bibinfo{person}{Yingbo Zhou}.} \bibinfo{year}{2023}\natexlab{a}.
\newblock \bibinfo{title}{CodeGen2: Lessons for Training LLMs on Programming and Natural Languages}.
\newblock
\newblock
\showeprint[arxiv]{2305.02309}~[cs.LG]


\bibitem[Nijkamp et~al\mbox{.}(2023b)]%
        {N76}
\bibfield{author}{\bibinfo{person}{Erik Nijkamp}, \bibinfo{person}{Bo Pang}, \bibinfo{person}{Hiroaki Hayashi}, \bibinfo{person}{Lifu Tu}, \bibinfo{person}{Huan Wang}, \bibinfo{person}{Yingbo Zhou}, \bibinfo{person}{Silvio Savarese}, {and} \bibinfo{person}{Caiming Xiong}.} \bibinfo{year}{2023}\natexlab{b}.
\newblock \showarticletitle{CodeGen: An Open Large Language Model for Code with Multi-Turn Program Synthesis}. In \bibinfo{booktitle}{\emph{The Eleventh International Conference on Learning Representations, {ICLR} 2023, Kigali, Rwanda, May 1-5, 2023}}. \bibinfo{publisher}{OpenReview.net}.
\newblock
\urldef\tempurl%
\url{https://openreview.net/pdf?id=iaYcJKpY2B\_}
\showURL{%
\tempurl}


\bibitem[Palacio et~al\mbox{.}(2023)]%
        {N125}
\bibfield{author}{\bibinfo{person}{David~N Palacio}, \bibinfo{person}{Alejandro Velasco}, \bibinfo{person}{Daniel Rodriguez-Cardenas}, \bibinfo{person}{Kevin Moran}, {and} \bibinfo{person}{Denys Poshyvanyk}.} \bibinfo{year}{2023}\natexlab{}.
\newblock \bibinfo{title}{Evaluating and Explaining Large Language Models for Code Using Syntactic Structures}.
\newblock
\newblock
\showeprint[arxiv]{2308.03873}~[cs.SE]


\bibitem[Pan et~al\mbox{.}(2023)]%
        {N126}
\bibfield{author}{\bibinfo{person}{Rangeet Pan}, \bibinfo{person}{Ali~Reza Ibrahimzada}, \bibinfo{person}{Rahul Krishna}, \bibinfo{person}{Divya Sankar}, \bibinfo{person}{Lambert~Pouguem Wassi}, \bibinfo{person}{Michele Merler}, \bibinfo{person}{Boris Sobolev}, \bibinfo{person}{Raju Pavuluri}, \bibinfo{person}{Saurabh Sinha}, {and} \bibinfo{person}{Reyhaneh Jabbarvand}.} \bibinfo{year}{2023}\natexlab{}.
\newblock \bibinfo{title}{Understanding the Effectiveness of Large Language Models in Code Translation}.
\newblock
\newblock
\showeprint[arxiv]{2308.03109}~[cs.SE]


\bibitem[Pearce et~al\mbox{.}(2022)]%
        {N145}
\bibfield{author}{\bibinfo{person}{Hammond Pearce}, \bibinfo{person}{Benjamin Tan}, \bibinfo{person}{Baleegh Ahmad}, \bibinfo{person}{Ramesh Karri}, {and} \bibinfo{person}{Brendan Dolan-Gavitt}.} \bibinfo{year}{2022}\natexlab{}.
\newblock \bibinfo{title}{Examining Zero-Shot Vulnerability Repair with Large Language Models}.
\newblock
\newblock
\showeprint[arxiv]{2112.02125}~[cs.CR]


\bibitem[Pearce et~al\mbox{.}(2023)]%
        {N119}
\bibfield{author}{\bibinfo{person}{Hammond Pearce}, \bibinfo{person}{Benjamin Tan}, \bibinfo{person}{Baleegh Ahmad}, \bibinfo{person}{Ramesh Karri}, {and} \bibinfo{person}{Brendan Dolan{-}Gavitt}.} \bibinfo{year}{2023}\natexlab{}.
\newblock \showarticletitle{Examining Zero-Shot Vulnerability Repair with Large Language Models}. In \bibinfo{booktitle}{\emph{44th {IEEE} Symposium on Security and Privacy, {SP} 2023, San Francisco, CA, USA, May 21-25, 2023}}. \bibinfo{publisher}{{IEEE}}, \bibinfo{pages}{2339--2356}.
\newblock
\urldef\tempurl%
\url{https://doi.org/10.1109/SP46215.2023.10179420}
\showDOI{\tempurl}


\bibitem[Phind(2023)]%
        {N114}
\bibfield{author}{\bibinfo{person}{Phind}.} \bibinfo{year}{2023}\natexlab{}.
\newblock \bibinfo{booktitle}{\emph{{Phind-CodeLlama}}}.
\newblock
\urldef\tempurl%
\url{https://huggingface.co/Phind/Phind-CodeLlama-34B-v1}
\showURL{%
\tempurl}


\bibitem[Raffel et~al\mbox{.}(2020)]%
        {N91}
\bibfield{author}{\bibinfo{person}{Colin Raffel}, \bibinfo{person}{Noam Shazeer}, \bibinfo{person}{Adam Roberts}, \bibinfo{person}{Katherine Lee}, \bibinfo{person}{Sharan Narang}, \bibinfo{person}{Michael Matena}, \bibinfo{person}{Yanqi Zhou}, \bibinfo{person}{Wei Li}, {and} \bibinfo{person}{Peter~J. Liu}.} \bibinfo{year}{2020}\natexlab{}.
\newblock \showarticletitle{Exploring the Limits of Transfer Learning with a Unified Text-to-Text Transformer}.
\newblock \bibinfo{journal}{\emph{J. Mach. Learn. Res.}}  \bibinfo{volume}{21} (\bibinfo{year}{2020}), \bibinfo{pages}{140:1--140:67}.
\newblock
\urldef\tempurl%
\url{http://jmlr.org/papers/v21/20-074.html}
\showURL{%
\tempurl}


\bibitem[Raychev et~al\mbox{.}(2016)]%
        {N69}
\bibfield{author}{\bibinfo{person}{Veselin Raychev}, \bibinfo{person}{Pavol Bielik}, {and} \bibinfo{person}{Martin~T. Vechev}.} \bibinfo{year}{2016}\natexlab{}.
\newblock \showarticletitle{Probabilistic model for code with decision trees}. In \bibinfo{booktitle}{\emph{Proceedings of the 2016 {ACM} {SIGPLAN} International Conference on Object-Oriented Programming, Systems, Languages, and Applications, {OOPSLA} 2016, part of {SPLASH} 2016, Amsterdam, The Netherlands, October 30 - November 4, 2016}}, \bibfield{editor}{\bibinfo{person}{Eelco Visser} {and} \bibinfo{person}{Yannis Smaragdakis}} (Eds.). \bibinfo{publisher}{{ACM}}, \bibinfo{pages}{731--747}.
\newblock
\urldef\tempurl%
\url{https://doi.org/10.1145/2983990.2984041}
\showDOI{\tempurl}


\bibitem[Ren et~al\mbox{.}(2020)]%
        {N103}
\bibfield{author}{\bibinfo{person}{Shuo Ren}, \bibinfo{person}{Daya Guo}, \bibinfo{person}{Shuai Lu}, \bibinfo{person}{Long Zhou}, \bibinfo{person}{Shujie Liu}, \bibinfo{person}{Duyu Tang}, \bibinfo{person}{Neel Sundaresan}, \bibinfo{person}{Ming Zhou}, \bibinfo{person}{Ambrosio Blanco}, {and} \bibinfo{person}{Shuai Ma}.} \bibinfo{year}{2020}\natexlab{}.
\newblock \showarticletitle{CodeBLEU: a Method for Automatic Evaluation of Code Synthesis}.
\newblock \bibinfo{journal}{\emph{CoRR}}  \bibinfo{volume}{abs/2009.10297} (\bibinfo{year}{2020}).
\newblock
\showeprint[arXiv]{2009.10297}
\urldef\tempurl%
\url{https://arxiv.org/abs/2009.10297}
\showURL{%
\tempurl}


\bibitem[replit(2023)]%
        {N116}
\bibfield{author}{\bibinfo{person}{replit}.} \bibinfo{year}{2023}\natexlab{}.
\newblock \bibinfo{booktitle}{\emph{{replit-code-v1-3b}}}.
\newblock
\urldef\tempurl%
\url{https://huggingface.co/replit/replit-code-v1-3b}
\showURL{%
\tempurl}


\bibitem[Rodriguez-Cardenas et~al\mbox{.}(2023)]%
        {N53}
\bibfield{author}{\bibinfo{person}{Daniel Rodriguez-Cardenas}, \bibinfo{person}{David~N. Palacio}, \bibinfo{person}{Dipin Khati}, \bibinfo{person}{Henry Burke}, {and} \bibinfo{person}{Denys Poshyvanyk}.} \bibinfo{year}{2023}\natexlab{}.
\newblock \bibinfo{title}{Benchmarking Causal Study to Interpret Large Language Models for Source Code}.
\newblock
\newblock
\showeprint[arxiv]{2308.12415}~[cs.SE]


\bibitem[Rozière et~al\mbox{.}(2023)]%
        {N52}
\bibfield{author}{\bibinfo{person}{Baptiste Rozière}, \bibinfo{person}{Jonas Gehring}, \bibinfo{person}{Fabian Gloeckle}, \bibinfo{person}{Sten Sootla}, \bibinfo{person}{Itai Gat}, \bibinfo{person}{Xiaoqing~Ellen Tan}, \bibinfo{person}{Yossi Adi}, \bibinfo{person}{Jingyu Liu}, \bibinfo{person}{Tal Remez}, \bibinfo{person}{Jérémy Rapin}, \bibinfo{person}{Artyom Kozhevnikov}, \bibinfo{person}{Ivan Evtimov}, \bibinfo{person}{Joanna Bitton}, \bibinfo{person}{Manish Bhatt}, \bibinfo{person}{Cristian~Canton Ferrer}, \bibinfo{person}{Aaron Grattafiori}, \bibinfo{person}{Wenhan Xiong}, \bibinfo{person}{Alexandre Défossez}, \bibinfo{person}{Jade Copet}, \bibinfo{person}{Faisal Azhar}, \bibinfo{person}{Hugo Touvron}, \bibinfo{person}{Louis Martin}, \bibinfo{person}{Nicolas Usunier}, \bibinfo{person}{Thomas Scialom}, {and} \bibinfo{person}{Gabriel Synnaeve}.} \bibinfo{year}{2023}\natexlab{}.
\newblock \bibinfo{title}{Code Llama: Open Foundation Models for Code}.
\newblock
\newblock
\showeprint[arxiv]{2308.12950}~[cs.CL]


\bibitem[Scao et~al\mbox{.}(2022)]%
        {N88}
\bibfield{author}{\bibinfo{person}{Teven~Le Scao}, \bibinfo{person}{Angela Fan}, \bibinfo{person}{Christopher Akiki}, \bibinfo{person}{Ellie Pavlick}, \bibinfo{person}{Suzana Ili{\'c}}, \bibinfo{person}{Daniel Hesslow}, \bibinfo{person}{Roman Castagn{\'e}}, \bibinfo{person}{Alexandra~Sasha Luccioni}, \bibinfo{person}{Fran{\c{c}}ois Yvon}, \bibinfo{person}{Matthias Gall{\'e}}, {et~al\mbox{.}}} \bibinfo{year}{2022}\natexlab{}.
\newblock \showarticletitle{Bloom: A 176b-parameter open-access multilingual language model}.
\newblock \bibinfo{journal}{\emph{arXiv preprint arXiv:2211.05100}} (\bibinfo{year}{2022}).
\newblock


\bibitem[Schäfer et~al\mbox{.}(2023)]%
        {N3}
\bibfield{author}{\bibinfo{person}{Max Schäfer}, \bibinfo{person}{Sarah Nadi}, \bibinfo{person}{Aryaz Eghbali}, {and} \bibinfo{person}{Frank Tip}.} \bibinfo{year}{2023}\natexlab{}.
\newblock \bibinfo{title}{An Empirical Evaluation of Using Large Language Models for Automated Unit Test Generation}.
\newblock
\newblock
\showeprint[arxiv]{2302.06527}~[cs.SE]


\bibitem[Shen et~al\mbox{.}(2023)]%
        {N23}
\bibfield{author}{\bibinfo{person}{Bo Shen}, \bibinfo{person}{Jiaxin Zhang}, \bibinfo{person}{Taihong Chen}, \bibinfo{person}{Daoguang Zan}, \bibinfo{person}{Bing Geng}, \bibinfo{person}{An Fu}, \bibinfo{person}{Muhan Zeng}, \bibinfo{person}{Ailun Yu}, \bibinfo{person}{Jichuan Ji}, \bibinfo{person}{Jingyang Zhao}, \bibinfo{person}{Yuenan Guo}, {and} \bibinfo{person}{Qianxiang Wang}.} \bibinfo{year}{2023}\natexlab{}.
\newblock \bibinfo{title}{PanGu-Coder2: Boosting Large Language Models for Code with Ranking Feedback}.
\newblock
\newblock
\showeprint[arxiv]{2307.14936}~[cs.CL]


\bibitem[Shirafuji et~al\mbox{.}(2023)]%
        {N12}
\bibfield{author}{\bibinfo{person}{Atsushi Shirafuji}, \bibinfo{person}{Yutaka Watanobe}, \bibinfo{person}{Takumi Ito}, \bibinfo{person}{Makoto Morishita}, \bibinfo{person}{Yuki Nakamura}, \bibinfo{person}{Yusuke Oda}, {and} \bibinfo{person}{Jun Suzuki}.} \bibinfo{year}{2023}\natexlab{}.
\newblock \bibinfo{title}{Exploring the Robustness of Large Language Models for Solving Programming Problems}.
\newblock
\newblock
\showeprint[arxiv]{2306.14583}~[cs.CL]


\bibitem[Siddiq et~al\mbox{.}(2023a)]%
        {N26}
\bibfield{author}{\bibinfo{person}{Mohammed~Latif Siddiq}, \bibinfo{person}{Beatrice Casey}, {and} \bibinfo{person}{Joanna C.~S. Santos}.} \bibinfo{year}{2023}\natexlab{a}.
\newblock \bibinfo{title}{A Lightweight Framework for High-Quality Code Generation}.
\newblock
\newblock
\showeprint[arxiv]{2307.08220}~[cs.SE]


\bibitem[Siddiq et~al\mbox{.}(2023b)]%
        {N43}
\bibfield{author}{\bibinfo{person}{Mohammed~Latif Siddiq}, \bibinfo{person}{Joanna C.~S. Santos}, \bibinfo{person}{Ridwanul~Hasan Tanvir}, \bibinfo{person}{Noshin Ulfat}, \bibinfo{person}{Fahmid~Al Rifat}, {and} \bibinfo{person}{Vinicius~Carvalho Lopes}.} \bibinfo{year}{2023}\natexlab{b}.
\newblock \bibinfo{title}{Exploring the Effectiveness of Large Language Models in Generating Unit Tests}.
\newblock
\newblock
\showeprint[arxiv]{2305.00418}~[cs.SE]


\bibitem[Sridhara et~al\mbox{.}(2023)]%
        {N148}
\bibfield{author}{\bibinfo{person}{Giriprasad Sridhara}, \bibinfo{person}{Ranjani~H. G.}, {and} \bibinfo{person}{Sourav Mazumdar}.} \bibinfo{year}{2023}\natexlab{}.
\newblock \bibinfo{title}{ChatGPT: A Study on its Utility for Ubiquitous Software Engineering Tasks}.
\newblock
\newblock
\showeprint[arxiv]{2305.16837}~[cs.SE]


\bibitem[Sun et~al\mbox{.}(2023a)]%
        {N33}
\bibfield{author}{\bibinfo{person}{Ruoxi Sun}, \bibinfo{person}{Sercan~O. Arik}, \bibinfo{person}{Hootan Nakhost}, \bibinfo{person}{Hanjun Dai}, \bibinfo{person}{Rajarishi Sinha}, \bibinfo{person}{Pengcheng Yin}, {and} \bibinfo{person}{Tomas Pfister}.} \bibinfo{year}{2023}\natexlab{a}.
\newblock \bibinfo{title}{SQL-PaLM: Improved Large Language Model Adaptation for Text-to-SQL}.
\newblock
\newblock
\showeprint[arxiv]{2306.00739}~[cs.CL]


\bibitem[Sun et~al\mbox{.}(2023b)]%
        {N35}
\bibfield{author}{\bibinfo{person}{Weisong Sun}, \bibinfo{person}{Chunrong Fang}, \bibinfo{person}{Yudu You}, \bibinfo{person}{Yun Miao}, \bibinfo{person}{Yi Liu}, \bibinfo{person}{Yuekang Li}, \bibinfo{person}{Gelei Deng}, \bibinfo{person}{Shenghan Huang}, \bibinfo{person}{Yuchen Chen}, \bibinfo{person}{Quanjun Zhang}, \bibinfo{person}{Hanwei Qian}, \bibinfo{person}{Yang Liu}, {and} \bibinfo{person}{Zhenyu Chen}.} \bibinfo{year}{2023}\natexlab{b}.
\newblock \bibinfo{title}{Automatic Code Summarization via ChatGPT: How Far Are We?}
\newblock
\newblock
\showeprint[arxiv]{2305.12865}~[cs.SE]


\bibitem[Svyatkovskiy et~al\mbox{.}(2020)]%
        {N57}
\bibfield{author}{\bibinfo{person}{Alexey Svyatkovskiy}, \bibinfo{person}{Shao~Kun Deng}, \bibinfo{person}{Shengyu Fu}, {and} \bibinfo{person}{Neel Sundaresan}.} \bibinfo{year}{2020}\natexlab{}.
\newblock \showarticletitle{Intellicode compose: Code generation using transformer}. In \bibinfo{booktitle}{\emph{Proceedings of the 28th ACM Joint Meeting on European Software Engineering Conference and Symposium on the Foundations of Software Engineering}}. \bibinfo{pages}{1433--1443}.
\newblock


\bibitem[Tanaka et~al\mbox{.}(2023)]%
        {N17}
\bibfield{author}{\bibinfo{person}{Toma Tanaka}, \bibinfo{person}{Naofumi Emoto}, {and} \bibinfo{person}{Tsukasa Yumibayashi}.} \bibinfo{year}{2023}\natexlab{}.
\newblock \bibinfo{title}{Inductive-bias Learning: Generating Code Models with Large Language Model}.
\newblock
\newblock
\showeprint[arxiv]{2308.09890}~[cs.LG]


\bibitem[Tang et~al\mbox{.}(2023)]%
        {N118}
\bibfield{author}{\bibinfo{person}{Xiangru Tang}, \bibinfo{person}{Bill Qian}, \bibinfo{person}{Rick Gao}, \bibinfo{person}{Jiakang Chen}, \bibinfo{person}{Xinyun Chen}, {and} \bibinfo{person}{Mark Gerstein}.} \bibinfo{year}{2023}\natexlab{}.
\newblock \bibinfo{title}{BioCoder: A Benchmark for Bioinformatics Code Generation with Contextual Pragmatic Knowledge}.
\newblock
\newblock
\showeprint[arxiv]{2308.16458}~[cs.LG]


\bibitem[Thakur et~al\mbox{.}(2023)]%
        {N21}
\bibfield{author}{\bibinfo{person}{Shailja Thakur}, \bibinfo{person}{Baleegh Ahmad}, \bibinfo{person}{Hammond Pearce}, \bibinfo{person}{Benjamin Tan}, \bibinfo{person}{Brendan Dolan-Gavitt}, \bibinfo{person}{Ramesh Karri}, {and} \bibinfo{person}{Siddharth Garg}.} \bibinfo{year}{2023}\natexlab{}.
\newblock \bibinfo{title}{VeriGen: A Large Language Model for Verilog Code Generation}.
\newblock
\newblock
\showeprint[arxiv]{2308.00708}~[cs.PL]


\bibitem[Thoppilan et~al\mbox{.}(2022)]%
        {N80}
\bibfield{author}{\bibinfo{person}{Romal Thoppilan}, \bibinfo{person}{Daniel~De Freitas}, \bibinfo{person}{Jamie Hall}, \bibinfo{person}{Noam Shazeer}, {et~al\mbox{.}}} \bibinfo{year}{2022}\natexlab{}.
\newblock \showarticletitle{LaMDA: Language Models for Dialog Applications}.
\newblock \bibinfo{journal}{\emph{CoRR}}  \bibinfo{volume}{abs/2201.08239} (\bibinfo{year}{2022}).
\newblock
\showeprint[arXiv]{2201.08239}
\urldef\tempurl%
\url{https://arxiv.org/abs/2201.08239}
\showURL{%
\tempurl}


\bibitem[THUDM(2023)]%
        {N115}
\bibfield{author}{\bibinfo{person}{THUDM}.} \bibinfo{year}{2023}\natexlab{}.
\newblock \bibinfo{booktitle}{\emph{{CodeGeeX2: A More Powerful Multilingual Code Generation Model}}}.
\newblock
\urldef\tempurl%
\url{https://github.com/THUDM/CodeGeeX2}
\showURL{%
\tempurl}


\bibitem[Touvron et~al\mbox{.}(2023)]%
        {N94}
\bibfield{author}{\bibinfo{person}{Hugo Touvron}, \bibinfo{person}{Thibaut Lavril}, \bibinfo{person}{Gautier Izacard}, \bibinfo{person}{Xavier Martinet}, \bibinfo{person}{Marie{-}Anne Lachaux}, {et~al\mbox{.}}} \bibinfo{year}{2023}\natexlab{}.
\newblock \showarticletitle{LLaMA: Open and Efficient Foundation Language Models}.
\newblock \bibinfo{journal}{\emph{CoRR}}  \bibinfo{volume}{abs/2302.13971} (\bibinfo{year}{2023}).
\newblock
\urldef\tempurl%
\url{https://doi.org/10.48550/arXiv.2302.13971}
\showDOI{\tempurl}
\showeprint[arXiv]{2302.13971}


\bibitem[Vaswani et~al\mbox{.}(2017)]%
        {N67}
\bibfield{author}{\bibinfo{person}{Ashish Vaswani}, \bibinfo{person}{Noam Shazeer}, \bibinfo{person}{Niki Parmar}, \bibinfo{person}{Jakob Uszkoreit}, \bibinfo{person}{Llion Jones}, \bibinfo{person}{Aidan~N. Gomez}, \bibinfo{person}{Lukasz Kaiser}, {and} \bibinfo{person}{Illia Polosukhin}.} \bibinfo{year}{2017}\natexlab{}.
\newblock \showarticletitle{Attention is All you Need}. In \bibinfo{booktitle}{\emph{Advances in Neural Information Processing Systems 30: Annual Conference on Neural Information Processing Systems 2017, December 4-9, 2017, Long Beach, CA, {USA}}}. \bibinfo{pages}{5998--6008}.
\newblock


\bibitem[Wang et~al\mbox{.}(2023a)]%
        {N101}
\bibfield{author}{\bibinfo{person}{Guan Wang}, \bibinfo{person}{Sijie Cheng}, \bibinfo{person}{Qiying Yu}, {and} \bibinfo{person}{Changling Liu}.} \bibinfo{year}{2023}\natexlab{a}.
\newblock \bibinfo{booktitle}{\emph{{opencoderplus}}}.
\newblock
\urldef\tempurl%
\url{https://huggingface.co/openchat/opencoderplus}
\showURL{%
\tempurl}


\bibitem[Wang et~al\mbox{.}(2022)]%
        {N95}
\bibfield{author}{\bibinfo{person}{Yizhong Wang}, \bibinfo{person}{Yeganeh Kordi}, \bibinfo{person}{Swaroop Mishra}, \bibinfo{person}{Alisa Liu}, \bibinfo{person}{Noah~A. Smith}, \bibinfo{person}{Daniel Khashabi}, {and} \bibinfo{person}{Hannaneh Hajishirzi}.} \bibinfo{year}{2022}\natexlab{}.
\newblock \showarticletitle{Self-Instruct: Aligning Language Model with Self Generated Instructions}.
\newblock \bibinfo{journal}{\emph{CoRR}}  \bibinfo{volume}{abs/2212.10560} (\bibinfo{year}{2022}).
\newblock
\urldef\tempurl%
\url{https://doi.org/10.48550/arXiv.2212.10560}
\showDOI{\tempurl}
\showeprint[arXiv]{2212.10560}


\bibitem[Wang et~al\mbox{.}(2023b)]%
        {N36}
\bibfield{author}{\bibinfo{person}{Yue Wang}, \bibinfo{person}{Hung Le}, \bibinfo{person}{Akhilesh~Deepak Gotmare}, \bibinfo{person}{Nghi D.~Q. Bui}, \bibinfo{person}{Junnan Li}, {and} \bibinfo{person}{Steven C.~H. Hoi}.} \bibinfo{year}{2023}\natexlab{b}.
\newblock \bibinfo{title}{CodeT5+: Open Code Large Language Models for Code Understanding and Generation}.
\newblock
\newblock
\showeprint[arxiv]{2305.07922}~[cs.CL]


\bibitem[Wang et~al\mbox{.}(2023c)]%
        {N89}
\bibfield{author}{\bibinfo{person}{Yue Wang}, \bibinfo{person}{Hung Le}, \bibinfo{person}{Akhilesh~Deepak Gotmare}, \bibinfo{person}{Junnan Li}, {and} \bibinfo{person}{Steven Hoi}.} \bibinfo{year}{2023}\natexlab{c}.
\newblock \bibinfo{title}{CodeT5Mix: A Pretrained Mixture of Encoder-decoder Transformers for Code Understanding and Generation}.
\newblock
\newblock
\urldef\tempurl%
\url{https://openreview.net/forum?id=VPCi3STZcaO}
\showURL{%
\tempurl}


\bibitem[Wang et~al\mbox{.}(2021)]%
        {N75}
\bibfield{author}{\bibinfo{person}{Yue Wang}, \bibinfo{person}{Weishi Wang}, \bibinfo{person}{Shafiq Joty}, {and} \bibinfo{person}{Steven~CH Hoi}.} \bibinfo{year}{2021}\natexlab{}.
\newblock \showarticletitle{Codet5: Identifier-aware unified pre-trained encoder-decoder models for code understanding and generation}.
\newblock \bibinfo{journal}{\emph{arXiv preprint arXiv:2109.00859}} (\bibinfo{year}{2021}).
\newblock


\bibitem[Xia et~al\mbox{.}(2022)]%
        {N123}
\bibfield{author}{\bibinfo{person}{Chunqiu~Steven Xia}, \bibinfo{person}{Yuxiang Wei}, {and} \bibinfo{person}{Lingming Zhang}.} \bibinfo{year}{2022}\natexlab{}.
\newblock \bibinfo{title}{Practical Program Repair in the Era of Large Pre-trained Language Models}.
\newblock
\newblock
\showeprint[arxiv]{2210.14179}~[cs.SE]


\bibitem[Xu et~al\mbox{.}(2023)]%
        {N96}
\bibfield{author}{\bibinfo{person}{Can Xu}, \bibinfo{person}{Qingfeng Sun}, \bibinfo{person}{Kai Zheng}, \bibinfo{person}{Xiubo Geng}, \bibinfo{person}{Pu Zhao}, \bibinfo{person}{Jiazhan Feng}, \bibinfo{person}{Chongyang Tao}, {and} \bibinfo{person}{Daxin Jiang}.} \bibinfo{year}{2023}\natexlab{}.
\newblock \showarticletitle{WizardLM: Empowering Large Language Models to Follow Complex Instructions}.
\newblock \bibinfo{journal}{\emph{CoRR}}  \bibinfo{volume}{abs/2304.12244} (\bibinfo{year}{2023}).
\newblock
\urldef\tempurl%
\url{https://doi.org/10.48550/arXiv.2304.12244}
\showDOI{\tempurl}
\showeprint[arXiv]{2304.12244}


\bibitem[Xu et~al\mbox{.}(2022)]%
        {N82}
\bibfield{author}{\bibinfo{person}{Frank~F. Xu}, \bibinfo{person}{Uri Alon}, \bibinfo{person}{Graham Neubig}, {and} \bibinfo{person}{Vincent~Josua Hellendoorn}.} \bibinfo{year}{2022}\natexlab{}.
\newblock \showarticletitle{A Systematic Evaluation of Large Language Models of Code}. In \bibinfo{booktitle}{\emph{Proceedings of the 6th ACM SIGPLAN International Symposium on Machine Programming}} (San Diego, CA, USA) \emph{(\bibinfo{series}{MAPS 2022})}. \bibinfo{publisher}{Association for Computing Machinery}, \bibinfo{address}{New York, NY, USA}, \bibinfo{pages}{1–10}.
\newblock
\showISBNx{9781450392730}
\urldef\tempurl%
\url{https://doi.org/10.1145/3520312.3534862}
\showDOI{\tempurl}


\bibitem[Yadav et~al\mbox{.}(2023)]%
        {N28}
\bibfield{author}{\bibinfo{person}{Prateek Yadav}, \bibinfo{person}{Qing Sun}, \bibinfo{person}{Hantian Ding}, \bibinfo{person}{Xiaopeng Li}, \bibinfo{person}{Dejiao Zhang}, \bibinfo{person}{Ming Tan}, \bibinfo{person}{Xiaofei Ma}, \bibinfo{person}{Parminder Bhatia}, \bibinfo{person}{Ramesh Nallapati}, \bibinfo{person}{Murali~Krishna Ramanathan}, \bibinfo{person}{Mohit Bansal}, {and} \bibinfo{person}{Bing Xiang}.} \bibinfo{year}{2023}\natexlab{}.
\newblock \bibinfo{title}{Exploring Continual Learning for Code Generation Models}.
\newblock
\newblock
\showeprint[arxiv]{2307.02435}~[cs.LG]


\bibitem[Yang et~al\mbox{.}(2023)]%
        {N15}
\bibfield{author}{\bibinfo{person}{Zhou Yang}, \bibinfo{person}{Zhipeng Zhao}, \bibinfo{person}{Chenyu Wang}, \bibinfo{person}{Jieke Shi}, \bibinfo{person}{Dongsun Kim}, \bibinfo{person}{DongGyun Han}, {and} \bibinfo{person}{David Lo}.} \bibinfo{year}{2023}\natexlab{}.
\newblock \bibinfo{title}{What Do Code Models Memorize? An Empirical Study on Large Language Models of Code}.
\newblock
\newblock
\showeprint[arxiv]{2308.09932}~[cs.SE]


\bibitem[Yu et~al\mbox{.}(2023)]%
        {N130}
\bibfield{author}{\bibinfo{person}{Hao Yu}, \bibinfo{person}{Bo Shen}, \bibinfo{person}{Dezhi Ran}, \bibinfo{person}{Jiaxin Zhang}, \bibinfo{person}{Qi Zhang}, \bibinfo{person}{Yuchi Ma}, \bibinfo{person}{Guangtai Liang}, \bibinfo{person}{Ying Li}, \bibinfo{person}{Tao Xie}, {and} \bibinfo{person}{Qianxiang Wang}.} \bibinfo{year}{2023}\natexlab{}.
\newblock \bibinfo{title}{CoderEval: A Benchmark of Pragmatic Code Generation with Generative Pre-trained Models}.
\newblock
\newblock
\showeprint[arxiv]{2302.00288}~[cs.SE]


\bibitem[Yuan et~al\mbox{.}(2023)]%
        {N19}
\bibfield{author}{\bibinfo{person}{Zhiqiang Yuan}, \bibinfo{person}{Junwei Liu}, \bibinfo{person}{Qiancheng Zi}, \bibinfo{person}{Mingwei Liu}, \bibinfo{person}{Xin Peng}, {and} \bibinfo{person}{Yiling Lou}.} \bibinfo{year}{2023}\natexlab{}.
\newblock \bibinfo{title}{Evaluating Instruction-Tuned Large Language Models on Code Comprehension and Generation}.
\newblock
\newblock
\showeprint[arxiv]{2308.01240}~[cs.CL]


\bibitem[Zan et~al\mbox{.}(2023a)]%
        {N22}
\bibfield{author}{\bibinfo{person}{Daoguang Zan}, \bibinfo{person}{Bei Chen}, \bibinfo{person}{Yongshun Gong}, \bibinfo{person}{Junzhi Cao}, \bibinfo{person}{Fengji Zhang}, \bibinfo{person}{Bingchao Wu}, \bibinfo{person}{Bei Guan}, \bibinfo{person}{Yilong Yin}, {and} \bibinfo{person}{Yongji Wang}.} \bibinfo{year}{2023}\natexlab{a}.
\newblock \bibinfo{title}{Private-Library-Oriented Code Generation with Large Language Models}.
\newblock
\newblock
\showeprint[arxiv]{2307.15370}~[cs.SE]


\bibitem[Zan et~al\mbox{.}(2022)]%
        {N62}
\bibfield{author}{\bibinfo{person}{Daoguang Zan}, \bibinfo{person}{Bei Chen}, \bibinfo{person}{Dejian Yang}, \bibinfo{person}{Zeqi Lin}, {et~al\mbox{.}}} \bibinfo{year}{2022}\natexlab{}.
\newblock \showarticletitle{{CERT:} Continual Pre-training on Sketches for Library-oriented Code Generation}. In \bibinfo{booktitle}{\emph{Proceedings of the Thirty-First International Joint Conference on Artificial Intelligence, {IJCAI} 2022, Vienna, Austria, 23-29 July 2022}}, \bibfield{editor}{\bibinfo{person}{Luc~De Raedt}} (Ed.). \bibinfo{publisher}{ijcai.org}, \bibinfo{pages}{2369--2375}.
\newblock
\urldef\tempurl%
\url{https://doi.org/10.24963/ijcai.2022/329}
\showDOI{\tempurl}


\bibitem[Zan et~al\mbox{.}(2023b)]%
        {N56}
\bibfield{author}{\bibinfo{person}{Daoguang Zan}, \bibinfo{person}{Bei Chen}, \bibinfo{person}{Fengji Zhang}, \bibinfo{person}{Dianjie Lu}, \bibinfo{person}{Bingchao Wu}, \bibinfo{person}{Bei Guan}, \bibinfo{person}{Yongji Wang}, {and} \bibinfo{person}{Jian-Guang Lou}.} \bibinfo{year}{2023}\natexlab{b}.
\newblock \bibinfo{title}{Large Language Models Meet NL2Code: A Survey}.
\newblock
\newblock
\showeprint[arxiv]{2212.09420}~[cs.SE]


\bibitem[Zeng et~al\mbox{.}(2021)]%
        {N87}
\bibfield{author}{\bibinfo{person}{Wei Zeng}, \bibinfo{person}{Xiaozhe Ren}, \bibinfo{person}{Teng Su}, \bibinfo{person}{Hui Wang}, {et~al\mbox{.}}} \bibinfo{year}{2021}\natexlab{}.
\newblock \showarticletitle{PanGu-{\(\alpha\)}: Large-scale Autoregressive Pretrained Chinese Language Models with Auto-parallel Computation}.
\newblock \bibinfo{journal}{\emph{CoRR}}  \bibinfo{volume}{abs/2104.12369} (\bibinfo{year}{2021}).
\newblock
\showeprint[arXiv]{2104.12369}
\urldef\tempurl%
\url{https://arxiv.org/abs/2104.12369}
\showURL{%
\tempurl}


\bibitem[Zhang et~al\mbox{.}(2023a)]%
        {N24}
\bibfield{author}{\bibinfo{person}{Jiyang Zhang}, \bibinfo{person}{Pengyu Nie}, \bibinfo{person}{Junyi~Jessy Li}, {and} \bibinfo{person}{Milos Gligoric}.} \bibinfo{year}{2023}\natexlab{a}.
\newblock \bibinfo{title}{Multilingual Code Co-Evolution Using Large Language Models}.
\newblock
\newblock
\showeprint[arxiv]{2307.14991}~[cs.SE]


\bibitem[Zhang et~al\mbox{.}(2023b)]%
        {N42}
\bibfield{author}{\bibinfo{person}{Kechi Zhang}, \bibinfo{person}{Huangzhao Zhang}, \bibinfo{person}{Ge Li}, \bibinfo{person}{Jia Li}, \bibinfo{person}{Zhuo Li}, {and} \bibinfo{person}{Zhi Jin}.} \bibinfo{year}{2023}\natexlab{b}.
\newblock \bibinfo{title}{ToolCoder: Teach Code Generation Models to use API search tools}.
\newblock
\newblock
\showeprint[arxiv]{2305.04032}~[cs.SE]


\bibitem[Zhang et~al\mbox{.}(2022)]%
        {N141}
\bibfield{author}{\bibinfo{person}{Yuntong Zhang}, \bibinfo{person}{Xiang Gao}, \bibinfo{person}{Gregory~J. Duck}, {and} \bibinfo{person}{Abhik Roychoudhury}.} \bibinfo{year}{2022}\natexlab{}.
\newblock \showarticletitle{Program vulnerability repair via inductive inference}. In \bibinfo{booktitle}{\emph{{ISSTA} '22: 31st {ACM} {SIGSOFT} International Symposium on Software Testing and Analysis, Virtual Event, South Korea, July 18 - 22, 2022}}, \bibfield{editor}{\bibinfo{person}{Sukyoung Ryu} {and} \bibinfo{person}{Yannis Smaragdakis}} (Eds.). \bibinfo{publisher}{{ACM}}, \bibinfo{pages}{691--702}.
\newblock
\urldef\tempurl%
\url{https://doi.org/10.1145/3533767.3534387}
\showDOI{\tempurl}


\bibitem[Zhao et~al\mbox{.}(2023)]%
        {N136}
\bibfield{author}{\bibinfo{person}{Wayne~Xin Zhao}, \bibinfo{person}{Kun Zhou}, \bibinfo{person}{Junyi Li}, \bibinfo{person}{Tianyi Tang}, \bibinfo{person}{Xiaolei Wang}, \bibinfo{person}{Yupeng Hou}, \bibinfo{person}{Yingqian Min}, \bibinfo{person}{Beichen Zhang}, \bibinfo{person}{Junjie Zhang}, \bibinfo{person}{Zican Dong}, {et~al\mbox{.}}} \bibinfo{year}{2023}\natexlab{}.
\newblock \showarticletitle{A survey of large language models}.
\newblock \bibinfo{journal}{\emph{arXiv preprint arXiv:2303.18223}} (\bibinfo{year}{2023}).
\newblock


\bibitem[Zheng et~al\mbox{.}(2023c)]%
        {N51-1}
\bibfield{author}{\bibinfo{person}{Qinkai Zheng}, \bibinfo{person}{Xiao Xia}, \bibinfo{person}{Xu Zou}, \bibinfo{person}{Yuxiao Dong}, \bibinfo{person}{Shan Wang}, \bibinfo{person}{Yufei Xue}, \bibinfo{person}{Zihan Wang}, \bibinfo{person}{Lei Shen}, \bibinfo{person}{Andi Wang}, \bibinfo{person}{Yang Li}, \bibinfo{person}{Teng Su}, \bibinfo{person}{Zhilin Yang}, {and} \bibinfo{person}{Jie Tang}.} \bibinfo{year}{2023}\natexlab{c}.
\newblock \bibinfo{title}{CodeGeeX: A Pre-Trained Model for Code Generation with Multilingual Evaluations on HumanEval-X}.
\newblock
\newblock
\showeprint[arxiv]{2303.17568}~[cs.LG]


\bibitem[Zheng et~al\mbox{.}(2023a)]%
        {N143}
\bibfield{author}{\bibinfo{person}{Zibin Zheng}, \bibinfo{person}{Kaiwen Ning}, \bibinfo{person}{Jiachi Chen}, \bibinfo{person}{Yanlin Wang}, \bibinfo{person}{Wenqing Chen}, \bibinfo{person}{Lianghong Guo}, {and} \bibinfo{person}{Weicheng Wang}.} \bibinfo{year}{2023}\natexlab{a}.
\newblock \showarticletitle{Towards an Understanding of Large Language Models in Software Engineering Tasks}.
\newblock \bibinfo{journal}{\emph{arXiv preprint arXiv:2308.11396}} (\bibinfo{year}{2023}).
\newblock


\bibitem[Zheng et~al\mbox{.}(2023b)]%
        {N154}
\bibfield{author}{\bibinfo{person}{Zibin Zheng}, \bibinfo{person}{Kaiwen Ning}, \bibinfo{person}{Jiachi Chen}, \bibinfo{person}{Yanlin Wang}, \bibinfo{person}{Wenqing Chen}, \bibinfo{person}{Lianghong Guo}, {and} \bibinfo{person}{Weicheng Wang}.} \bibinfo{year}{2023}\natexlab{b}.
\newblock \bibinfo{title}{Towards an Understanding of Large Language Models in Software Engineering Tasks}.
\newblock
\newblock
\showeprint[arxiv]{2308.11396}~[cs.SE]


\bibitem[Zhong and Wang(2023)]%
        {N124}
\bibfield{author}{\bibinfo{person}{Li Zhong} {and} \bibinfo{person}{Zilong Wang}.} \bibinfo{year}{2023}\natexlab{}.
\newblock \bibinfo{title}{A Study on Robustness and Reliability of Large Language Model Code Generation}.
\newblock
\newblock
\showeprint[arxiv]{2308.10335}~[cs.CL]


\bibitem[Zhong et~al\mbox{.}(2022)]%
        {N117}
\bibfield{author}{\bibinfo{person}{Maosheng Zhong}, \bibinfo{person}{Gen Liu}, \bibinfo{person}{Hongwei Li}, \bibinfo{person}{Jiangling Kuang}, \bibinfo{person}{Jinshan Zeng}, {and} \bibinfo{person}{Mingwen Wang}.} \bibinfo{year}{2022}\natexlab{}.
\newblock \bibinfo{title}{CodeGen-Test: An Automatic Code Generation Model Integrating Program Test Information}.
\newblock
\newblock
\showeprint[arxiv]{2202.07612}~[cs.SE]


\bibitem[Zhou et~al\mbox{.}(2023b)]%
        {N152}
\bibfield{author}{\bibinfo{person}{Andy Zhou}, \bibinfo{person}{Kai Yan}, \bibinfo{person}{Michal Shlapentokh-Rothman}, \bibinfo{person}{Haohan Wang}, {and} \bibinfo{person}{Yu-Xiong Wang}.} \bibinfo{year}{2023}\natexlab{b}.
\newblock \bibinfo{title}{Language Agent Tree Search Unifies Reasoning Acting and Planning in Language Models}.
\newblock
\newblock
\showeprint[arxiv]{2310.04406}~[cs.AI]


\bibitem[Zhou et~al\mbox{.}(2023a)]%
        {N121}
\bibfield{author}{\bibinfo{person}{Shuyan Zhou}, \bibinfo{person}{Uri Alon}, \bibinfo{person}{Sumit Agarwal}, {and} \bibinfo{person}{Graham Neubig}.} \bibinfo{year}{2023}\natexlab{a}.
\newblock \bibinfo{title}{CodeBERTScore: Evaluating Code Generation with Pretrained Models of Code}.
\newblock
\newblock
\showeprint[arxiv]{2302.05527}~[cs.SE]


\bibitem[Zhuo(2023a)]%
        {N139}
\bibfield{author}{\bibinfo{person}{Terry~Yue Zhuo}.} \bibinfo{year}{2023}\natexlab{a}.
\newblock \bibinfo{title}{Large Language Models Are State-of-the-Art Evaluators of Code Generation}.
\newblock
\newblock
\showeprint[arxiv]{2304.14317}~[cs.AI]


\bibitem[Zhuo(2023b)]%
        {N44}
\bibfield{author}{\bibinfo{person}{Terry~Yue Zhuo}.} \bibinfo{year}{2023}\natexlab{b}.
\newblock \bibinfo{title}{Large Language Models Are State-of-the-Art Evaluators of Code Generation}.
\newblock
\newblock
\showeprint[arxiv]{2304.14317}~[cs.AI]


\bibitem[Zhuo et~al\mbox{.}(2023)]%
        {N146}
\bibfield{author}{\bibinfo{person}{Terry~Yue Zhuo}, \bibinfo{person}{Zhuang Li}, \bibinfo{person}{Yujin Huang}, \bibinfo{person}{Fatemeh Shiri}, \bibinfo{person}{Weiqing Wang}, \bibinfo{person}{Gholamreza Haffari}, {and} \bibinfo{person}{Yuan-Fang Li}.} \bibinfo{year}{2023}\natexlab{}.
\newblock \showarticletitle{On Robustness of Prompt-based Semantic Parsing with Large Pre-trained Language Model: An Empirical Study on Codex}. In \bibinfo{booktitle}{\emph{Proceedings of the 17th Conference of the European Chapter of the Association for Computational Linguistics}}. \bibinfo{publisher}{Association for Computational Linguistics}, \bibinfo{address}{Dubrovnik, Croatia}, \bibinfo{pages}{1090--1102}.
\newblock


\end{thebibliography}
